\documentclass{article}
\usepackage[utf8]{inputenc}
\usepackage{authblk}
\usepackage{setspace}
\usepackage[margin=1.25in]{geometry}
\usepackage{graphicx}
\graphicspath{ {./figures/} }
\usepackage{subcaption}
\usepackage{amsmath,amssymb,bm}
\usepackage{lineno}
\usepackage{algorithm}
\usepackage{longtable}
\usepackage{booktabs}
\usepackage{diagbox}
\usepackage{array}
\usepackage{caption}
\usepackage{multicol}
\usepackage{multirow}
\usepackage{makecell}
\usepackage{commath}
\usepackage{xcolor,bm}
\usepackage{eucal}
\usepackage[style=ieee, 
citestyle=numeric-comp,
sorting=none]{biblatex}
\title{Bayesian CART models for aggregate \zzz{claim} modeling}
\author[1]{Yaojun Zhang 
\thanks{zyaojun@swufe.edu.cn}}
\author[2]{Lanpeng Ji 
\thanks{l.ji@leeds.ac.uk}}
\author[2]{Georgios Aivaliotis 
\thanks{g.aivaliotis@leeds.ac.uk}}
\author[2]{Charles C. Taylor 
\thanks{c.c.taylor@leeds.ac.uk}}
\affil[1]{\footnotesize School of Statistics, Southwestern University of Finance and Economics}
\affil[2]{\footnotesize Department of Statistics, University of Leeds}
\date{}

\newtheorem{remark}{Remark}

\newcommand{\COM}[1]{}

\newcommand{\vk}[1]{\bm{#1}}
\def\mT{\CMcal{T}}

\def\P{\mathbb{P}}
\def\E{\mathbb{E}}

\newcommand{\ct}[1]{\textcolor{black}{#1}}
\newcommand{\yz}[1]{\textcolor{black}{#1}}
\newcommand{\zz}[1]{\textcolor{black}{#1}}
\newcommand{\ji}[1]{\textcolor{black}{#1}}
\newcommand{\lj}[1]{\textcolor{black}{#1}}
\newcommand{\yao}[1]{\textcolor{black}{#1}}
\newcommand{\lji}[1]{\textcolor{black}{#1}}
\newcommand{\llj}[1]{\textcolor{black}{#1}}
\newcommand{\ljj}[1]{\textcolor{black}{#1}}

\newcommand{\jj}[1]{\textcolor{black}{#1}}
\newcommand{\zzz}[1]{\textcolor{black}{#1}}

\newcommand{\jjj}[1]{\textcolor{black}{#1}}
\newcommand{\zyj}[1]{\textcolor{black}{#1}}
\newcommand{\jlp}[1]{\textcolor{black}{#1}}
\newcommand{\yyy}[1]{\textcolor{black}{#1}}
\newcommand{\zyy}[1]{\textcolor{black}{#1}}
\newcommand{\llji}[1]{\textcolor{black}{#1}}
\newcommand{\ga}[1]{\textcolor{black}{#1}}

\allowdisplaybreaks[4]

\begin{document}

\maketitle

%%%%%% Abstract %%%%%%
\begin{abstract}
\noindent 
This paper proposes three types of Bayesian CART (or BCART) models for aggregate \zzz{claim} amount, namely, frequency-severity models, sequential models and joint models. We propose a general framework for BCART models applicable to data with multivariate responses, which is particularly useful for the joint BCART models \yyy{with} a bivariate response: the number of claims and the aggregate claim amount. To facilitate frequency-severity modeling, we investigate BCART models for the right-skewed and heavy-tailed claim severity data using various distributions. 
We discover that the Weibull distribution is superior to gamma and lognormal distributions, \ga{due} to its ability to capture different tail characteristics in tree models. \ct{Additionally,} we find that sequential BCART models and joint BCART models, which can incorporate more complex dependence between the number of claims and severity, are beneficial and thus preferable to the frequency-severity BCART models in which %\ga{delete some} some 
%independence between the number of claims and \zzz{average} severity is assumed 
independence is commonly assumed. The effectiveness of these models' performance is illustrated by carefully designed simulations and real insurance data.
%Specifically, in joint BCART models, we implement BCART models for the zero-inflated Compound Poisson Gamma (ZICPG) distribution, by using one joint tree for bivariate response modelling. One joint tree offers advantages in interpretability and enables information sharing between frequency and severity, as opposed to using two separate trees. \zzz{We also make the deviance information criterion (DIC) adaptable to more scenarios for tree model selection.} %We also develop another new type of deviance information criterion (DIC) for tree model selection. 

%\zzz{write joint partition and ARI here?}

\medskip
\noindent
\textbf{Keywords:} %\ct{OMIT keywords which appear in the title: Bayesian CART; aggregate claim amount;} 
\jj{average severity}; dependence; DIC; information sharing; zero-inflated compound Poisson gamma distribution.

\noindent
%\zzz{Question: there is no DIC, ZICPG and information sharing mentioned in the abstract, and maybe delete these words. LJ- I think it's good to keep them here.}
\end{abstract}

%%%%%% Main Text %%%%%%

\section{Introduction}
%%%%%%%%%%%%%%%%%%%%%%%%%old introduction %%%%%%%%%%%%%%%%%

\zyy{Aggregate claim amount estimation in non-life insurance is an important task of actuaries for  the calculation of premiums, pricing of insurance contracts and management of risk. A common model for aggregate claim amount is the so-called {\em collective risk model} defined as
 \begin{equation}\label{eq:NY}
S=\left\{
 \begin{array}{ll}
  \sum_{i=1}^N Y_i, & \hbox{if \ $N>0$} \\
0, & \hbox{if \ $ N=0$}
  \end{array},
\right.
\end{equation}
where $N$ is a nonnegative integer valued random variable for the {\em number of claims} and $\{Y_i\}_{i=1}^\infty$ are positive random variables for the {\em individual severity}. Note that the above aggregate claim amount can also be expressed as 
 \begin{equation}\label{eq:NS}
 S=N\bar S, \ \ \ \ \text{with }\ \ \bar S=\left\{
 \begin{array}{ll}
  \frac{1}{N}\sum_{i=1}^N Y_i, & \hbox{if \ $N>0$} \\
0, & \hbox{if \ $ N=0$}
  \end{array},
\right.
\end{equation}
where $\bar S$ is the so-called {\em average severity}. Following the \ga{convention} in \cite{oh2021copula}, we call $(N, Y_1,\ldots,Y_N)$ micro-level data and  $(N, \bar S)$ (or $(N,  S)$) summarized data.}

\zyy{The classical theory in actuarial science is based on the full independence assumption \ga{between} the number of claims and the individual severities, i.e., $N$ is independent of $\{Y_i\}_{i=1}^\infty$ which is a sequence of independent and identically distributed (IID) random variables.} 
%With the assumed independence, 
In a regression setting given risk factors (or covariates), the {\em frequency-severity models} treat these two components separately by using generalized linear models (GLMs),
 assuming distributions from the exponential family; see e.g., \cite{ohlsson2010non}. 
 The frequency study focuses on the occurrences of claims, and the severity study --- provided that a claim has occurred --- investigates the claim amount.  \zyy{In the current literature, the severity component has been discussed from two perspectives:
 by modeling individual severity $Y_i$ or by modeling the average severity $\bar S$, depending on the format  of   data  available (micro-level or summarized).} Under the full independence assumptions, the expected aggregate claim (or {\em pure premium}) can  be determined by multiplying the estimated frequency with the estimated  severity or conditional average severity. %, due to the following fact:
%\begin{eqnarray}
%    \E(S) =\E(N)\E(Y_1)= \E(N) \E(\bar S | N>0).\label{eq:FS-0}
%\end{eqnarray}

In recent years, a growing body of literature emphasizes the importance of understanding the interrelated nature of claim occurrences and their associated claim amounts to improve model applicability. \zyy{Since the scope of the current paper is on the regression setting, hereafter we  review some of the relevant literature. We refer to e.g., \cite{cossette2019collective, blier2024collective}  for recent developments on this topic under a non-regression setting where different aspects of the distribution of the aggregate claim $S$ are investigated. Research has shown that ignoring the dependence between number of claims and claim severities may lead to serious bias in inference and thus the evaluation of risks; see \cite{oh2021copula, shi2020regression} and references therein. There are mainly three strategies to address the dependence between (i) the number of claims and the individual severities (including dependence among individual severities) and  (ii) the number of claims and the average severity; namely, using a coupla, using shared random effects and using a two-part model where the severity component directly depends on the frequency.}

\zyy{Copula models for both micro-level data and summarized data have been discussed;  see, e.g., \cite{ahn2021copula, shi2015dependent, czado2012mixed, kramer2013total, shi2020regression, oh2021copula}. Estimation of parameters in these models is usually obtained using a likelihood-based method which requires a tractable formulation of the likelihood. Working with micro-level data, \cite{oh2021copula} models the full dependence of number of claims and individual severities  using multivariate Gaussian and $t$-copula functions which is also extended to include a vine copula. In \cite{shi2020regression}, the authors  assume conditional independence of the individual severities given the number of claims so that they only require a single bivariate copula to construct the likelihood function, by doing so a general class of copulas can be adopted and incomplete data due to censoring or truncation can also be accounted for. Working with summarized data, \cite{czado2012mixed, kramer2013total, shi2015dependent} analyse different copula models for the number of claims and average severity, where the challenge persists in selecting the appropriate copula family. As noticed in \cite{shi2015dependent, shi2020regression}, for regression copula models, the association between the frequency and severity is introduced by both the covariates and the copula used; the dependence introduced by copula (interpreted as {\em residual dependence}) should be seen as an extra layer of association in addition to that introduced by the covariates. It has been noted that in all these copula models the dependence parameter in the copula is assumed to be constant and does not vary across covariates (i.e., {\em portfolio-level dependence}). To overcome this drawback, a regression approach using shared random effects is introduced in \cite{baumgartner2015bayesian} where the dependence between the number of claims and average severity is induced by shared random effects. By analyzing a German car insurance portfolio, they show that the proposed shared random effects model can reflect the varying dependence characteristics across different geographical regions.} %However, the linear additive structure of the regression approach makes it hard to incorporate variable selection and detect interaction between covariates.

%theoretical studies on the properties of the models, density of the joint distribution and the calculation of E(S) and var(S). In the regression setting, classical regression is used for zero, frequency and severity. Weakness, sometimes likelihood is hardly simple, maximum likelihood method or other method (computationally expensive), hard to add interaction of covariates, the dependence modeled are residual dependence (not interpreted by covariates).  the shared random effect model in ... [see Oh,AhnLee 2021; review Baumgartner etal 2015]. 

The two-part models enable the severity component to depend on the frequency component more explicitly. Specifically, the number of claims is introduced as a covariate in the (average) severity modeling; see, e.g., \cite{frees2011predicting, garrido2016generalized,gschlossl2007spatial, shi2015dependent}. \zyy{It is noted that when dealing with the average severity, \cite{garrido2016generalized,gschlossl2007spatial} include the number of claims as weights as well as covariate\zyy{s} assuming the individual severities are conditionally IID and from the exponential family. Whereas, \cite{shi2015dependent} imposes a generalized gamma distribution for the average severity where the number of claims is only included as \zyy{a} covariate. % for its location parameter.  In both cases, there is no independence between the number of claims and average severity. 
%makes it hard to incorporate variable selection and detect interaction between covariates (including introduced number of claims as covariate), and 
Under the classical Poisson-gamma risk model with full independence, there is an intrinsic dependence between the number of claims %$N$ 
and the average severity, %$\bar S$, 
which can be difficultly modeled by any copula as illustrated in \cite{oh2021copula}
using a simulation example. \zyy{This difficulty} %which
may suggest a preference for simple two-part models for  summarized data. It is also noticed that in these two-part models the dependence introduced by the coefficient of the number of claims for the severity component can only interpret a portfolio-level dependence which has the same issue as pointed out above for the current copula models.}

As a comparable alternative, the aggregate claim amount can be directly modeled using Tweedie's %model
distribution which assumes a Poisson sum of gamma variables for the aggregate claim amount. This modeling approach simplifies the analysis by accommodating discrete claim numbers and continuous claim amounts in one distribution; see, e.g., \cite{jorgensen1994fitting}. Concurrently, discussions regarding the suitability of GLMs for aggregate claim amount analysis have focused on the trade-off between model complexity and predictive performance, emphasizing the benefits and contexts where Tweedie's model excels and where alternative methodologies may be better; see, e.g., \cite{quijano2015generalised, delong2021making}. Particularly, \cite{delong2021making} has investigated two parametrization approaches and provided theoretical evidence supporting the industry preference for the the Poisson-gamma parametrization over the Tweedie’s compound Poisson parametrization.
%\jj{Recently, a novel approach  is introduced to reduce computational costs for Tweedie's parameter estimation within GLMs, %and evidence is presented to support the industry preference,  as seen in \cite{delong2021making}.} 
%\zyj{In a related advance,  \cite{gao2024fitting} proposed fitting the Tweedie distribution within GLMs through the Expectation-Maximization (EM) algorithm, which is conceptually analogous to iteratively re-weighted Poisson-gamma modeling on an augmented dataset. This approach simplifies the problem by leveraging expectations of latent variables. Numerical examples indicate that the EM-based method outperforms the traditional likelihood maximization approach, particularly in terms of computational efficiency and accuracy.}
%which directly estimates parameters from observed data using likelihood maximization, the EM algorithm .}  %\jj{!I wonder what the differences are between the EM and GLM..???}
%,} \zzz{highlighting the advantages of the EM algorithm in terms of computational efficiency and accuracy.} %[I am not clear, what is the point here? Do you want to address the advantages of Tweedie's model or the opposite?]
We refer to \cite{zhou2022tweedie, gao2024fitting, lian2023tweedie} for recent developments of Tweedie's model. %\ct{. OMIT the rest: for insurance claim modeling.} 
%While both are considered in the literature, the insurance industry typically favours the separated frequency-severity models. \zyy{can we delete this sentence?} %  Poisson and Gamma parametrization. 

\zyy{The above literature review seems to indicate that almost all the models rely on likelihood-based approaches for parameter estimation, assuming a linear additive structure when incorporating covariates. This  makes it difficult for these models to achieve variable selection and interaction detection  automatically. Moreover, the above discussions show that there is no simple answer to the question on the association/dependence between number of claims and severities, because it can be introduced by the covariates or model dependence or a combination of them.}

More recently, machine learning methods %\ct{ OMIT examples: such as neural networks, regression trees, bagging techniques, random forests and boosting machines} 
have been introduced in the context of insurance by adopting actuarial
loss distributions %\ct{OMIT: in these models} 
to capture the characteristics of insurance claims. We refer to  \cite{blier2020machine, denuit2019effective, WuthrichMerz2022b,WuthrichBuser2022} for recent discussions. Insurance pricing models are heavily regulated and %they 
must meet specific requirements before being deployed in practice, which poses %some 
challenges for most %\ct{OMIT: of the aforementioned} 
machine learning methods. 
As discussed in \cite{henckaerts2021boosting, zhang2024bayesian}, tree-based models are considered appropriate for insurance rate-making due to their transparent nature. \llj{Furthermore, tree-based models are known to be able to capture nonlinearities and complex higher order interactions among covariates, and automatically implement variable selection}. In our previous work  \cite{zhang2024bayesian}, we have demonstrated the superiority of  Bayesian classification and regression tree (BCART) models in \zzz{claim} frequency analysis. In this sequel, \llj{working with summerized data $(N, \bar S)$ (or $(N,S)$}),  we %\ct{OMIT: As a continuation of that work, in this paper, we } 
construct some novel insurance pricing models using BCART for both % good to keep it, as this emphasizes the contribution 
\zzz{average} severity \zzz{and aggregate claim amount}. 

\zzz{Specifically}, inspired by the claim loss models discussed in the literature,
%\ct{OMIT that is, the frequency-severity model, the severity model with the number of claims as a covariate, and Tweedie's model,} 
we introduce and investigate three types of BCART models.  We first discuss a benchmark frequency-severity BCART model, where the number of claims and \zzz{average} severity are modeled separately using BCART. Furthermore, 
%we propose BCART models for insurance claims prediction. Instead of making an ensemble of trees, we look for one \yz{good}  tree, which can improve the prediction ability whilst \ji{ensuring} model transparency, by adopting a Bayesian approach applied to CART.
%Based on these discussions, we introduce the joint models, which utilize Compound Poisson Gamma (CPG) and Zero-Inflated Compound Poisson Gamma (ZICPG) distributions for bivariate response (number of \zzz{claim} and aggregate claim amount) modelling; see, e.g.,   \cite{smyth2002fitting}. In contrast to the frequency-severity and sequential models, joint models construct one joint tree to directly model the aggregate claim amount.
We propose two other types of BCART models, with an aim to better
incorporate the underlying complex association  between the number of claims and average severity. These are sequential BCART models (motivated by \cite{frees2011predicting, garrido2016generalized,gschlossl2007spatial}) % by including $N$ \jj{(or its prediction $\hat N$)} as a covariate when modelling the average severity in \eqref{eq:FS-0}. 
 and joint BCART models (motivated by \cite{jorgensen1994fitting, smyth2002fitting, delong2021making}). %by considering $(N,S)$ as a bivariate response.
 In contrast to the frequency-severity and sequential BCART models which result in two separate trees for the number of claims and \zzz{average} severity, %and thus the aggregate claim amount
 the joint BCART models generate one joint tree for \llj{the bivariate response $(N, S)$ which suggests a joint effect of covariates to both parts.} % aggregate claim amount.
 \zyy{In the real data analysis below, we observe that different cells of the resulting tree partition can exhibit varying (positive or negative) conditional dependencies between the number of claims and average severity, overcoming the drawback of most existing models which are only able to capture a unique portfolio-level dependence; see Remark \ref{Rem:PNCor}. %This finding demonstrates that our proposed tree model effectively captures the influence of covariates on these dependencies, providing a promising direction for further exploration. %[I think that it may not belong here and should not be listed as a main contribution. It may be more appropriately placed in the summary?]
 }

%In previous work (see \cite{zhang2024bayesian}), we demonstrated the superiority of BCART models in \zzz{claim} frequency analysis. In this paper, we extend our previous research by proposing BCART models for \zzz{\jj{average severity} and aggregate claim amount.} %insurance pricing that take into account \ct{this part is not new, is it? special features of insurance data, such as the large number of zeros, the involvement of exposure for \zzz{claim} frequency, and} the right-skewed and heavy-tailed nature of \jj{average severity} \ct{Does severity need to be defined here?}.
The main contributions of this paper are as follows: %\llji{[It would be interesting and necessary to check how/whether our model can overcome some of the drawbacks of other models mentioned above, by providing some evidence from our data analysis]}
\begin{itemize}
  
    %\item To accommodate various scenarios, such as incorporating the data augmentation technique and treating certain parameters as known, we introduce another new type of DIC \ct{needs a reference/explanation?} for the tree model selection; \zzz{not just cite the paper...give more instructions...} \ga{if it is the main contribution we cannot just refer to previous paper} adaptation.
 
    \item We implement BCART models for \jj{average severity} including gamma, lognormal and \jj{Weibull} distributions and for aggregate \zzz{claim} amount including compound \ga{Poisson} gamma (CPG) and zero-inflated compound Poisson gamma (ZICPG) distributions. These are not currently available in any \textsf{R} package. %In particular, we introduce three different ways of incorporating exposure in the ZICPG models, \zzz{which has been validated for its superiority in \cite{zhang2024bayesian}.}
    
    %following \cite{lee2020delta} and \cite{lee2021addressing} \ct{ omit this? who focused on \zzz{claim} frequency modelling using delta boosting}. 
    
     \item  To explore the %potential dependence 
     complex association between the number of claims  and  average severity, we propose novel sequential BCART models that treat the number of \zzz{claims} (or its \zzz{estimate}) as a covariate in average severity modeling. \llji{For these models, the aggregate claim amount cannot be obtained analytically due to the assumed dependence, for which a Monte Carlo method will be used.} The effectiveness 
     %\ct{OMIT: of these models} 
     is illustrated using simulated and real insurance data. %This approach builds on the work of \cite{garrido2016generalized}, who considered this dependence in GLMs.

     %We draw on some ideas from \cite{garrido2016generalized} to address the dependence between the number of \zzz{claim} and claim amount, expanding their application to sequential models. 
 
\item 
We present a general framework for the BCART models applicable for %\ct{for OMIT: to data with} 
multivariate responses, extending the MCMC algorithms discussed in 
\cite{zhang2024bayesian}. %\ct{. OMIT for univariate cases.}  %where a data augmentation may be needed. %In doing so, we follow some ideas in \cite{meng1999seeking, van2001art}. \cite{um2023bayesian}
\ga{To the best of our knowledge,} there have been very few discussions on Bayesian tree models with multivariate responses in the current literature, with the only exception \cite{um2023bayesian}. % \ga{ Delete:as we are aware of}. 
As a particular application, we propose novel joint BCART models with a bivariate response to simultaneously model the number of claims and aggregate claim amount. In doing so, we employ the commonly used distributions such as CPG and ZICPG. The potential advantages of information sharing using one joint tree compared with two separate trees %\ct{OMIT: from the frequency-severity or sequential BCART models} 
are also illustrated by simulated and real \jlp{insurance} data. % (see related discussion in \cite{linero2020semiparametric}).

\item \jj{For the comparison of one joint tree (generated from the joint BCART models) with two separate trees (generated from the frequency-severity or sequential BCART models), we propose some evaluation metrics which %it is %crucial to create a joint partition from the two trees. This involves combining all the cuts from each tree to refine the segmentation of the predictor space, capturing the collective behaviour of these two trees. We refer to \cite{rockova2020posterior} for a detailed discussion. 
involve a combination of trees using an idea of \cite{rockova2020posterior}.} %in and introduce some specific performance measures for testing tree models.
We also propose an application of the adjusted Rand Index (ARI) in assessing the similarity between trees. Although ARI is widely used in cluster analysis, its application to tree comparisons seems to be a novel idea. The use of ARI enhances the understanding of the necessity of information sharing, an aspect not covered in relevant literature; see, e.g., \cite{linero2020semiparametric}.  %\zzz{Is it okay to write ARI in the main contribution directly?}

\end{itemize}

\noindent\textbf{Outline of the rest of the paper}: In Section 2, we briefly review the BCART framework, introducing a more general MCMC algorithm for BCART models with multivariate responses. Section 3 introduces the notation for insurance \zzz{claim} data and investigates three types of BCART models for the aggregate claim amount. %\ct{. OMIT: namely, the frequency-severity models, the sequential models and the joint models.} 
Section 4 develops a performance assessment of the proposed aggregate \zzz{claim} models using simulation examples. In Section 5, we present a detailed analysis of real insurance data using the proposed models. Section 6 concludes the paper. \llji{Some of the technical details are included in an online Supplementary Material.}

\section{Bayesian CART: \ljj{A general framework}} \label{Sec_BCART}

\ljj{The BCART models, as introduced in the seminal papers \cite{chipman1998bayesian, denison1998bayesian}, provide a Bayesian perspective on CART models. In this section, we  give a brief review of the BCART model using a \ga{more} general framework that \jj{applies to multivariate response data; see \cite{zhang2024bayesian} \ct{for the univariate case. %OMIT for detailed discussion under a framework with univariate response variable.
}
}
}

\subsection{\ljj{Data, model and training algorithm}}

Consider a \jj{matrix-form} dataset $(\vk{X},\vk{Y})=\big((\vk{x}_1, \vk{y}_1),(\vk{x}_2, \vk{y}_2), \ldots, (\vk{x}_n, \vk{y}_n)\big)^\top$ with $n$ \ct{indepedent} observations. For the $i$-th observation, $\vk{x}_i=(x_{i1}, x_{i2},\ldots,x_{ip})$ is a vector of $p$ explanatory variables (or covariates) sampled from a space \ji{$\CMcal{X}$,} %=\CMcal{X}_1\times \CMcal{X}_2\times\cdots\times \CMcal{X}_p$, 
 while \ljj{$\vk{y}_i=(y_{i1}, y_{i2},\ldots,y_{iq})$ is a vector of $q$ response variables} sampled from a space $\CMcal{Y}$. \ljj{For the severity (or frequency) modeling, $\CMcal{Y}$ is a space of real positive (or integer) values. For aggregate \zzz{claim} modeling, %\ct{OMIT: we discuss models where} 
 $\CMcal{Y}$ is a space of 2-dimensional vectors with two components: 
 an integer  number of \zzz{claims} and a real valued aggregate \zzz{claim} amount. %\ct{OMIT since included in the first line: Throughout the paper,  observations are assumed to be independent.}
 }
%\ga{Why do you need to define the space X? It is not defined and it might give the impression vector of covariates is not fixed. Ok, I can see what you want to do below.}

%A CART model describes the conditional distribution of the response variable $y_{i}$ given $\vk{x}_i$

A CART has two main components: a binary tree $\CMcal{T}$ with $b$ terminal nodes which induces a partition of the covariate space $\CMcal{X}$, denoted by $\left\{\CMcal{A}_{1}, \ldots, \CMcal{A}_{b}\right\}$, and a parameter \ji{$\vk{\theta}=\left(\vk{\theta}_{1}, \vk{\theta}_{2}, \ldots, \vk{\theta}_{b}\right)$}  which associates the parameter value $\vk{\theta}_t$ with the $t$-th terminal node. Note that here we do not specify the dimension \ji{and range} of the parameter $\vk{\theta}_t$ which should be clear \ct{from the context. %OMIT in the considered context below.
}%\ga{but it looks like $\vk{\theta}$ is a vector, maybe better denoted with set notation?}
%A tree and its associated decision rules induce a partition of the covariate space $\left\{\CMcal{X}_{1}, \ldots, \CMcal{X}_{k}\right\}$, where each element of the partition corresponds to a terminal node in the tree. 
If $\boldsymbol{x_{i}}$ is located in %the partition corresponding to 
the $ t$-th terminal node (i.e., $\vk{x}_i\in \CMcal{A}_t$), then $\vk{y}_{i}$ %$\mid\boldsymbol{x_{i}}$ 
has a (joint) distribution $f\left(\vk{y}_{i}\mid\vk{\theta}_t\right)$, where $f$ represents a parametric family indexed by $\vk{\theta}_t$. By associating observations with the $b$ terminal nodes in the tree $\CMcal{T}$, we can re-order the $n$ observations such that 
$$
(\vk{X},\boldsymbol{Y}) =\big((\vk{X}_1,\boldsymbol{Y}_{1}),(\vk{X}_2,\boldsymbol{Y}_{2}) \ldots, (\vk{X}_b,\boldsymbol{Y}_{b})\big)^\top,
$$
where $\boldsymbol{Y}_{t}=\left(\vk{y}_{t 1}, \ldots \vk{y}_{t n_{t}}\right)^\top$ \jj{is an $n_t\times q$ matrix} with $n_{t}$ denoting the number of observations and $\vk{y}_{ti}$ denoting the $i$-th \ljj{observed response} in the $t$-th terminal node, and $\boldsymbol{X}_t$ is an analogously defined $n_t\times p$ design matrix. %For a CART model, it is typically 
We make the typical assumption that %for a CART model, that is,
conditionally on $(\vk{\theta}, \CMcal{T})$, response variables %within a terminal node 
are independent and identically distributed (IID). 
%\ct{. OMIT the rest since we have already stated they are independent: and they are also independent across terminal nodes.} %In this case, the CART model likelihood will be of the form
\yao{The CART model likelihood \ct{is then %OMIT rest: in this case will take the form
}
}
\begin{equation}\label{17}
p(\boldsymbol{Y} \mid \boldsymbol{X}, \vk{\theta}, \CMcal{T})=\prod_{t=1}^{b} f\left(\boldsymbol{\vk{Y}}_{t} \mid \vk{\theta}_t\right)=\prod_{t=1}^{b} \prod_{i=1}^{n_{t}} f\left(\vk{y}_{ti} \mid \vk{\theta}_t\right).
\end{equation}
%Since a CART model is identified by $(\vk{\theta}, \CMcal{T}),$ a Bayesian analysis of the problem proceeds by specifying a prior distribution 、$p(\vk{\theta}, \CMcal{T}).$ 
%\ct{Omit this sentence?} \lji{It is worth noting that instead of the IID assumption within the terminal nodes more general models can be considered, see, e.g., \cite{chipman2002bayesian}, \cite{ chipman2003bayesian} and the references therein.%despite the fact that we emphasise this case as the first step of the study. More general models might be taken into account at the terminal node such as considering modelling the mean of the response variable using piecewise linear or quadratic functions as opposed to constant functions given the IID assumption.}

\yao{Given $(\vk{\theta}, \CMcal{T})$, a Bayesian analysis involves specifying a prior distribution $p(\vk{\theta}, \CMcal{T})$,} \lji{and inference about $\vk{\theta}$ and $\CMcal{T}$ is based on the joint posterior $p(\vk{\theta}, \CMcal{T}| \vk{Y} , \ljj{\vk{X}})$ using a suitable MCMC algorithm.% outlined below.
} Since $\vk{\theta}$ indexes the parametric model whose dimension depends on the number of terminal nodes of the tree, it is usually convenient to %use 
apply the relationship
%\begin{equation}  % equation not cross-referenced so remove display?
$p(\vk{\theta}, \CMcal{T})=p(\vk{\theta}\mid\CMcal{T}) p(\CMcal{T})$,
%\end{equation}
and specify the tree prior distribution $p(\CMcal{T})$ and the terminal node parameter prior distribution $p(\vk{\theta}\mid\CMcal{T})$, respectively. %This strategy offers the advantage that the choice of prior for $\CMcal{T}$ does not depend on the form of the parametric family indexed by $\vk{\theta}$, and thus the same specification of $\CMcal{T}$ can be used for any type of response variable.
%This strategy, introduced by \cite{george1998}, offers several advantages for Bayesian model selection as outlined in \cite{chipman1998bayesian}. \zyy{can we delete this sentence?}% that the specification of $\CMcal{T}$ can be the same for any type of response variable because it does not depend on the form of the parametric family indexed by $\vk{\theta}$.

\ljj{The prior distribution $p(\CMcal{T})$ has two components: a tree topology and a decision rule for each of the internal %\ct{OMIT: /branch}
nodes. \ct{We follow%OMIT: According to the most popular tree topology generating scheme in
}
\cite{chipman1998bayesian}, \ct{in which}
a draw of the tree is obtained by generating, for each node  at depth $d$ (with $d=0$ for the root node), two child nodes with probability
\begin{equation}\label{eq:qd}
p(d)=\gamma\left(1+d\right)^{-\rho}, 
\end{equation}
where $\gamma\jj{\in(0,1]}, \rho \geq 0$ are parameters %that control
controlling the \yz{structure} and size of the tree. 
This process iterates for $d=0,1, \ldots$ until we reach a depth at which all the nodes
cease growing. After the tree topology is generated, each internal node is associated with a decision rule which will be drawn uniformly among all the possible decision rules for that node. 
We refer to   \cite{zhang2024bayesian} for  detailed discussion on the choice of the prior distribution $p(\CMcal{T})$.}
%\ct{OMIT: ** When choosing %a terminal node parameter prior given tree, $p(\boldsymbol{\theta}\mid\CMcal{T})$, it is %important \yz{vital} to realize that %using \yz{employing} priors that allow for analytical simplification can greatly reduce the computational burden of posterior calculation and exploration. This is especially true for the choice of **} 
\ct{
It is important to choose the form $p(\boldsymbol{\theta}\mid\CMcal{T})$ for which it is possible to analytically margin out $\vk{\theta}$ to obtain the integrated likelihood}
\begin{align}
\label{eq:int_lik_1}
p(\boldsymbol{Y} \mid \boldsymbol{X}, \CMcal{T})&=\int p(\boldsymbol{Y} \mid \boldsymbol{X}, \vk{\theta}, \CMcal{T}) p(\boldsymbol{\theta}\mid\CMcal{T}) d {\vk{\theta}}= \prod_{t=1}^{b}  \int f\left(\vk Y_{t} \mid \vk{\theta}_t\right) p(\vk{\theta}_t) d\vk{\theta}_t\nonumber\\
&=   \prod_{t=1}^{b} \int \prod_{i=1}^{n_{t}} f\left(\vk y_{ti} \mid \vk{\theta}_t\right) p(\vk{\theta}_t) d\vk{\theta}_t,
\end{align}
where in the second equality we assume that conditional on the tree $\CMcal{T}$ with $b$ terminal nodes as above, the parameters $\vk{\theta}_t, t=1,2,\ldots,b$, have IID priors $p(\vk{\theta}_t)$, which is a common assumption. \ljj{Examples where this integration has \yz{a} closed-form expression can be found in, e.g., \cite{chipman1998bayesian, linero2017review}.}

When there is no obvious prior distribution $p(\vk{\theta}_t)$  such that the integration in \eqref{eq:int_lik_1} is of closed-form, particularly, for non-Gaussian distributed data $\vk Y$, \ljj{a data augmentation method is usually utilized %in implementing the MCMC algorithm, 
 in the literature, e.g., \cite{murray2021log, zhang2024bayesian}. \ct{Here,%OMIT In this paper,
 }
 we present a general framework in which apart from including a data augmentation, some components of $\vk \theta_t$ are assumed to be known a priori, but some others are assumed to be unknown. More precisely, we assume $\vk{\theta}_t=(\vk{\theta}_{t,M},\vk{\theta}_{t,B})$, where $\vk{\theta}_{t,M}$ are the parameters that are treated as known and computed using \zzz{Method of Moments Estimation (MME), or Maximum Likelihood Estimation (MLE),} and $\vk{\theta}_{t,B}$ are the unknown parameters that need to be estimated in the Bayesian framework. This newly proposed framework aims to reduce the overall computational time of the algorithm and overcome the difficulty of finding an appropriate prior for some parameters even with the data augmentation (that is why $\vk{\theta}_{t,M}$ is assumed known a priori). 
 %It is thus applicable when it involves not only the data augmentation technique but also includes both known and unknown parameters that need to be estimated using different methods, thereby extending the scope of the application of BCART models, 
 }
Under this framework, we augment the data $\vk{Y}$ by introducing a latent variable $\vk{Z}=(\vk z_1,\vk z_2,\ldots,\vk z_n)^{\jj{\top}}$ so that the integration \jlp{in \eqref{eq:int_lik_2}} is computable for \ga{the} augmented data $(\vk{Y}, \vk{Z})$. The integrated likelihood is given as
\begin{eqnarray*} %\label{eq:int_lik_2_0}
p(\boldsymbol{Y} \mid \boldsymbol{X}, \jj{\vk{\theta}_{M},}\CMcal{T})=
\int    p(\boldsymbol{Y}, \vk{Z} \mid \boldsymbol{X}, \jj{\vk{\theta}_{M},} \CMcal{T})  d\vk{Z},
\end{eqnarray*}
where %\ct{OMIT middle line below to save space??} % the augmented integrated likelihood is
\begin{align} \label{eq:int_lik_2}
    p(\boldsymbol{Y}, \vk{Z} \mid \boldsymbol{X}, \jj{\vk{\theta}_{M},} \CMcal{T})&=\int p(\boldsymbol{Y}, \vk{Z} \mid \boldsymbol{X}, \vk{\theta}_M, \vk{\theta}_B,\CMcal{T}) p(\boldsymbol{\theta}_B\mid\CMcal{T}) d {\vk{\theta}_B}\\
    %&&=\prod_{t=1}^{b}  \int f\left(\vk Y_{t}, \vk{Z}_t \mid \vk{\theta}_{t,M}, \vk{\theta}_{t,B}\right) p(\vk{\theta}_{t,B}) d\vk{\theta}_{t,B} 
    &=   \prod_{t=1}^{b} \int \prod_{i=1}^{n_{t}} f\left(\vk y_{ti}, \vk z_{ti} \mid \vk{\theta}_{t,M}, \vk{\theta}_{t,B}\right) p(\vk{\theta}_{t,B}) d\vk{\theta}_{t,B},\nonumber
\end{align}
with $\vk{Z}_t=(\vk z_{t1},\vk z_{t2},\ldots,\vk z_{tn_t})^{\jj{\top}}$ defined according to the partition of $\CMcal{X}$. %\ct{. OMIT as already stated: and with independence assumed.} %\yz{it is not very clear about what kind of independence; lji-this should be clear from the product form}.
\COM{ %%%%%%%%%%%%%5
Besides, to reduce the overall computational time of the algorithm and avoid the difficulty of finding an appropriate prior for some parameters with corresponding data augmentation (see \cite{zhang2024bayesian}), we shall treat some parameters as known in the Bayesian framework which can be estimated upfront by using, e.g., the Method of Moments Estimation (MME) method or Maximum Likelihood Estimation (MLE) method, and we shall treat other parameters as uncertain and use conjugate priors with corresponding data augmentation. Based on these discussions, we extend Algorithm 1 to a new Algorithm \ref{Alg:NB} to simulate a Markov chain sequence of pairs $(\vk{\theta}^{(1)}, \CMcal{T}^{(1)}),(\vk{\theta}^{(2)}, \CMcal{T}^{(2)}),\ldots,$ starting from the root node. }%%%%%%%%%%%%%%%%

%Examples where this integration has \yz{a} closed-form expression can be found in, e.g., \cite{chipman1998bayesian, linero2017review}, particularly for Gaussian-distributed data $\vk{y}$. When no such priors can be found, we have to resort to the technique of data augmentation (see, e.g., \cite{kindo2016multinomial, linero2020semiparametric, murray2021log}) which will  be discussed later.
Combining the augmented integrated likelihood $p(\boldsymbol{Y}, \vk Z \mid \boldsymbol{X}, \jj{\vk{\theta}_{M},} \CMcal{T})$ with tree prior $p(\CMcal{T})$, allows us to calculate the posterior of $\CMcal{T}$ 
\begin{equation}\label{eq:post_T}
p(\CMcal{T} \mid \boldsymbol{X}, \boldsymbol{Y}, \jj{\vk{\theta}_{M},} \vk Z) \propto p(\boldsymbol{Y}, \vk Z \mid \boldsymbol{X}, \jj{\vk{\theta}_{M},} \CMcal{T}) p(\CMcal{T}).
\end{equation}
% up to a normalizing constant. 

When using MCMC to conduct Bayesian inference, $\CMcal{T}$ can be updated using a Metropolis-Hastings (MH) algorithm with the right-hand side of \eqref{eq:post_T} used to compute the acceptance \yz{ratio}. %These MH simulations %\yz{\#samples} 
%can be used to stochastically search the posterior space over trees to determine the high posterior probability trees from which we can choose \yz{a best} one. %An additional Gibbs sampler is then used to obtain a posterior sequence for $\vk{\theta}$. 
%\yz{The posterior sequence for $\vk{\theta}$ is then obtained using an additional Gibbs sampler.} %It is worth noting that by integrating out $\vk\theta$ in \eqref{eq:int_lik_1} we avoid the possible complexities associated with reversible jumps between continuous spaces of varying dimensions \cite{chipman2010bart, green1995reversible}.
Starting from the root node, the MCMC algorithm for simulating a Markov chain sequence of pairs $(\vk{\theta}^{(1)}, \CMcal{T}^{(1)}), (\vk{\theta}^{(2)}, \CMcal{T}^{(2)}), \ldots,$ using the posterior given in \eqref{eq:post_T}, is given in Algorithm \ref{Alg:2} \ct{in which %OMIT rest:.
%
%In Algorithm \ref{Alg:2},
} commonly used proposals (or transitions) for $q(\cdot,\cdot)$ include grow, prune, change and swap (see \cite{chipman1998bayesian}). %, which are usually selected with equal probability (i.e., $1/4$ each). 
\jj{See \cite{zhang2024bayesian} for further details.}

\COM{ %%%%%%%%%%%%%%%%%%%
\begin{algorithm} 
	\caption{One step of the MCMC algorithm for updating \yz{the} BCART parameterized by $(\vk{\theta}, \CMcal{T})$} 
	\hspace*{0.02in} {\bf Input:}
	Data $(\vk{X}, \vk{y})$ and current values $\left(\vk{\theta}^{(m)}, \CMcal{T}^{(m)}\right)$ \\
	\hspace*{0.3in} {\bf 1:}
	Generate a candidate value \(\CMcal{T}^{*}\) with probability distribution \(q\left(\mT^{(m)}, \mT^{*}\right)\)\\
	\hspace*{0.3in} {\bf 2:}
	Set the acceptance ratio $\alpha\left(\mT^{(m)}, \mT^{*}\right)=\min \left\{\frac{q\left(\mT^{*}, \mT^{(m)}\right)}{q\left(\mT^{(m)}, \mT^{*}\right)} \frac{p\left(\vk{y} \mid \vk{X}, \mT^{*}\right)}{p\left(\vk{y} \mid \vk{X}, \mT^{(m)}\right)}
\frac{p\left(\mT^{*}\right)}{p\left(\mT^{(m)}\right)}, 1\right\}$\\
	\hspace*{0.3in} {\bf 3:}
	Update \(\mT^{(m+1)}=\mT^{*}\) with probability  $\alpha\left(\mT^{(m)}, \mT^{*}\right)$, otherwise, set $\mT^{(m+1)}=\mT^{(m)}$\\
	\hspace*{0.3in} {\bf 4:}
	Sample $\vk\theta^{(m+1)} \sim p\left(\vk\theta \mid \mT^{(m+1)},\vk{X}, \vk{y}\right)$\\
 \hspace*{0.02in} {\bf Output:} 
	New values  $\left(\vk{\theta}^{(m+1)}, \CMcal{T}^{(m+1)}\right)$

\label{Alg:1}
\end{algorithm}
}%%%%%%%%%%%%%%%%%%%%%%%%%%%%%%%%%%%%%%%
%%%%%%%%%%%%%%%%%%%%%%%%%%%%%%%%%%%%%%%%%%%

%To implement \textbf{Algorithm 1}, we consider \(q\left(\mT, \mT^{*}\right)\) which generates \(\mT^{*}\) from \(\mT\) by randomly choosing among following four tree moves:

\begin{algorithm}
	\caption{One step of the MCMC algorithm for the BCART models parameterized by $(\vk{\theta}_M,\vk{\theta}_B,\CMcal{T})$ using data augmentation with both known and unknown parameters}
	\hspace*{0.02in} {\bf Input:}
	Data $(\vk{X},\vk{Y})$ and current values $(\hat{\vk{\theta}}_{M}^{(m)}, \vk{\theta}_B^{(m)}, \vk{Z}^{(m)}, \CMcal{T}^{(m)})$ \\
	\hspace*{0.2in} {\bf 1:}
	Generate a candidate value \(\CMcal{T}^{*}\) with probability distribution \(q(\CMcal{T}^{(m)}, \CMcal{T}^{*})\)\\
        \hspace*{0.2in} {\bf 2:}
	Estimate $\hat{\vk{\theta}}_{M}^{(m+1)}$, using MME (or MLE) \\
 %$ \sim p(\vk{\xi}\mid \vk{X}, \vk{v},\vk{N}, \vk{\hat{\kappa}_t}^{(m)}, \vk{\lambda}^{(m)}, \CMcal{T}^{(m)})$\\
        \hspace*{0.2in} {\bf 3:}
	Sample $\vk{Z}^{(m+1)} \sim p(\vk{Z}\mid \vk{X}, \vk{Y}, \hat{\vk{\theta}}_{M}^{(m+1)}, \vk{\theta}_B^{(m)}, \CMcal{T}^{(m)})$\\
	\hspace*{0.2in} {\bf 4:}
	Set the acceptance ratio
 $$\alpha(\CMcal{T}^{(m)}, \CMcal{T}^{*}) =\min \left\{\frac{q(\CMcal{T}^{*}, \CMcal{T}^{(m)})p(\vk{Y},\vk{Z}^{(m+1)} \mid \vk{X},  \hat{\vk{\theta}}_{M}^{(m+1)}, \CMcal{T}^{*})p(\CMcal{T}^{*})}{q(\CMcal{T}^{(m)}, \CMcal{T}^{*})p(\vk{Y},\vk{Z}^{(m)} \mid \vk{X}, \hat{\vk{\theta}}_{M}^{(m)}, \CMcal{T}^{(m)})p(\CMcal{T}^{(m)})}, 1\right\}
$$
	\hspace*{0.2in} {\bf 5:}
 	Update \(\CMcal{T}^{(m+1)}=\CMcal{T}^{*}\) with probability  $\alpha(\CMcal{T}^{(m)}, \CMcal{T}^{*})$, otherwise, set $\CMcal{T}^{(m+1)}= \CMcal{T}^{(m)}$\\
	\hspace*{0.2in} {\bf 6:}
 	Sample $\vk{\theta}_B^{(m+1)} \sim p(\vk{\theta}_B \mid \vk{X}, \vk{Y}, \hat{\vk{\theta}}_{M}^{(m+1)}, \vk{Z}^{(m+1)}, \CMcal{T}^{(m+1)})$\\
 \hspace*{0.02in} {\bf Output:} 
	New values $(\hat{\vk{\theta}}_{M}^{(m+1)}, \vk{\theta}_B^{(m+1)}, \vk{Z}^{(m+1)}, \CMcal{T}^{(m+1)})$
\label{Alg:2}
\end{algorithm}

\begin{remark}

(a). In Algorithm \ref{Alg:2}, the sampling steps should be done only as required. For example, in Step 2, $\hat{\vk{\theta}}_{M}^{(m+1)}$ needs to be estimated only for those nodes that were involved in the proposed move from $\CMcal T^{(m)}$ to $\CMcal T^*$.

(b). % It should be noted that Algorithm \ref{Alg:NB} can be easily extended to accommodate multivariate parameters for both $\vk{\theta_M}$ and $\vk{\theta_B}$.
\ljj{Algorithm \ref{Alg:2} is a general algorithm from which we can retrieve all the algorithms discussed in \cite{zhang2024bayesian}, e.g., the algorithm for the zero-inflated Poisson model therein can be retrieved by assuming there is no component $\vk{\theta}_{M}$.}

%(c). \jj{The above is for multivariate, why do you make the following remark? }
%\zzz{We propose a bivariate response BCART model here. In scenarios where multiple outcomes require simultaneous prediction, such as in financial risk modelling (predicting different types of financial risks concurrently) and environmental studies (modelling various environmental factors), it becomes essential to generalize the bivariate response BCART model to multivariate response BCART models. We refer to \cite{um2023bayesian} for further discussion.}
(c).  \zyy{The proposed \llji{parameter separation framework $\vk{\theta}_t=(\vk{\theta}_{t,M},\vk{\theta}_{t,B})$} is designed for simplicity. %to simplify calculations and adopt conjugate priors for all assumed unknown parameters. %Conjugate priors ensure analytical tractability, allowing explicit posterior forms and efficient parameter estimation. 
While general non-conjugate priors could offer greater flexibility in modeling complex data, they would require computationally intensive methods %, such as MCMC or other numerical methods, to
in approximating the posteriors, which significantly increases computational cost. Exploring non-conjugate priors,   using, \llji{e.g.,} Laplace approximation  as in \cite{chipman2003bayesian}, may be a direction for future research. %We refer also to \cite{chipman2000hierarchical} for an interesting incorporation of some hierarchical priors.
} 
\end{remark} 

\subsection{\ljj{Model selection and prediction}}
The MCMC algorithm described in Algorithm \ref{Alg:2} can be used to search for desirable trees, and we use the three-step approach proposed in \cite{zhang2024bayesian} based on deviance information criterion (DIC) to select an  ``optimal'' tree among those visited trees; see Table 1. %\ref{table_SS} 
%therein. %To this end,  we let $m_s<m_e$ be two user input integers which represent the belief of where the optimal number of terminal nodes of the tree might fall into. 
%Here we recall the DIC for tree models with data augmentation. %Depending on whether the latent variable $\vk{Z}$ is treated as a parameter or not, there are several types of likelihoods leading to eight versions of DIC as discussed in \cite{celeux2006deviance}. %, among which the fourth DIC is justified to be the most reliable. 
%Due to the complexity of implementing any of those eight and motivated by the idea that 
Note that in the following sections, we  introduce the DIC for different models based on the idea that
DIC$=$``goodness of fit''$+$``complexity''. \ct{See%OMIT:We refer to
} 
\cite{spiegelhalter2002bayesian,celeux2006deviance} for discussion on DIC in a general Bayesian framework.

\COM{ %%%%%%%%
\begin{table}[!t] 

 \centering

 \caption{Three-step approach for ``optimal'' tree selection}
 \begin{tabular}{p{1.2cm}|p{12.5cm}}  

\toprule   

{\bf Step 1:} & Set a sequence of hyper-parameters $(\gamma_j, \rho_j), j=m_s, \cdots, m_e,$ such that for $(\gamma_j, \rho_j)$, the MCMC algorithm converges to a region of trees with $j$ terminal nodes.\\
\hline\\[-10pt]
{\bf Step 2:} &
For each $j$ in Step 1, select the tree with maximum likelihood $p(\vk{Y}\mid \vk{X}, \vk{\overline{\vk{\theta}}}, \CMcal{T})$ from the convergence region, \jj{where $\vk{\overline{\vk{\theta}}}$ is an output from Algorithm 1.} \\  
\hline
%\midrule   

{\bf Step 3:} & From the trees obtained in Step 2, select the optimal one using \zzz{deviance information criterion (DIC)}.\\
  \bottomrule  
 % \hline
  
\end{tabular}

  \label{table_SS}
\end{table}
}%%%%%%%%%%%%%%

%In addition to the three DICs discussed in \cite{zhang2024bayesian}, one more specific DIC will be introduced in the next section for other models.
%\zzz{put the table three-step approach}

Suppose $\CMcal{T}$, with $b$ terminal nodes and parameter 
$\overline{\vk{\theta}}$, is the \jjj{selected} tree from the above %\ct{OMIT: three-step}
approach. For %\ct{OMIT: a given} 
new $\vk{x}$ the predicted  $\hat{\vk y}$ %\ct{OMIT: using this tree model} 
is defined as
\begin{equation}\label{eq:yhat}
    \hat{\vk y}\mid \vk{x} = \sum_{t=1}^b E(\vk y\mid \bar{\vk{\theta}}_t) I_{(\vk{x}\in \CMcal{A}_t)},
\end{equation}
where $I_{(\cdot)}$ denotes the indicator function \lj{and $\{\CMcal{A}_t\}_{t=1}^b$ is the partition of $\CMcal{X}$ by $\CMcal{T}$}.

%\subsection{\ljj{Model Performance Evaluation}}\label{sec:eva}

%\textcolor{red}{make all the evaluation metrics together... make a table about them...\\
%LJI- I think this can be moved to somewhere below, since all the metrics are defined for univariate response, in our case, we are mainly concerned with the aggregated \zzz{claim} amount.}

\section{Aggregate claim amount modeling with Bayesian CART}\label{Sec:CF}

%\jj{Question/discussion - whether we should call it "aggregate claims modelling" or "aggregate claim modelling"? claims frequency or claim frequency? Or \jj{average severity} or claim severity or average severity?}

This section introduces the BCART models for aggregate claim amount by specifying the response distribution within the framework outlined in Section 2. % To facilitate this, w
We begin by introducing the type of insurance \zzz{claim} data that will be discussed in this paper. A \zzz{claim} dataset with $n$ policyholders can be described by $(\vk{X}, \vk{v}, \vk{N}, \vk{S})=\big((\vk{x}_1,v_1, N_1, S_1), \ldots, (\vk{x}_n,v_n, N_n, S_n)\big)^\top$, where 
$\vk{x}_i=(x_{i 1}, \ldots, x_{i p})\in \CMcal{X}$ represents rating variables (e.g., driver age, age of the car and car brand in car insurance); $v_{i}\in(0,1]$ is the exposure in yearly units, quantifying the duration the policyholder $i$ is exposed to risk; $N_{i}$ is the number of \zzz{claims} reported during exposure time of the policyholder, and $S_{i}$ is the aggregate (total) claim amount.  %\ct{Is $Y$ (as mentioned below) part of the dataset, or unobserved?}
\llji{Following the convention in \cite{oh2021copula}, we have a summarized dataset as the individual severities are not accessible.}

%%%%%%%%%5
 \ct{Before describing our BCART models,%OMIT: We shall introduce three types of models for the data described above, namely, frequency-severity BCART models, sequential BCART models, and joint BCART models.  Below
 } 
 we briefly recall some basics on the aggregate claim amount \eqref{eq:NY}-\eqref{eq:NS}; % as motivation; %OMIT: to motivate these proposed models;
  see, e.g., \cite{Wuthrich2022, garrido2016generalized,  frees2016multivariate} for %\ct{OMIT relevant} 
 discussions.
%\subsubsection{Classical model for aggregate claim amount}
%\ljj{Some basic calculations for the classical aggregate loss model, which should be helpful to clarify the dependence and our motivation. The position of these calculations should be discussed later.}
Consider a given (generic) policyholder, and assume unit exposure (i.e., $v=1$), for simplicity. 
\COM{The aggregate claim amount of the policyholder can be expressed as
\begin{eqnarray}
    S = N \bar{S},\label{eq:NS}
\end{eqnarray}
where $N$ is the number of \zzz{claims} within a year and $\bar{S}$ denotes the \jjj{{\it average severity} %(or amount)
} of the claims. By convention, $\bar{S}=0$ when $N=0$.} We are primarily interested in estimating the pure premium  defined as $\E(S)$ (we remark that generally the pure premium should be defined as $\E(S)/v$, i.e., the expected \jj{claim} amount per year). 
%It can be easily derived that
%\zyy{When the number of claims  and the average severity  are assumed to be independent, it follows that:}
Under the classical collective risk model with full independence between $N$ and $\{Y_i\}_{i\ge 1}$, we have
\begin{eqnarray}
   \E(S) %=\E(N)\E(Y_1)
   = 
   \E(N) \E(\bar S | N>0).\label{eq:FS-0}
\end{eqnarray}
%\begin{eqnarray}
   % \E(S) = \E(N) \E(\bar S |N>0).\label{eq:FS-0}
%\end{eqnarray}
When a vector $\vk x=(x_1,\ldots,x_p)$ of covariates for this policyholder is available, it can be incorporated into separate GLMs   for frequency $N$ and (conditional) average severity $\bar S | N>0$, this is the classical frequency-severity model. %Traditionally, both frequency and average severity components are modeled by GLMs. %Since in our data it is the number of \zzz{claims} $N$ and the total claim amount $S$ rather than the severity $Y$ %\ct{do we need subscript $i$?}that are available, we shall use BCART models for frequency and average severity, resulting in the {\it frequency-severity BCART models}. Note that BCART models for frequency have been discussed in \cite{zhang2024bayesian}. 
%In the literature (see, e.g., \cite{henckaerts2021boosting, omari2018modeling, frees2014predictive}), two primary approaches are used to model the conditional average severity $\bar S |N>0$. First, 
In particular, a distribution from exponential distribution family (EDF) with $N$ as \zyy{a} weight is used for the average severity $\bar S |N>0$; see e.g., \cite{garrido2016generalized, gschlossl2007spatial}. %Let $Y_j (j=1,2,\ldots, N)$ be a generic random variable that denotes \textit{severity} of the $j$-th claim. Assuming $Y_{1}, Y_{2},..., Y_{N}$, given $N$, are IID positive random variables, independent of $N$, the average severity is given by $$\bar S = \sum_{j=1}^{N} Y_{j} / N \quad \text{for $N>0$}.$$ In this case, $$\E(S) = \E(N) \E(Y).$$ 
%This method often assumes that \zzz{$Y_j$} follows an exponential distribution family (EDF) due to the convolution property.
\llji{More specifically, assuming $Y_j\sim \text{EDF}(\mu,\phi)$, with mean $\mu$ and dispersion $\phi$, in \eqref{eq:NY}, we have, $\bar S|N=n \sim \text{EDF}(\mu, \phi/n)$ due to the convolution property, which means that modeling individual severity is equivalent to modeling the average severity where $N$ is included as \zyy{a} weight. %because $\mu=\E(Y)=\E(\bar S | N>0)$. %a weight in the model for $\bar S|N>0.$ 
A common distribution for $Y_j$ is the gamma distribution. This approach introduces an intrinsic functional dependence of the distribution of average severity on $N$ and can be used for summarized data. The pure premium can then be calculated by multiplying the estimations for the two parts by \eqref{eq:FS-0}.
Another approach when dealing with summarized data $(N,\bar S)$ is to use only the first formulation of \eqref{eq:NS}, imposing a distribution (independent of $N$) directly for the average severity $\bar S |N>0$;  see e.g., \cite{czado2012mixed, baumgartner2015bayesian,shi2015dependent, kramer2013total}.  In \cite{baumgartner2015bayesian} a shared random effects model is used to induce a conditional independence between the number of claims and average severity. In \cite{shi2015dependent, czado2012mixed} copulas are used to model the  dependence between the two parts, and the performance of these models is compared with the independence case, that is, technically speaking, $N$ independent of $\bar S$ given $N>0$.} It is clear that for this independence case the pure premium can still be estimated based on the product form  \eqref{eq:FS-0}. However, for the general dependence case, the product form as in \eqref{eq:FS-0} is not valid any more, and instead we apply a Monte Carlo method for pure premium estimation.

%%%%%%%%%%%%%%%%%%%%%%%
%\zyy{Nevertheless,} as discussed in \cite{garrido2016generalized}, \zyy{a negative correlation is often observed}
%\zzz{the number of claims} and average severity are often negatively correlated in %many \zyy{certain} types of insurance data, particularly, in collision automobile insurance data. \zzz{In our real data analysis below, we observe a similar negative correlation, following the approach in \ct{\cite{garrido2016generalized}.}} %OMIT the paper mentioned above. %[\ljj{we should check our data and see if this is the case, following the idea as in the paper mentioned above}] To better \jj{
%model \zzz{this type of} data and capture the underlying} dependence between \zzz{the number of claims} and average severity, we introduce two other types of BCART models. 
Based on these discussions, we shall introduce the frequency-severity BCART models as a benchmark, where the frequency part and severity part are dealt with independently; see Subsection \ref{sec:FS_BCART}. 
%This observation motivates us
In order to explore methods to capture the association between $\bar{S}$ 
and $N$ more comprehensively, 
%rather than restricting attention to specific correlation types. To this end, 
we propose two other types of BCART models.
First, following the idea of \cite{frees2011predicting, garrido2016generalized} we introduce the {\it sequential BCART models} by including $N$ \jj{(or its estimate $\hat N$)} as a covariate when modeling the average severity.
%in \eqref{eq:FS-0}. \zyy{However, the distribution of the aggregate claim amount is not analytically tractable as (\ref{eq:FS-0}) no longer holds due to the introduced dependence. To address this, we approximate %the posterior predictive distribution of the aggregate claim amount 
%it using Monte Carlo simulations based on separate models for the number of claims and the average severity, inspired by \cite{gschlossl2007spatial}.} 
Second, motivated by \cite{jorgensen1994fitting, smyth2002fitting, delong2021making}, we introduce {\it joint BCART models} by considering $(N,S)$ as a bivariate response. %inherently capturing the dependence between the number of claims and average severity. 
In this framework, association between the number of claims and average severity induced by potentially shared information (through covariates) is naturally incorporated within a selected single tree structure. %\jj{By its nature, the dependence between \zzz{the number of claims} and \zzz{average severity} is directly incorporated in the sequential BCART models.} In the joint BCART models, the dependence caused by potentially shared information (through covariates) can be captured in a %\jjj{selected} single tree. 
These %\ct{ OMIT two types of} 
BCART models will be discussed in detail in Subsections \ref{sec:sequential} and \ref{sec:joint}, respectively.

\subsection{Frequency-severity BCART models}\label{sec:FS_BCART}

%This subsection describes a standard model of insurance \zzz{claim}, which independently models \zzz{claim} frequency and \jj{average severity} using two trees. The ultimate goal of insurance pricing is to select the premium. Under the assumption that \zzz{claim} frequency and \jj{average severity} are independent, the pure premium can be calculated as:
%$$
%\text{Pure Premium = \zzz{claim} Frequency $\times$ \jj{average severity}.} 
%$$
%Previously, 
\COM{In this subsection, we shall discuss the frequency-severity BCART models.} Recall that the BCART models for the frequency component $\E(N)$ of \eqref{eq:FS-0} have been discussed in \cite{zhang2024bayesian}. Here we shall focus on the BCART modeling of the average severity component $\E(\bar S | N>0)$ of \eqref{eq:FS-0}.  More precisely, we will discuss a gamma distribution (as an example in the EDF) with $N$ included as a weight, and three other distributions to directly model the average severity without including $N$ as a weight, namely, gamma, lognormal, and Weibull.
For this purpose, we will only consider a data subset with $N_i>0$, and denote by $\bar n$ $(\le n)$ the size of this subset. The subset of average severity data will be denoted by
$(\vk{X}, \vk{N}, \vk{\bar S})=\big((\vk{x}_1, N_1, \bar S_1), \ldots, (\vk{x}_{\bar n}, N_{\bar n}, \bar S_{\bar n})\big)^\top$.

%we have introduced BCART models for \zzz{claim} frequency (see  Now we shall introduce BCART models for \jj{average severity}, which can be directly used for the premium calculation.

\subsubsection{ Average severity modeling using gamma distribution with $N$ as a weight}\label{sec:gamma_ind_bcart}

Assume the generic average severity $\bar S|N>0$ follows a gamma distribution with parameters being multipliers of $N$, i.e., %\ct{do not display following equation to save space?}
%\begin{eqnarray*}
    $\bar{S} |N>0 \sim \text{Gamma} \ (N\alpha, N \beta),$
%\end{eqnarray*}
with $\alpha,\beta>0$. Note that %the above distribution for $\bar S|N>0$ 
\jjj{this is} equivalent to assuming that the individual severity $Y_j$ follows Gamma($\alpha, \beta$) distribution, due to the convolution property. Recall that the probability density function (pdf) of the Gamma($\alpha, \beta$) distribution and its mean and variance are given as
\begin{eqnarray}\label{eq:fGx}
    f_G(x) = \frac{ \beta^{\alpha} x^{\alpha-1} }{\Gamma(\alpha)} e^{-\beta x},\ x>0, \ \ \ \  \mu_G=\frac{\alpha}{\beta}, \ \ \sigma_G^2=\frac{\alpha}{\beta^2},
\end{eqnarray}
where $\Gamma(\cdot)$ is the gamma function. %\ct{OMIT? Is it light-tailed? 
\jjj{It is known that gamma distribution is right-skewed and relatively light-tailed.}

\COM{ %%%%%%%%%%%%
In previous subsection, we model $\bar{S}_i$ using three different distributions (Gamma, LogNormal, and Weibull). Now, we propose another way that uses the Gamma distribution to model the individual claim amount $Y_{ij}$ instead of modeling $\bar{S}_i$ directly, %because of the nice additive property of the Gamma distribution, and it will become clear in the following context; see, e.g., 
following \cite{henckaerts2021boosting} and \cite{frees2014predictive} (also see Remark 2 (b)). Assume the insurance policyholder $i=1,2,\ldots,n$ have independent claim amounts $Y_{ij}$ ($j=1,2,\ldots,N_i$) given $N_i>0$. And they follow a common Gamma distribution with parameters $\alpha>0$ and $\beta>0$. Based on the additive property of the Gamma distribution, we can obtain the distribution for $\bar{S}_i$, i.e., }
%%%%%%%%%%%%%%%%%%%%%5555

According to the general BCART framework in Section \ref{Sec_BCART}, considering a tree $\CMcal{T}$ with $b$ terminal nodes and $\vk \theta_t=(\alpha_t, \beta_t)$  the two-dimensional parameter for the $t$-th terminal node, we assume %\ct{do not display following equation to save space?}
$
\bar{S}_i| \vk x_i, N_i \sim \text{Gamma} \ \left(N_i \alpha(\vk{x}_i), N_i \beta(\vk{x}_i)\right)
$ for the $i$-th observation,
where $\alpha(\vk{x}_i) = \sum_{t=1}^{\jj{b}} \alpha_t I_{(\vk x_i \in \CMcal{A}_t)}$ and $\beta(\vk{x}_i) = \sum_{t=1}^b \beta_t I_{(\vk x_i \in \CMcal{A}_t)}$, with $\{\CMcal{A}_t\}_{t=1}^b$ being the corresponding partition of $\CMcal{X}$. Specifically, for $i$-th observation such that $\vk{x}_i\in \CMcal{A}_t$, we have
(with $N_i$ compressed in $f_G$)

%Consider a tree $\CMcal{T}$ with $b$ terminal nodes, following a similar procedure to Subsection \ref{sec:sev_bcart}, for terminal node $t$, \yz{we denote the associated data as %$\left(\vk{X}_t, \vk{N}_t,\vk{v}_t)=(({X}_{t1}, N_{t1},v_{t1}), \ldots, ({X}_{tn_t}, N_{tn_t},v_{tn_t})\right)^\top$. 
%$(\vk{X}_t, \vk{N}_t,\vk{\bar{S}_t})=(({X}_{t1}, N_{t1}, \bar{S}_{t1}), \ldots, ({X}_{tn_t}, N_{tn_t}, \bar{S}_{tn_t}))^\top$. We then have} 

\begin{equation*} %\label{eq:f_Gamma2}
	f_{\text{G}}\left(\bar{S}_i| \alpha_t, \beta_t \right)= \frac{ (N_i \beta_t)^{N_i \alpha_t} \bar{S}_i^{N_i \alpha_t-1} }{\Gamma(N_i \alpha_t)} e^{-N_i \beta_t \bar{S}_i}.
\end{equation*}
\COM{ %%%%%%%%%%%%
for the $i$-th observation such that $\vk{x}_i\in \CMcal{A}_t$, where $N_i$ can be obtained directly from the data within the node.}
The mean and variance of $\bar{S}_{i}$ are thus given by \ct{$\alpha_t/\beta_t$ and 
$\alpha_t/(N_{i}\beta_t^2)$, respectively.  %OMIT equation
}
%\begin{equation}\label{eq:S_mean_N}
	%\mathbb{E}(\bar{S}_{i}\mid \alpha_t,\beta_t) %= \frac{\alpha_t}{\beta_t}, \quad \quad
	%\text{Var}(\bar{S}_{i}\mid N_i, %\alpha_t,\beta_t)= \frac{\alpha_t}
 %{N_{i}\beta_t^2}.
%\end{equation}
%Note that, for simplicity, here and hereafter, $
%\vk{x}_{i}$ will be compressed in some notation.  %\yz{Do we need to write $\vk{x}_{i}$ in the Poisson probability density function}.
%When choosing a prior distribution for $\lambda_{t}$ given $T$, a conjugate Gamma prior is selected 
%Based on the discussions in subsection \ref{subsec_para_prior}, %\ga{I would refer to Section 2.2 instead} below \eqref{eq:int_lik_1}, 
%Similar to Subsection \ref{sec:sev_bcart}, 

For each terminal node $t$, we treat $\alpha_t$ as known and $\beta_t$ as unknown and shall not apply any data augmentation. According to the notation used in Section \ref{Sec_BCART} this means $\vk \theta_{t,M}= \alpha_t$ and $\vk\theta_{t,B}=\beta_t$.
Here $\alpha_t$ will be estimated using MME, \ct{i.e., %OMIT according to \eqref{eq:S_mean_N}
} 
%we have
\begin{equation}\label{eq:alpha_gamma2}
	\hat{\alpha}_t=\frac{(\bar{S})_t^2}{\text{Var}(\bar{S})_t \Bar{N}_t},
\end{equation}
where $(\bar{S})_t$ and $\text{Var}(\bar{S})_t$ are the empirical mean and variance of the average severity, respectively, and $\Bar{N}_t$ is the average claim number of the data in the $t$-th terminal node.
 We treat $\beta_{t}$ as uncertain and use a conjugate gamma prior with hyper-parameters $\alpha_\pi, \beta_\pi>0$.  Denote the associated data in terminal node $t$ as $\left(\vk{X}_t, \vk{N}_t,\vk{\bar S}_t)=(({X}_{t1}, N_{t1},\bar S_{t1}), \ldots, ({X}_{t \bar n_t}, N_{t \bar n_t},\bar S_{t \bar n_t})\right)^\top$.
The integrated likelihood for the terminal node $t$ can then be obtained as
%\ct{perhaps omit one or two lines in the calculation?}
\begin{align}
\label{gamma2_integrated}
		p_{\text {G}}(\vk{\bar{S}}_{t} \mid \vk{X}_{t},\vk{N}_t,\hat\alpha_t )
		&=  \int_0^\infty f_{\text{G}}(\vk{\bar{S}}_{t} \mid \vk{N}_t, \hat{\alpha}_{t},\beta_t ) p(\beta_{t} ) d \beta_{t} \nonumber\\ 
		&= \int_0^\infty \prod_{i=1}^{\bar n_{t}} \left( \frac{ (N_{ti} \beta_t)^{N_{ti} \hat{\alpha}_t} \bar{S}_{ti}^{N_{ti} \hat{\alpha}_t-1} e^{-N_{ti} \beta_t \bar{S}_{ti}}}{\Gamma(N_{ti} \hat{\alpha}_t)} \right)
		\frac{\beta_\pi^{\alpha_\pi} {\beta_{t}}^{\alpha_\pi-1} e^{-\beta_\pi \beta_{t}}}{\Gamma(\alpha_\pi)} d \beta_{t} \\ \nonumber
		%&= \int_0^\infty  \frac{\lambda_{t}^{\sum_{i=1}^{n_{t}} N_{ti}}  \prod_{i=1}^{n_{t}} v_{ti}^{N_{ti}} e^{-\sum_{i=1}^{n_{t}} \lambda_{t} v_{ti}}}{\prod_{i=1}^{n_{t}} N_{ti} !}   \frac{\beta^{\alpha} {\lambda_{t}}^{\alpha-1} e^{-\beta \lambda_{t}}}{\Gamma(\alpha)} d \lambda_{t}\\
		%&= \frac{\beta_\pi^{\alpha_\pi}\prod_{i=1}^{\bar n_{t}} \left(N_{ti}^{N_{ti}\hat{\alpha}_{t}}\bar{S}_{ti}^{N_{ti} \hat{\alpha}_{t}-1}\right)}{\Gamma(\alpha_\pi) \prod_{i=1}^{\bar n_{t}}\Gamma(N_{ti}\hat{ \alpha}_{t})} \int_0^\infty \beta_{t}^{ \sum_{i=1}^{\bar n_{t}} N_{ti} \hat{\alpha}_{t}+\alpha_\pi-1} e^{-(\sum_{i=1}^{\bar n_{t}}N_{ti} \bar{S}_{ti}+\beta_\pi)\beta_{t}} d \beta_{t}\\ \nonumber
		&= \frac{\beta_\pi^{\alpha_\pi}\prod_{i=1}^{\bar n_{t}} \left(N_{ti}^{N_{ti}\hat{\alpha}_{t}}\bar{S}_{ti}^{N_{ti} \hat{\alpha}_{t}-1}\right)}{\Gamma(\alpha_\pi) \prod_{i=1}^{\bar n_{t}}\Gamma(N_{ti}\hat{ \alpha}_{t})} 
		\frac{\Gamma(\sum_{i=1}^{\bar n_{t}} N_{ti} \hat{\alpha}_{t}+\alpha_\pi)}{(\sum_{i=1}^{\bar n_{t}}N_{ti} \bar{S}_{ti}+\beta_\pi)^{\sum_{i=1}^{\bar n_{t}} N_{ti} \hat{\alpha}_{t}+\alpha_\pi}}.
\end{align}
%\yz{The first equality lacks $v_{ti}$ ?}
Clearly, from \eqref{gamma2_integrated}, we see that the posterior distribution of $\beta_{t}$ conditional on data ($\vk{N}_t$, $\vk{\bar{S}}_t$)  \jjj{and the estimated parameter $\hat{\alpha}_t$}, %,(and $\vk{X}_t, \vk{v}_t$, which are, obviously,  compressed for notational simplicity) \yz{is it necessary? since before we have mentioned it}, 
is given by
$$
	\beta_{t} \mid \vk{N}_t, \vk{\bar{S}}_t, \zzz{\hat{\alpha}_t}  \ \   \sim \ \ \text{Gamma}\left(\sum_{i=1}^{\bar n_{t}} N_{ti} \hat{\alpha}_{t}+\alpha_\pi,\sum_{i=1}^{\bar n_{t}}N_{ti} \bar{S}_{ti}+\beta_\pi\right).
$$
The integrated likelihood for the tree $\CMcal{T}$ is thus given by
\begin{equation*} %\label{eq:gamma2_whole}
	p_{\text{G}}\left(\vk{\bar{S}} \mid \vk{X},\vk{N}, \vk{\hat{\alpha}},\CMcal{T} \right)=\prod_{t=1}^{b} p_{\text{G}}\left(\vk{\bar{S}}_{t} \mid \vk{X}_{t},\vk{N}_t, \hat{\alpha}_t \right).
\end{equation*}
Next, we discuss the DIC for this tree. %Since we only consider uncertainty for $\vk{\beta}$ but not for $\vk{\alpha}$ and there is no data augmentation involved, the three different DICs defined in \cite{zhang2024bayesian} cannot be adopted directly. Thus, using 
\zzz{%OMIT (repeated): 
Following \cite{zhang2024bayesian}, 
%we shall use the idea that 
%DIC$=$``goodness of fit''$+$``complexity''.
} 
a DIC$_t$ for terminal node $t$  can %\ct{OMIT thus} 
be defined as %\ct{do not display following equation to save space?}
%\begin{eqnarray*}
	$\mathrm{DIC}_t= D(\bar{\beta}_t)+2p_{Dt},$
%\end{eqnarray*}
%\ljj{Shall we change all these effective numbers to $p_{Dt}$?}
where the posterior mean of $\beta_t$ is given by
\begin{equation}
	\label{eq:gamma2_mean_N}
	\bar{\beta_t}= \frac{\sum_{i=1}^{\bar n_{t}} N_{ti} \hat{\alpha}_{t}+\alpha_\pi}{\sum_{i=1}^{\bar n_{t}}N_{ti} \bar{S}_{ti}+\beta_\pi}\yyy{;}
\end{equation}
the goodness-of-fit is given as
\begin{equation*}
\begin{aligned}
%\label{eq:gamma2_D}
	D(\bar{\beta_{t}})&= -2 \sum_{i=1}^{\bar n_t} \log f_{\text{G}}(\bar{S}_{ti}\mid N_i,\hat{\alpha}_t, \bar{\beta}_t)\\
 &=-2 \sum_{i=1}^{\bar n_t} \left[N_{ti}\hat{\alpha}_t\log(N_{ti}\bar{\beta_t})+(N_{ti}\hat{\alpha}_t-1)\log(\bar{S}_{ti})-\bar{\beta_t}N_{ti}\bar{S}_{ti}-\log \left(\Gamma(N_{ti}\hat{\alpha}_t)\right)\right],
 \end{aligned}
\end{equation*}
and the effective number of parameters $p_{Dt}$ is defined by 
\begin{equation}
	\begin{aligned}
		%\label{eq:sDt_gamma}
		p_{Dt}&=1+\overline{D(\beta_t)}-D(\bar{\beta}_t)\\
		%&= 1- 2 \mathbb{E}_{\text{post}}\left(\log(f(\vk{y}_t \mid \vk{\theta}_t))\right) +2 \log(f(\vk{y}_t \mid \bar{\vk{\theta}}_t))\\
		% \nonumber\\
		&=1+ 2 \sum_{i=1}^{\bar n_t} \left\{\log \left(f_{\text{G}}(\bar{S}_{ti} \mid \hat{\alpha}_t, \bar{\beta}_t)\right)-  \mathbb{E}_{\text{post}}\left(\log(f_{\text{G}}\left(\bar{S}_{ti} \mid \hat{\alpha}_t, \beta_t)\right)\right) \right\},
	\end{aligned}
\end{equation}
where $1$ is added for the parameter $\alpha_t$ which was %\ct{OMIT assumed to be known and}
estimated upfront, and the difference of the last two terms on the right-hand side of the first line is the effective number for the unknown parameter $\beta_t$. %See \cite{spiegelhalter2002bayesian} for general discussions on the effective number of \zzz{parameters in} Bayesian models.
Some direct calculations yield that %the effective number of parameters for 
%terminal node $t$ is given by
\begin{equation*}%\label{eq:gamma2_sdt}
	\begin{aligned}
		p_{D{t}}=1+ 2  \left(\log \left(\sum_{i=1}^{\bar n_{t}} N_{ti}  \hat{\alpha}_t+\alpha_\pi \right)-\psi\left(\sum_{i=1}^{\bar n_{t}} N_{ti} \hat{\alpha}_t+\alpha_\pi\right)\right) \sum_{i=1}^{\bar n_{t}} N_{ti} \hat{\alpha}_t,
	\end{aligned}
\end{equation*}
with $\psi(x)=\Gamma'(x)/\Gamma(x)$ being the digamma function,
and thus $\text{DIC}_t$ can be derived.
\COM{
$$
\begin{aligned}
\text{DIC}_t 
%&=D(\bar{\beta_t})+2 p_{Dt} \\
	&=-2 \sum_{i=1}^{\bar n_t} \left[N_{ti}\hat{\alpha}_t\log(N_{ti}\bar{\beta_t})+(N_{ti}\hat{\alpha}_t-1)\log(\bar{S}_{ti})-\bar{\beta_t}N_{ti}\bar{S}_{ti}-\log\left(\Gamma(N_{ti}\hat{\alpha}_t)\right)\right] \\
	&\ \ \  ~~~+2+4 \sum_{i=1}^{\bar n_{t}} \left(\log \left(\sum_{i=1}^{\bar n_{t}} N_{ti}  \hat{\alpha}_t+\alpha_\pi \right)-\psi\left(\sum_{i=1}^{\bar n_{t}} N_{ti} \hat{\alpha}_t+\alpha_\pi\right)\right) \sum_{i=1}^{\bar n_{t}} N_{ti} \hat{\alpha}_t.
\end{aligned}
$$
}Consequently, the DIC of the whole tree $\CMcal{T}$ is obtained as
\begin{eqnarray}\label{eq:DIC_chapter4_1}
	\mathrm{DIC}=\sum_{t=1}^b \mathrm{DIC}_t.
\end{eqnarray}
%\begin{equation}\label{eq:DIC_pois}
%   \text{DIC} =\sum_{t=1}^b \text{DIC}_t.
%\end{equation}

With the above formulas derived for the gamma case, we can use the %\ct{OMIT three-step}
approach presented in Table 1 of \cite{zhang2024bayesian}, %\ref{table_SS}, 
together with Algorithm \ref{Alg:2}, to search for an optimal tree which can then be used to predict new data \ct{such that%OMIT. Given an optimal tree,
} the estimated average severity $\hat{\alpha}_t/\bar{\beta}_t$ in each terminal node $t$ can be determined using \eqref{eq:alpha_gamma2} and \eqref{eq:gamma2_mean_N}. %Some of the evaluation metrics based on Gamma distribution are provided in Table \ref{table:EM_Gamma}.

\subsubsection{ Average severity modeling using distributions without $N$ as a weight}\label{sec:sev_bcart}

\COM{In this subsection we shall discuss BCART models for the average severity using distributions not including $N$ as a weight. Precisely, t}

\COM{
\begin{table}[!t]  
	
	\centering
 \captionsetup{width=.8\linewidth}
	\caption{Lognormal and Weibull distributions % for average severity
                }
	%\begin{tabular}{lp{1.7cm}p{2.4cm}p{1.9cm}p{2.1cm}p{1.5cm}} 
	\begin{tabular}{l|c|c}
		
		\toprule   
		
		Distribution  & Lognormal ($\mu\in \mathbb{R},\sigma>0$) & Weibull ($\alpha,\beta>0$)  \\ [4pt] 
		
		%\midrule   
\hline&&\\[-8pt]
	Pdf &  $f_{\text{LN}}\left(x \right)=
(x \sigma \sqrt{2 \pi})^{-1} \exp\left(-\frac{(\log (x)-\mu)^2}{2 \sigma^2}\right) $ &  $f_{\text{Weib}}\left(x \right)=
\frac{\alpha}{\beta} x^{\alpha-1} \exp\left(-x^{\alpha}/\beta\right)$
		 \\[6pt]
    %\hline
	Mean	 & $\exp\left(\mu+\sigma^2/2\right)$ & $\beta^{\zzz{1/\alpha}} \Gamma(1+1 / \alpha) $  \\[4pt]
  %\hline
	Variance  & $\left(\exp(\sigma^2)-1\right) \exp(2 \mu+\sigma^2)$  &  $\beta^{\zzz{2/\alpha}}\left[\Gamma(1+2/\alpha)-\left(\Gamma(1+1/\alpha)\right)^2\right]$\\[5pt]
 %\hline
		Characteristics & \zzz{positively skewed,} & \zzz{positively skewed,}  \\
		 & \ljj{ heavy-tailed} & \ljj{versatile tail behaviour}  \\
		\bottomrule  
		
	\end{tabular}
	
	\label{table-severity-distr}
\end{table}
}
Three distributions (gamma, lognormal and Weibull) will be used to model the average severity $\bar S|N>0$. %\jjj{%OMIT this sentence and the Table? 
%See Table \ref{table-severity-distr} for basic properties of lognormal and Weibull distributions, recalling also \eqref{eq:fGx}.}
Selecting among these three distributions for certain data may pose a considerable challenge, and scholars have extensively explored this topic; see, e.g., \cite{siswadi1982selecting}. In \jj{average severity} modeling, insurers want to gain more insights into the right tail.
%\ga{OMIT?, which describes the behaviour of the distribution at large values}\ct{agreed}. 
The gamma distribution would be a suitable model for losses that are not catastrophic, such as auto insurance. The lognormal distribution is more suitable for fire insurance, which may exhibit more extreme values than auto insurance. Moreover, the Weibull distribution has the ability to handle different scenarios by tuning the shape parameter to adapt to different tail characteristics.

We demonstrate how to apply these distributions in BCART models for the average severity data.
The idea, as in the previous section, is to specify the distributions/parameters in the general BCART framework of Section \ref{Sec_BCART}. We only give some key information below, and defer some detailed calculations to \zzz{Section SM.A of the Supplementary Material}.

%%%%%%%%%%%%%%%%%%%%%%%%%%%%%%%%%%%%%%%%%%%%%%%

%%%%%%%%%%%%%%%%%%%%%%%%%%%%%%%%%%%%%%%%%%%555%%

%\noindent\textbf{Gamma-Bayesian CART}

Consider a tree $\CMcal{T}$ with $b$ terminal nodes %\ct{OMIT? and the corresponding partition $\{\CMcal{A}_t\}_{t=1}^b$ of $\CMcal{X}$} 
for the average severity data. In %a gamma model and Weibull model, we respectively assume %all 
gamma and Weibull models, we respectively assume
%insurance policyholders $i=1,2,\ldots,n$ have independent average claim amounts $\bar{S}_i$ with 
%\begin{align*}
$    \bar{S}_{i} \mid \vk{x}_i  \sim 
	\operatorname{Gamma}\left(\alpha(\vk{x}_i), \beta(\vk{x}_i)\right)$, and 
 $\bar{S}_{i} \mid \vk{x}_i   \sim 
	\operatorname{Weib}\left(\alpha(\vk{x}_i), \beta(\vk{x}_i)\right)$,
%\end{align*}
where $\alpha(\vk{x}_i)= \sum_{t=1}^b  \alpha_t I_{(\vk{x}_i\in \CMcal{A}_t)}$, $\beta(\vk{x}_i)=  \sum_{t=1}^b  \beta_t I_{(\vk{x}_i\in \CMcal{A}_t)}$.
In a lognormal model, we assume that
%\begin{equation*}
$	\bar{S}_{i} \mid \vk{x}_i \ \stackrel{\text { }}{\sim} 
	\ \operatorname{LN}\left(\mu(\vk{x}_i), \sigma(\vk{x}_i)\right)$,
%\end{equation*}
where $\mu(\vk{x}_i)= \sum_{t=1}^b  \mu_t I_{(\vk{x}_i\in \CMcal{A}_t)}$, and $\sigma(\vk{x}_i)=  \sum_{t=1}^b  \sigma_t I_{(\vk{x}_i\in \CMcal{A}_t)}$.

For each terminal node $t$, we treat one parameter as known and the other as unknown, that is, according to the notation in Section \ref{Sec_BCART}, $\vk\theta_{t,M}$ is $\alpha_t$ for the gamma and Weibull models and is $\sigma_t$ for the lognormal model, and  $\vk\theta_{t,B}$ is $\beta_t$ for the gamma and Weibull models and is $\mu_t$ for the lognormal model. Furthermore, we apply a conjugate prior for $\beta_t$ and $\mu_t$, namely, a Gamma($\alpha_\pi, \beta_\pi$) prior for the $\beta_t$ in the gamma model, a Normal($\mu_\pi, \sigma_\pi$) prior for the $\mu_t$ in the lognormal model, and an inverse-Gamma($\alpha_\pi, \beta_\pi$) prior for the $\beta_t$ in the Weibull model, i.e.,
\begin{equation}\label{eq:inverse_gamma}
	p(\beta_t)=\frac{\beta_\pi^{\alpha_\pi}}{\Gamma(\alpha_\pi)} \beta_t^{-\alpha_\pi-1} 
 \exp(-\beta_\pi/\beta_t),
\end{equation}
with $\alpha_\pi, \beta_\pi>0.$ Estimates for the unknown parameters, calculations of the integrated likelihood and DIC$_t$ for these three models are given in \zzz{Section SM.A of the Supplementary Material}.  We can then use the \ct{above procedure leading to the%OMIT three-step approach presented in Table \ref{table_SS}, together with \eqref{eq:DIC_chapter4_1} and Algorithm \ref{Alg:2}, to search for an optimal tree which can then be used to predict new data. The
} predictions obtained using \eqref{eq:yhat} from different models, as displayed in 
Table~\ref{table_Sev_pred}.

\begin{table}[!t]  
	
	\centering
 \captionsetup{width=.8\linewidth}
	\caption{Estimations for average severity in terminal node $t$. Here $(\bar{S})_t$ and $\operatorname{Var}(\bar{S})_t$ denote the \zzz{empirical mean and variance} of the \jj{average severity} in the $t$-th node, respectively. See \zyy{Supplementary Material SM.A} for  details.}
	%\begin{tabular}{lp{1.7cm}p{2.4cm}p{1.9cm}p{2.1cm}p{1.5cm}} 
	\begin{tabular}{l|c|c|c}
		
		\toprule   
		
		Distribution & Gamma($\alpha_t,\beta_t$) & LN($\mu_t,\sigma_t$) & Weib($\alpha_t,\beta_t$)  \\  
		
		%\midrule   
\hline
	Prediction $\hat{\bar S}_t$ & $\hat{\alpha}_t/\bar{\beta}_t$  & $\exp (\bar{\mu}_t+\hat{\sigma}_t^2/2)$ &  $\bar{\beta}_t^{\zzz{1/\hat{\alpha}_t}}  \Gamma(1+1 / \hat{\alpha}_t)$
		 \\
    \hline&&&\\[-9pt]
		Parameter & $\hat{\alpha}_t =\frac{(\bar{S})_t^2}{\text{Var}(\bar{S})_t}$& $\hat \sigma_t$ obtained using MME %software \textsf{R} 
  & $\hat \alpha_t$ obtained using MME %software \textsf{R}
  \\
	estimation  & $\bar{\beta}_t=\frac{\bar n_t \hat{\alpha}_t+\alpha_\pi}{\sum_{i=1}^{\bar n_t} \bar{S}_{ti}+\beta_\pi}$& $\bar\mu_t=\frac{\hat{\sigma}_{t}^2 \sigma_\pi^2}{\bar n_t \sigma_\pi^2+\hat{\sigma}_{t}^2}\left(\frac{\mu_\pi}{\sigma_\pi^2}+\frac{\sum_{i=1}^{\bar n_t} \log(\bar{S}_{ti})}{\hat{\sigma}_{t}^2}\right)$  & $\bar{\beta}_t = \frac{\sum_{i=1}^{\bar n_t} \bar{S}_{ti}^{\hat{\alpha}_{t}} + \beta_\pi}{\bar n_t+\alpha_\pi-1}$\\
		
		\bottomrule  
		
	\end{tabular}
	
	\label{table_Sev_pred}
\end{table}

%%%%%%%%%%%%%%%%
\begin{remark}

(a). \zzz{There are different ways to parameterize the Weibull distribution, either with two or three parameters; see, e.g., \cite{rinne2008weibull}. For simplicity, we adopt the common parameterization with two parameters; see, e.g., \cite{fink1997compendium}.}

(b). In the above BCART models for average severity we have assumed that one parameter of the distribution is treated as known and the other is treated as unknown which is given a conjugate prior. We note that this is not the only way to implement the BCART algorithms. There are other ways to treat the parameters. For example,  for the gamma distribution, the following two alternative approaches can be considered:
\begin{itemize}
    \item Treat the parameter $\beta_t$ as known and use a prior for $\alpha_t$, i.e., $ p(\alpha_t) \propto a_0^{\jlp{\alpha_t}-1} \beta_t^{\alpha_t c_0}/\Gamma(\alpha_t)^{b_0}$ where $a_0,b_0,c_0$ are prior hyper-parameters.
\item Treat both  $\alpha_t$ and $\beta_t$ as unknown and use a joint prior for them, i.e., $p(\alpha_t,\beta_t) \propto 1/(\Gamma(\alpha_t)^{c_0} \beta_t^{-\alpha_t d_0})$ ${a_0}^{\alpha_t-1} e^{-\beta_t b_0}$ where $a_0,b_0,c_0,d_0$ are prior hyper-parameters; see, e.g., \cite{fink1997compendium}.
\end{itemize}
Although the joint prior can be used to obtain estimators for $\alpha_t$ and $\beta_t$ simultaneously in the Bayesian framework, it is not formulated as an exact distribution, leading to less accurate estimators. The first way also has this shortcoming. 
For the lognormal distribution, %in the Bayesian framework by treating different parameters as known and assuming corresponding conjugate priors. \yz{For example, 
a normal and inverse-gamma joint prior can be used for the parameters $\mu$ and $\sigma^2$; see, e.g., \cite{fink1997compendium}. These more complicated cases are \jj{not} considered in our current implementation. 

%We choose the most commonly used one to introduce here and implement. 

(c). Many other distributions can also be used to model \jj{average severity}, such as Pareto, generalized gamma, generalized Pareto distributions, and so on. However, they either have too many parameters or are challenging to make explicit calculations in the Bayesian framework. We believe further research into the selection of these distributions is worth exploring; see, e.g., \cite{shi2015dependent, mehmet2005bonus, farkas2021cyber} for some insights on the application of these distributions to insurance pricing.

\end{remark}

In the frequency-severity BCART models, we obtain two trees for frequency and \zzz{average} severity respectively. The pure premium can be calculated using the predictions from these two trees together with the pricing formula \eqref{eq:FS-0}. There can be many different combinations of predictions for the frequency-severity models, i.e., any model discussed in \cite{zhang2024bayesian} for frequency and any model introduced above for average severity can be adopted. 

 One benefit of modeling frequency and \zzz{average} severity separately using two  trees is that the \jjj{important} risk factors associated with each component can be discovered separately. However, it can be challenging to interpret two trees as a whole, since several policyholders may be classified in one cell by the frequency tree but in a different cell by the \zzz{average} severity tree. In the next section, we discuss the combination of two trees for prediction and interpretation.

\subsubsection{Evaluation metrics for frequency-severity BCART models}\label{sec:eva_two_trees}
\zzz{In this section, we begin by exploring some performance evaluation metrics  for average severity BCART models, %Understanding these metrics is crucial for assessing the accuracy and reliability of our models in predicting insurance \jj{average severity}. 
then we introduce the idea of combining two trees to derive evaluation metrics for the frequency-severity BCART models.}
\jj{Application of these evaluation metrics \zzz{will} be discussed %in the simulation examples and real data analysis
in Sections \ref{sec:sim} and \ref{sec:rd}.}

%\jj{(Note - To shorten the text, we may just use terminology like ``average severity models" rather than ``average severity BCART models", etc.)}

\subsubsection*{\underline{\jj{Evaluation} metrics for average severity trees}}

\ct{We use%OMIT: To evaluate the model performance of the average severity BCART models, we use
} the same performance measures that were introduced in \cite{zhang2024bayesian}. %for prediction comparisons. Here we provide a summary of them. 
Suppose we have obtained a tree with $b$ terminal nodes and the corresponding %parameter estimates which we will use to obtain the 
predictions $\hat{\bar S}_t$ $(t=1,\ldots,b)$ given in Table \ref{table_Sev_pred}. Consider a test dataset with $\bar m$ observations. Denote the number of test data in terminal node $t$  by $\bar m_t$, and denote the associated data in terminal node $t$ as $\left(\vk{X}_t, \vk{N}_t,\vk{\bar S}_t)=((\jlp{\vk{x}_{t1}}, N_{t1},\bar S_{t1}), \ldots, (\jlp{\vk{x}_{t \bar m_t}}, N_{t \bar m_t},\bar S_{t \bar m_t})\right)^\top$ $(t=1,\ldots,b)$. The evaluation metrics are listed below.

\begin{itemize}
    \item[{\bf M1:}] The residual sum of squares (RSS) is given by 
    $\text{RSS}(\vk{\bar{S}})=\sum_{t=1}^{b} \sum_{i=1}^{\bar m_t} %\sum_{i=1}^{n_{t}}
	(\bar{S}_{ti} - \hat{\bar S}_{ti})^2.$
  
%$$ \text{RSS}(\vk{y})=\sum_{i=1}^{m} %\sum_{i=1}^{n_{t}}
%(y_{i}-\hat{y}_i)^2.$$  

%Although RSS is a commonly used metric with its advantages (sensitivity to model fit, simple interpretation, and good mathematical properties), it is important to be aware of its limitations and consider other evaluation metrics, especially when dealing with datasets that contain large outliers. For example, in insurance \jj{average severity} data, RSS places considerable emphasis on outliers, i.e., extremely large claim amounts. This can result in an excessively large RSS, which is undesirable.

    \item[{\bf M2:}] 
%Given the discussion above, for the tree models, we propose a specifically designed metric called 
The squared error (SE), based on a sub-portfolio  (i.e., those instances in the same terminal node) level, %which 
is defined by
$ \text{SE}(\vk{\bar{S}})=\sum_{t=1}^{b} \left(\sum_{i=1}^{\bar m_t}S_{ti}/\sum_{i=1}^{\bar m_t}N_{ti}-\hat{\bar S}_t\right)^2.
$
%where $\epsilon_t$/$\hat{\epsilon}_t$ is the empirical/estimated \zzz{claim} cost (or \jj{average severity}) respectively for terminal node $t$ depending on the type of \zzz{claim} model considered. This estimation $\hat{\epsilon}_t$ is obtained using \eqref{eq:yhat}, assuming unit exposure for \zzz{claim} cost, and unit claim number for \jj{average severity}. \ga{it may be helpful to also say how $\epsilon$ is computed.}

    \item[{\bf M3:}] 
    
    Discrepancy statistic (DS) is defined as a weighted version of SE, given by
%$$ \text{DS}(\vk{\epsilon})=\sum_{t=1}^{b} \frac{1}{\hat{\sigma}^2_t}\left(\epsilon_t-\hat{\epsilon}_t\right)^2,$$
%where $\epsilon_t$ and $\hat{\epsilon}_t$ are the same as in M2, and $\hat{\sigma}^2_t$ is the estimated variance of \zzz{claim} cost (or \jj{average severity}) for terminal node $t$. 
$ \text{DS}(\vk{\bar{S}})= \\ \sum_{t=1}^{b}  \left(\sum_{i=1}^{\bar m_t}S_{ti}/\sum_{i=1}^{\bar m_t}N_{ti}-\hat{\bar S}_t\right)^2/ \jjj{\hat V_t},
$
where \zzz{$\hat V_t$} for different models are given in Table \ref{table_Sev_Var}. %\ct{Check - is this intended to be the square of the variance?}

    \item[{\bf M4:}] Model Lift indicates the ability to differentiate between cells of policyholders with low and high %\zzz{claim} cost (or \zzz{claim} 
    risks (average severity here), and is defined by using the data and their predicted values in the most and least risky cells. %terminal nodes \jj{taking into account the weights (what do you mean by weights? I wonder how you have done this, I don't think you need to consider any weights in these calculations, or just make sure you compare the same number of PHs in both groups)}. % without assuming an underlying distribution. A higher lift illustrates that the model is more capable of separating the extreme values from the average. Since lift focuses on the tails of the distribution, this metric enables actuaries to effectively construct risk mitigation plans using mechanisms beyond pricing, such as underwriting, reinsurance, and enforcement of safety measures. 
    We use a similar approach as in \cite{zhang2024bayesian} to calculate Lift for the \zzz{average} severity tree models; \jj{more details on these calculations can be found in \cite{zhang2024insurance}. }

    %\item[{\bf M5:}] Negative log-likelihood (NLL) is calculated by using the assumed response distribution in the terminal node with the estimated parameters from training, see Table \ref{table_Sev_pred}. % estimated from the training. %compare the losses of the models’ corresponding distributions because they 
    
\end{itemize}

\begin{table}[!t]  
	
	\centering
 \captionsetup{width=.8\linewidth}
	\caption{Variance ($\hat V_t$) of the average severity distribution in terminal node $t$ using the BCART estimations. Below GammaN means the gamma model with $N$ as a weight, while Gamma means the gamma model without $N$ as a weight.}
	%\begin{tabular}{lp{1.7cm}p{2.4cm}p{1.9cm}p{2.1cm}p{1.5cm}} 
	\begin{tabular}{l|c|c|c|c}
		
		\toprule   
		
		Dist. & GammaN & Gamma & Lognormal & Weibull %(\jj{correct?}) 
  \\ 
		
		%\midrule   
\hline&&&&\\[-9pt]
	$\hat{V}_t$ & $\hat{\alpha}_t/(\bar N_t\bar{\beta}^2_t$) & $\hat{\alpha}_t/
 \bar{\beta}^2_t$ &  $\left(e^{\hat{\sigma}_t^2}-1\right) e^{2 \bar{\mu}_t+\hat{\sigma}_t^2}$
& $\bar{\beta}_t^{\zzz{2/\hat{\alpha}_t}}\left[\Gamma\left(1+2/\hat{\alpha}_t\right)-\left(\Gamma\left(1+1/\hat{\alpha}_t\right)\right)^2\right]$\\
		
		\bottomrule  
		
	\end{tabular}
	
	\label{table_Sev_Var}
\end{table}

%%%%%%%%%%%%%%%%%%%%%%%%%%

\subsubsection*{\underline{Evaluation metrics for two trees from the frequency-severity model}} \label{eva:two_trees}

The \zzz{frequency-severity BCART} model yields two trees. %(one for frequency and the other for average severity). 
We now explain how to combine these two trees %Now, we discuss ways that these two trees can be combined 
to evaluate model performance based on the aggregate \zzz{claim} amount
(or  pure premium) prediction $\hat {S}$ for  a test dataset with $m$ observations. % in the context of two trees in frequency-severity models (and subsequent sequential models in Section \ref{sec:sequential}) needs some further discussion. %Obtaining two separate trees with potentially distinct structures for \zzz{claim} frequency and \jj{average severity} makes it challenging to compare performance based on DIC between two separate trees and one tree (for joint models in Section \ref{sec:joint}) on training data with varying numbers of terminal nodes. Nevertheless, DIC can still assist in identifying the optimal tree for each model, enabling subsequent comparisons on the test data.
\jj{%To this end, we need to combine those two trees to form a joint partition of the covariate space $\CMcal{X}$. 
The idea is natural -- individual tree partitions are superimposed to form a joint} partition of the covariate space $\CMcal{X}$. This process evolves by merging all the splitting rules from both trees. %, creating what is known as a global partition. 
The splits of each tree contribute to a refined segmentation of the covariate space, resulting in a joint partition that represents the collective behaviour of the original two tree partitions\ct{; see \cite{rockova2020posterior}}. 

Suppose we have obtained a joint partition with $c$ cells. \llji{The corresponding prediction $\hat{S}_t$ for \zyy{cell} $t$ $(t=1,\ldots,c)$ is obtained by \eqref{eq:FS-0}, %by multiplying the estimated frequency and average severity derived from the joint partition. 
\begin{equation}\label{eq:SSi}
\hat{S}_{t}=\hat{N}_{s(t)}\hat{\bar{S}}_{l(t)}, 
\end{equation}
where $s(t)$ is the corresponding node index of $t$ in the frequency tree and $l(t)$ is the corresponding one in the severity tree, and $\hat{N}_{s(t)}$   and $\hat{\bar{S}}_{l(t)}$ are the corresponding estimates from these two individual trees. We further denote $m_t$ as the number of observations in \zyy{cell} $t.$
Using the notation above we have 
} 
\COM{ %%%%%%%%%
\jj{First, note that RSS($\vk{S}$), similarly defined as in the above M1,} can be easily employed. \jj{More precisely, given} the independence assumption between \zzz{the number of claims and  average severity} in the frequency-severity models, %RSS($\vk{S}$) is \yz{straightforwardly} obtained by multiplying $\hat{N}_i$ and $\hat{\bar{S}}_i$, i.e., 
we have,  \jj{by \eqref{eq:FS-0},} %\llji{[Q- what is the meaning of $t$ below?]}
\begin{equation}\label{eq:SSi}
\hat{S}_{ti}=\hat{N}_{ti}\hat{\bar{S}}_{ti}, \ \  \zyy{t=1,\ldots,c}, i=1\ldots, m_t,
\end{equation}
where $\hat{N}_{ti}$ is obtained from the \zzz{claim} frequency tree and $\hat{\bar{S}}_{ti}$ is obtained from the \jj{average severity} tree, \llji{with $m_t$ being the number of observations in \zyy{cell} $t$.}
%Similarly, NLL can be obtained by summing up the corresponding NLLs from the two trees. 
\zyy{While RSS can be used to evaluate the prediction error based on individual observations, it does not take the structure of the two involved trees into account. To address this, three additional tree structure-dependent metrics—SE, DS, and Lift (as defined in the above M2-M4)—for the aggregate claim amount $S$ are introduced.} The corresponding predictions $\hat{S}_t$ for \zyy{cell} $t$ $(t=1,\ldots,c)$ are obtained by multiplying the estimated frequency and average severity derived from the joint partition. See an example in Remark 3 (a) for further illustration.  
}%%% end COM %%%%

\begin{itemize}
    \item[{\bf M1':}] The residual sum of squares %(RSS) %is given by 
    $\text{RSS}(\vk{{S}})=\sum_{t=1}^{c} \sum_{i=1}^{ m_t} %\sum_{i=1}^{n_{t}}
	({S}_{ti} - \hat{ S}_{ti})^2.$
%\zyy{[maybe this can be directly written as $\text{RSS}(\vk{{S}})= \sum_{i=1}^{ m} %\sum_{i=1}^{n_{t}}({S}_{i} - \hat{ S}_{i})^2.$]}

    \item[{\bf M2':}] 
%Given the discussion above, for the tree models, we propose a specifically designed metric called 
The squared error %(SE), based on a sub-portfolio  (i.e., those instances in the same terminal node) level, %which  is defined by
$ \text{SE}(\vk{{S}})=\sum_{t=1}^{c} \left(\sum_{i=1}^{ m_t}S_{ti}/\sum_{i=1}^{ m_t}v_{ti}-\hat{ S}_t\right)^2.
$  
\COM{\zyy{The relationship between $\text{RSS}(\vk{{S}})$ and $ \text{SE}(\vk{{S}})$}:  \llji{[lj- be careful with the products, $S_t v_{t i}, \hat{S}_t v_{t i}$. Do these always make sense?]}
$$
\operatorname{RSS}(\vk{{S}})=\sum_{t=1}^c \sum_{i=1}^{m_t}\left(S_t v_{t i}-\hat{S}_t v_{t i}\right)^2
=\sum_{t=1}^c \sum_{i=1}^{m_t} v_{t i}^2\left(S_t-\hat{S}_t\right)^2,
$$

$$
\operatorname{SE}(\vk{{S}})=\sum_{t=1}^c\left(\frac{\sum_{i=1}^{m_t} S_t v_{t i}}{\sum_{i=1}^{m_t} v_{t i}}-\hat{S}_t\right)^2 =\sum_{t=1}^c\left(S_t \frac{\sum_{i=1}^{m_t} v_{t i}}{\sum_{i=1}^{m_t} v_{t i}}-\hat{S}_t\right)^2=\sum_{t=1}^c\left(S_t-\hat{S}_t\right)^2.
$$

Therefore,
$$
\operatorname{RSS}(\vk{{S}}) = \sum_{t=1}^c w_t \operatorname{SE}(\vk{{S}}),
$$
where $w_t=\sum_{i=1}^{m_t} v_{ti}^{2}$ represents weight coefficient.}

%where $\epsilon_t$/$\hat{\epsilon}_t$ is the empirical/estimated \zzz{claim} cost (or \jj{average severity}) respectively for terminal node $t$ depending on the type of \zzz{claim} model considered. This estimation $\hat{\epsilon}_t$ is obtained using \eqref{eq:yhat}, assuming unit exposure for \zzz{claim} cost, and unit claim number for \jj{average severity}. \ga{it may be helpful to also say how $\epsilon$ is computed.}

    \item[{\bf M3':}] 
    The discrepancy statistic % (DS) is defined as a weighted version of SE, given by
%$$ \text{DS}(\vk{\epsilon})=\sum_{t=1}^{b} \frac{1}{\hat{\sigma}^2_t}\left(\epsilon_t-\hat{\epsilon}_t\right)^2,$$
%where $\epsilon_t$ and $\hat{\epsilon}_t$ are the same as in M2, and $\hat{\sigma}^2_t$ is the estimated variance of \zzz{claim} cost (or \jj{average severity}) for terminal node $t$. 
$ \text{DS}(\vk{{S}})=\sum_{t=1}^{c}  \left(\sum_{i=1}^{m_t}S_{ti}/\sum_{i=1}^{ m_t}v_{ti}-\hat{ S}_t\right)^2/ \hat V_t,
$
where \jjj{$\hat V_t$} is the estimated model variance of $S$ in the $t$-th cell which is derived using the model specific assumptions and its parameter estimates. More specifically, if the average severity model is as in Subsection \ref{sec:gamma_ind_bcart}, we can rewrite %$S$ as in \eqref{eq:SYj} 
$S=\sum_{j=1}^N Y_j$ 
with $Y_j$ following independent gamma distribution with parameters $\alpha, \beta$ as in \eqref{eq:fGx}.
Thus, we use 
$
\text{Var}(S) = \mathbb{E}(N) \text{Var}(Y) + \left(\mathbb{E}(Y)\right)^2 \text{Var}(N) 
$
to derive an estimate of $\hat V_t$ for the $t$-th cell, together with corresponding estimated parameters for $\alpha, \beta$ given in Subection \ref{sec:gamma_ind_bcart} and for different frequency models in \cite{zhang2024bayesian}.  %combined with the estimated model parameters therein
 %\zzz{we estimate the model variance $\hat V_t^2$ with unit exposure. Initially, this requires deriving the following formulas based on the model's specific assumptions. Following this derivation, the estimated parameters are then substituted into the model to obtain the final variance estimate.} %$\hat V^2_t$ \zzz{can be \jj{estimated} using the following formulas for different cases.}
%is given in Table ?? (\jj{we may need to list the value for all the possible cases. Appendix?})
%\jj{ [It looks like your idea before was ok, but you just need to explain how the formulas were used. Check if the following is correct - what you want is to estimate the model variance $\hat V_t^2$ with unit exposure which should be derived first by their model assumptions and then insert the estimated parameters to get their estimation.]\\
%Case 1: For $S=\sum_{j=1}^N Y_j$, the formula to use for variance is
%\zzz{ , where $N$ is included as a model weight (see Subsection \ref{sec:gamma_ind_bcart}). These estimated parameters are then substituted into the formulas for calculating $Var(S)$. The same approach is applied to $N$.}\\
Further, if the average severity model is as in Subsection \ref{sec:sev_bcart}, assuming $N$ and $\bar S|N>0$ are independent, we use %\ct{do not display the following equation to save space?}
\COM{$$
\mathbb{E}(S)=\mathbb{E}(S \mid N>0) \mathbb{P}(N>0)+\mathbb{E}(S \mid N=0) \mathbb{P}(N=0).
$$
Given that $\mathbb{E}(S \mid N=0)=0$ and $\mathbb{E}(S \mid N>0)=\mathbb{E}(N \bar{S} \mid N>0)$, we can obtain:
$$
\mathbb{E}(S)=\mathbb{E}(N \bar{S} \mid N>0) \mathbb{P}(N>0)=\mathbb{E}(N \mid N>0)  \mathbb{E}(\bar{S} \mid N>0)\mathbb{P}(N>0)
$$
which uses the independence assumption in the second equality.

Similarly, we know
\begin{eqnarray*}
\mathbb{E}(S^2)&&=\mathbb{E}\left((N\bar{S})^2 \mid N>0\right) \mathbb{P}(N>0)+\mathbb{E}(S^2 \mid N=0)\mathbb{P}(N=0)\\
&&=\mathbb{E}\left((N \bar{S})^2 \mid N>0\right)\mathbb{P}(N>0)\\
&&=\mathbb{E}(N^2 \mid N>0) \mathbb{E}(\bar{S}^2 \mid N>0) \mathbb{P}(N>0)\\
&&=\left[\left((\mathbb{E}(N\mid N>0)\right)^2+\text{Var}(N\mid N>0)\right]\mathbb{E}(\bar{S}^2 \mid N>0) \mathbb{P}(N>0). 
\end{eqnarray*}
The variance can be computed as
$$
\begin{aligned}
\text{Var}(S)&=\mathbb{E}(S^2)-\left(\mathbb{E}(S)\right)^2 \\
&=\left[\left((\mathbb{E}(N\mid N>0)\right)^2+\text{Var}(N\mid N>0)\right]\mathbb{E}(\bar{S}^2 \mid N>0) \mathbb{P}(N>0)\\
& \ \ \ \ -\left(\mathbb{E}(N \mid N>0)  \mathbb{E}(\bar{S} \mid N>0)\mathbb{P}(N>0)\right)^2.
\end{aligned}
$$
} $\text{Var}(S)= \left(\text{Var}(N) +(\E(N))^2\right) \text{Var}(\bar S|N>0) + \text{Var}(N) (\E(\bar S|N>0))^2$
to derive an estimate for $\hat V_t$ for the $t$-th cell, together with corresponding estimated parameters for different average severity models in Tables \ref{table_Sev_pred}--\ref{table_Sev_Var} and for different frequency models in \cite{zhang2024bayesian}.

    \item[{\bf M4':}] Model Lift - \zzz{similarly defined %indicates the ability to differentiate between the group of policyholders with low and high %\zzz{claim} cost (or \zzz{claim} 
   % risk (average severity here), and is defined by using the data and their predicted values in the most and least risky terminal nodes taking into account the weights.  
    %We use a similar approach 
    as M4 and \cite{zhang2024bayesian}}.  %\jj{(Note - We need more details in M4, in order to be able to say this here.)} % in \cite{zhang2024bayesian}.
    %to calculate lift for the severity tree models. 
% \item[{\bf M5:}] Negative log-likelihood (NLL) is calculated by using the assumed response distribution in the terminal node with the estimated parameters from obtained training, see Table \ref{table_Sev_pred}. % estimated from the training. %compare the losses of the models’ corresponding distributions because they 

\end{itemize}

\begin{remark}\label{Rem:Measure}

%\zzz{(a). Although RSS is a commonly used metric with several advantages, such as simple interpretation and good mathematical properties. It is important to be aware of its limitations, especially when dealing with datasets that contain large outliers. For example, in insurance claim severity data, RSS places considerable emphasis on outliers (extremely large claim amounts). This can lead to an excessively large RSS, which is undesirable.}

%\zzz{(b). While lift can provide insights into the economic value of the model, it is a relative measure. Therefore, it may not provide a clear indication of the actual economic impact or value of a model without additional context.} 

 %\zyy{Suppose we have a claim frequency tree modeled using a Poisson distribution with parameter $\lambda$ and an average severity tree modeled using a Gamma distribution with parameters $\alpha$ and $\beta$. For the joint partition of these two trees, each partition $t (t=1,2,\ldots,c)$ has corresponding estimated parameters $\hat{\lambda}_t$, $\hat{\alpha}_t$ and $\hat{\beta}_t$. The estimated aggregate claim amount for partition $t$ is then given by $\hat{S}_t=\hat{\lambda}_t \hat{\alpha}_t/\hat{\beta}_t$. Other combinations of frequency-severity models can follow the same approach to obtain the estimated aggregate claim amount.} 

   We remark that for each of the  BCART models, we apply the three-step approach in Table 1 of \cite{zhang2024bayesian} % \ref{table_SS}
to select a tree model. The above evaluation metrics are then used to evaluate the performance of these tree models on test data, based on which we can select a best tree model among different types of BCART models. As observed in  \cite{zhang2024bayesian} when discussing different types of frequency BCART models, %among the four evaluation metrics discussed above, 
all four metrics yield the same type of tree model choice based on their performance on test data. It is worth pointing out that %both metrics SE and DS 
%both metrics RSS and lift tend to improve for larger trees, but with slighter improvement for trees larger than the one selected based on DIC, %more splits in tree models, 
%while 
within a certain type of BCART model, the test data performance using SE and DS  aligns with the tree model selected using the three-step approach and thus they are deemed to be preferred metrics for comparison. %will also be verified in the 
See also Sections \ref{sec:sim} and \ref{sec:rd} %simulation examples and real data analysis below 
for further discussion. 

%\ct{I think it is correct to choose the model based on SE and DS (on the training data) but to use RSS and lift on the test data for comparison}%  can assist in identifying the optimal model.  

\end{remark}

%\yz{Although it is possible to combine two trees to obtain a joint partition, if both separate trees are already large, the process of deriving their combined partition becomes significantly complex.} 
%and may be unnecessary. ###discuss it for the coming paper} 
%Therefore, we do not apply them in the subsequent comparisons. 

%\ljj{(maybe remove this comparison completely? This does not seem to be important in the era  of big data, and has been justified in our previous paper!) From another perspective, two additional indicators, time and memory usage, can be employed to examine the computational efficiency. For the frequency-severity models, time and memory usage are determined by the summation of corresponding values from the two trees.}

%\textcolor{red}{obtain the results for SE, DS, and lift for two trees...%emphasize more for two trees to introduce something new...more details about joint partitions...see another paper to see if we can formulate this...introduce something new here}
%\zzz{sequential models: how to do the production and NLLs... in theory, not ok, in practice, use it (not precise)
%}

\subsection{Sequential BCART models}\label{sec:sequential}

In this section, we introduce the sequential model to better capture the potential association between \zzz{the number of claims} and average severity. 
One popular approach in the literature, e.g., \cite{frees2011predicting, garrido2016generalized,gschlossl2007spatial}, is to %the dependence can be introduced by treating 
treat the number of \zzz{claims} $N$ as a covariate for the average severity modeling. % in \eqref{eq:FS-0}. 
Following this idea, our sequential BCART model consists of %\zyy{three} 
two steps: 1) model the frequency component of \eqref{eq:FS-0} using the BCART models developed in \cite{zhang2024bayesian}; 2) treat the number of \zzz{claims} $N$ as a covariate (also treated as \zzz{a} model weight in the GammaN model) for the average severity component in \eqref{eq:FS-0} using the BCART models introduced in Subsection \ref{sec:gamma_ind_bcart} or  \ref{sec:sev_bcart}. % \zyy{3) derive %the approximate posterior predictive distribution of 
%the aggregate claim amount using Monte Carlo simulations based on separate models for the number of claims and the average severity; see \cite{gschlossl2007spatial}.}
% This strategy has gained popularity due to the increased focus on the dependence in aggregate \zzz{claim} modelling. Recent studies have explored it extensively;

%including $N$ as a covariate when modelling the average severity.

 %see, e.g., \cite{frees2016multivariate}, \cite{shi2015dependent} and \cite{garrido2016generalized}.

%In this subsection, to address the dependence between the number of \zzz{claim} and \jj{average severity}, we shall treat the claim count $N_i$ as a covariate (\yz{also treated as model weights in some cases; see Subsection \ref{sec:gamma_ind_bcart}}) in the \jj{average severity} tree, keeping everything else the same as the frequency-severity models in the previous section. 
When modeling average severity with $N$ as a covariate, there are usually two ways to treat $N$, namely, either use $N$ as a numeric covariate (see \cite{garrido2016generalized}) or treat $N$ as a factor % with different levels 
(see \cite{gschlossl2007spatial}). %where $N_i=q$ $(q \in \mathbb{N^{+}})$ can be taken as a reference level. The latter approach is more suitable when there are not many different values for $N_i$; see, e.g., \cite{gschlossl2007spatial}. 
 In this \zzz{paper}, we propose %\ga{delete: another way of }
 including the information of claim count for the average severity modeling, %\ga{delete: that is, we} 
 using the estimation of claim count $\hat{N}$ from the frequency BCART model as a numeric covariate. The underlying ideas for this proposal are as follows. First,  for a new observation we do not know $N$ but can  only obtain its estimate $\hat{N}$ through frequency model. Second, %the rationale
the frequency tree will classify the policyholders with similar risk (in terms of claim frequency) into the same cell and assign similar estimations $\hat N$ (the value of them depends also on their exposure). If the claim count information is highly correlated to the average severity, then the estimated value $\hat N$ will be chosen as the splitting covariate and the policyholders in the same frequency cell % with similar risk in terms of claim frequency 
will be more likely (than using $N$) to be classified into the same cell by the average severity tree. %\zyy{[it is not exactly correct since the same $\hat{N}$ may come from two different frequency cells because of $v_i$; and I think we should explain the another reason we use $\hat{N}$ is that sometimes we do not know the real information about $N$. LJI- yes, it is not precisely correct, but we just need intuition here, considering that most of the exposures are 1, and there are many zeros.]} 
In doing so, we expect the sequential model would be able to better capture the potential dependence between \zzz{the number of claims} and average severity. This is \zzz{demonstrated} to be true by our \zzz{simulated and real data below.}
%simulation examples and real %\zzz{non-life} 
%insurance data below. %[\ljj{Is this the case?}].
% by the BCART models for average severity. 
%Considering that in real life, it is important to allow premium estimation for new customers when there is no observed claim count $N_i$, 

\COM{ %%%%%%%%%%%%%%%%%%%%%%%%%%%%%%%%%%%%%%%%%%%%%%%%%%%%
Using the tower property of probability expectation, it is easily seen that the expectation of the aggregate claim amount $S_i=\sum_{j=1}^{N_i} Y_{ij}=N_i \bar{S}_i$ for an individual policyholder $i$ can be given as (omitting the subscript $i$ for simplicity),
\begin{equation}\label{eq:S_mean_dependence}
\begin{aligned}
	\mathbb{E}(S) & =\mathbb{E}(N \bar{S})  =\mathbb{E}(\mathbb{E}(N \bar{S} \mid N))  =\mathbb{E}(N \mathbb{E}(\bar{S} \mid N)). %\neq \mathbb{E}(N) \mathbb{E}(\bar{S}).
\end{aligned}
\end{equation}
\yz{Because of this formulation, we can estimate the expected \jj{average severity} $\mathbb{E}(\bar{S} \mid N)$ using $N$ (or $\hat{N}$) as a covariate, as in \cite{garrido2016generalized}.} \zzz{check it} 
} %%%%%%%%%%%%%%%%%%%%%%%%%%%%%%%%%%%%%%%%%%
The sequential BCART models will also \lj{result in} two trees, one for frequency and the other for average severity. \llji{If the resulting severity tree does not include the number of claims $N$ (or $\hat N$) as its splitting variable, then the  evaluation metrics and estimates introduced in Subsection \ref{sec:eva_two_trees} can still be employed. %Clearly, if the number of claims $N$ (or $\hat N$) will be chosen as a splitting variable in the \zzz{average} severity tree, then 
Otherwise, the aggregate claim amount $\hat S$ cannot be  estimated directly by multiplying the estimates from these two trees due to lack of independence, %as in \eqref{eq:SSi}, 
for which we will apply a Monte Carlo method following the idea in  \cite{gschlossl2007spatial}. In this case, %The evaluation metrics introduced in Subsection \ref{sec:eva_two_trees} cannot be directly applied to sequential BCART models due to the
there are also challenges involved in combining two trees to form a joint partition of the covariate space $\CMcal{X}$, since now %Although the underlying idea remains the same, %to merge all the splitting rules, 
%a key issue is that we lack the information about the number of claims for a new policyholder, a 
%$N$ or ($\hat N$) covariate that would  
the average severity tree does not only involve splitting variables $\vk x$  but also involve $N$ (or $\hat N$). To overcome this, we propose to construct a joint partition using only the splitting rules introduced by $\vk x$ in these two trees, ignoring any ones based on $N$ (or $\hat N$).  % during the tree merging process. % since the frequency tree already incorporates that information. 
As a result, %while the formulas for the four evaluation metrics remain unchanged, the way to obtain estimates differs. Details on the  differences 
some changes for M1'-M4' are necessary, which are as follows. Assume the joint partition has $c$ cells.}

\begin{itemize}
    \item \llji{M1': we will use $\text{RSS}(\vk{{S}})= \sum_{i=1}^{ m} ({S}_{i} - \hat{ S}_{i})^2,$ where the prediction $\hat{S}_{i}$ is obtained using Monte Carlo simulation  for the $i$-th test data given $(\vk x_i,v_i)$. The simulation algorithm is included in Section SM.B of the Supplementary Material; see Algorithm SM.1.} %\zyy{[Should we modify the two previous formulas for RSS($\vk{S}$) to maintain the consistency? LJI-We can leave it as it is]}
    %It is applied to each individual in the test data, where $v_i$ is assumed to be known and should be used in the prediction.

    \item \llji{M2'-M4': we need estimates for $\hat{S}_t, \hat{V}_t$ ($t=1,\ldots, c$) which can be obtained using similar Monte Carlo method as for M1'. Specifically, we consider a generic data in node $t$, denoted by $\vk x$ with an exposure $v=1$. Inputting this data $(\vk x,v=1)$ into the Algorithm SM.1, %in the Supplementary Material, 
    we  obtain a sequence of realizations $S_{t}^{1},S_{t}^{2},\ldots,S_{t}^{R}$, from which we can obtain $\hat{S}_t$ and $\hat{V}_t$ using their empirical mean and variance.}
   % the approach differs slightly, as they require considering the combined tree structure. Specifically, We can assume $v=1$ in Algorithm SM.1 (see Supplementary Material SM.B) to obtain $S_{t}$ for each combined group $t=1,\ldots,c$.  The algorithm generates multiple realizations of the aggregate claim amounts, denoted as $S_{i}^{1},S_{i}^{2},\ldots,S_{i}^{R}$. 
   % Using each policyholder's information (i.e., risk factors), we can assign them to the combined groups $t=1,\ldots,c$, enabling us to obtain the aggregate claim amount for each group, i.e., $S_{ti}$ ($i=1,\ldots,n)$. We then compute $S_{t}=\sum_{j=1}^{n_{t}}S_{tj}/n_t$, where $j$ denotes a policyholder within the $t$-th group, and $n_t$ is the number of policyholders in that group. For the model variance of $S$ in the $t$-th terminal group, denoted as $\hat{V}_t$ (i.e., Var($S$)), we can directly compute it since the realizations of $S_{ti}$ are available. 
    
    \end{itemize}

\begin{remark}
\llji{In \cite{quan2023hybrid}, a hybrid approach is used for insurance pricing where the  terminal nodes of a frequency tree serve as the partition of $\CMcal{X}$ and a regression model is used in each terminal  node for claim severity. In contrast, our approach for partition incorporates information from both  frequency tree and  severity tree. %, which can sometimes identifies different relevant variables that are useful for pricing. 
We have validated the effectiveness and superiority of our method (over a partition using only frequency tree) in both simulated and real data analysis.} %\llji{May be you can also try to see and compare the result of M2'-M3', just assuming the frequency tree.}

\end{remark}

\subsection{Joint BCART models}\label{sec:joint}

%Following this, we introduce a third model, joint models, which utilize CPG and ZICPG distributions for bivariate response (number of \zzz{claim} and aggregate claim amount) modelling; see, e.g., \cite{quijano2015generalised} and \cite{smyth2002fitting}. In particular, for ZICPG distributions, we employ the data augmentation technique (see \cite{murray2021log}) and explore different ways to embed the exposure, as discussed in \cite{zhang2024bayesian}.

% motivated by \cite{jorgensen1994fitting, smyth2002fitting}, we also introduce the {\it joint BCART models} by considering

\ljj{
Different from the previous two types of BCART models where separate tree models are used for the frequency and average severity, in this section we introduce the third type of BCART models, called joint BCART models, where we consider  $(N,S)$ as a bivariate response; see \cite{jorgensen1994fitting, smyth2002fitting} for a similar treatment in GLMs. We discuss
%an alternative approach for aggregate \zzz{claim} modelling by using one joint tree, i.e., using a \textit{joint model} for bivariate response (number of \zzz{claim} $N_i$ and aggregate claim amount $S_i$) modelling. 
two commonly used distributions for aggregate claim amount $S$, namely, \jjj{CPG and ZICPG distributions}. %compound Poisson gamma distribution (CPG) and zero-inflated compound Poisson gamma distribution (ZICPG). 
The presence of a discrete mass at zero makes them suitable for modeling aggregate claim amount; see, e.g., \cite{quijano2015generalised, yang2018insurance, denuit2021autocalibration}. As in \cite{zhang2024bayesian}, for the ZICPG models we need to employ a data augmentation technique. We also explore different ways to embed exposure.%\ga{REMOVE? resulting in slightly different models.}\ct{agreed}
} %and a new data augmentation technique will be introduced in Subsection \ref{sec:ZICPG}.
\ljj{
The advantage of modeling frequency and (average) severity components separately has been recognized in the literature; see, e.g., \cite{quijano2015generalised, frees2016multivariate}. In particular, this separate treatment can reflect the situation when the covariates that affect the frequency and severity are very different. However, one disadvantage is that it takes more effort to combine the two resulting tree models, as we have already seen in Subsection \ref{sec:eva_two_trees}.
Compared to the use of two separate tree models, the advantage of joint modeling is that the resulting single tree is easier to interpret as it simultaneously gives estimates for frequency, pure premium and thus average severity. Additionally, for the situation where frequency and average severity are linked through shared covariates, using a parsimonious joint tree model % for the frequency and aggregate claim amount 
might be advantageous; this will be illustrated in the examples in Section \ref{sec:sim}.}

\zyy{Since the ZICPG model is an extension of the CPG model and accounts for the high proportion of zero claims in practice, we focus primarily on the ZICPG model and refer to Section SM.C of the Supplementary Material for details of the CPG model.}

\subsubsection{Zero-inflated compound Poisson gamma model}\label{sec:ZICPG}

 We consider a response $(N,S)$ \lj{with exposure $v=1$,} %\zyy{[why? this is a general model and we need to consider $v$. LJI-because at this point, we only need to explain the notation of ZICPG(...), and do not want to introduce the complicate "exposure" which is specified later.]} % in the framework of Section \ref{Sec_BCART},
 where $N$ is zero-inflated Poisson distributed with parameters $\mu$ and \lj{$\lambda$}, 
 and $S=\sum_{j=1}^N Y_j$  %defined in \eqref{eq:SYj}
with individual severity $Y_j$ following \zzz{independent} gamma distribution with parameters $\alpha, \beta>0$. In the following, $S$ is called a zero-inflated compound Poisson gamma random variable, denoted by $\text{ZICPG}(\mu, \lambda, \alpha, \beta)$. \lj{
According to the general BCART framework in Section \ref{Sec_BCART}, considering a tree $\CMcal{T}$ with $b$ terminal nodes and $\vk \theta_t=(\mu_t, \lambda_t,\alpha_t, \beta_t)$ $(t=1,\ldots,b)$, we assume $
N_i| \vk x_i \sim \text{{ZIP}} \left(\mu(\vk x_i), \lambda(\vk x_i) \right)$, and  $S_i| \vk x_i, N_i>0\sim \text{Gamma} \left( %\lambda(\vk x_i) v_i, 
N_i\alpha(\vk{x}_i),  \beta(\vk{x}_i)\right)
$ for the $i$-th observation,
where $\mu({x_i}) = \sum_{t=1}^{\jj{b}} \mu_t I_{(\vk x_i \in \CMcal{A}_t)}$, $\lambda(\vk{x_i}) = \sum_{t=1}^{\jj{b}} \lambda_t I_{(\vk x_i \in \CMcal{A}_t)}$, $\alpha(\vk{x_i}) = \sum_{t=1}^b \alpha_t I_{(\vk x_i \in \CMcal{A}_t)}$ and $\beta(\vk{x_i}) = \sum_{t=1}^b \beta_t I_{(\vk x_i \in \CMcal{A}_t)}$.
}
%\ct{OMIT: Since insurance data normally includes a large number of zero \zzz{claims}, it is natural to model the zero mass part separately to the non-zero part resulting in a zero-inflated model; see, e.g., \cite{zhou2022tweedie}.} 
%In this section, we again consider a bivariate response $(N,S)$, where $N$ is now zero-inflated Poisson distributed and  $S$ is defined in \eqref{eq:SYj} with individual severity $Y_j$ following a gamma distribution with parameters $\alpha, \beta>0$. In the following, $S$ is called a zero-inflated compound Poisson gamma random variable. 
%Unlike the CPG models, 

\zzz{For ZICPG models, we need to} introduce a data augmentation strategy as in \cite{zhang2024bayesian} to obtain a closed form expression for the integrated likelihood; see \eqref{eq:ZICPG3_integrated} below.
Motivated by the discussion on the ZIP-BCART models in \cite{zhang2024bayesian}, we construct three ZICPG models according to how the exposure is embedded into the modeling. We try to cover
all three ZICPG models in a general set-up, which requires some general notation for exposure. \lj{%\ct{. OMIT:, with $\{\CMcal{A}_t\}_{t=1}^b$ being the corresponding partition of $\CMcal{X}$.} 
Specifically, for $i$-th observation such that $\vk{x}_i\in \CMcal{A}_t$, we have the joint distribution (recalling that the associated data for terminal node $t$ is denoted by 
$(\vk{X}_t, \vk{v}_t, \vk{N}_t, \vk{S}_t)$)}
\begin{equation}
\begin{aligned}
\label{eq:f_ZICPG3}
f_{\text{ZICPG}}(N_{ti}, S_{ti} &\mid \mu_{t}, \lambda_{t}, \alpha_t, \beta_t)  
=  f_{\text{ZIP}}(N_{ti} \mid \mu_t, \lambda_t) f_{\text{G}}(S_{ti} \mid N_{ti}, \alpha_t, \beta_t)  \\
=& \begin{cases} \frac{1}{1+\mu_t w_{ti}}+ \frac{\mu_t w_{ti}}{1+\mu_t w_{ti}} e^{- \lambda_t u_{ti}} & \jj{(N_{ti}, S_{ti})=\ }(0,0), \\ 
\frac{\mu_t w_{ti}}{1+\mu_t w_{ti}}\frac{(\lambda_t u_{ti})^{N_{ti}} e^{-\lambda_t u_{ti}} }{N_{ti}!} \frac{\beta_t^{N_{ti} \alpha_t} S_{ti}^{N_{ti} \alpha{_t}-1}  e^{-\beta_t S_{ti}} }{\Gamma(N_{ti} \alpha{_t}) } &  \jj{(N_{ti}, S_{ti})\in\ }(\mathbb{N} \times \mathbb{R}^{+}),  \end{cases}
\end{aligned}
\end{equation}
where  we use $w_{ti}$ to denote the ``exposure" for the zero mass part and $u_{ti}$ to denote the ``exposure" for the Poisson part. The above general formulation can cover three different models as special cases. Namely, 1) setting $w_{ti}=1$ and $u_{ti}=v_{ti}$, \ct{then%OMIT we have a model where
} the exposure is only embedded in the Poisson part, \jj{yielding the ZICPG1 model}; 2) setting $w_{ti}=v_{ti}$ and $u_{ti}=1$ \ct{then%OMIT: we have a model where
} the exposure is only embedded in the zero mass part, yielding the ZICPG2 model; 3)  setting $w_{ti}=u_{ti}=v_{ti}$  
\ct{means%OMIT we have a model where
} the exposure is embedded in both parts, yielding the ZICPG3 model. 
Note that $1/(1+\mu_t w_{ti})\in(0,1)$ is the probability that zero is due to the point mass component.

For computational convenience, a data augmentation scheme is used. To this end, we
introduce two latent variables $\vk{\phi}_t=(\phi_{t1},\phi_{t2},\ldots,\phi_{tn_t}) \in(0, \infty)^{n_t}$ and $\vk{\delta}_t=(\delta_{t1},\delta_{t2}, \ldots, \delta_{tn_t}) \in\{0,1\}^{n_t}$, and define the data augmented likelihood for the $i$-th data instance in terminal node $t$ as
\begin{equation}
\label{f_ZICPG3_aug}
\begin{aligned}
f&_{\text{ZICPG}}( N_{ti}, S_{ti}, \delta_{t i}, \phi_{t i} \mid \mu_t,  \lambda_{t}, \alpha_t, \beta_t) \\
%\begin{cases} 
=& e^{-\phi_{ti}(1+\mu_{t}w_{ti})} \left( \frac{\mu_{t}w_{ti}\left(\lambda_{ t} u_{t i}\right)^{ N_{t i}} }{N_{t i} !} e^{-\lambda_{t} u_{t i}} \right)^{\delta_{t i}} \jj{\left(\left(\frac{\beta_t^{N_{ti} \alpha_t} S_{ti}^{N_{ti} \alpha{_t}-1}  e^{-\beta_t S_{ti}} }{\Gamma(N_{ti} \alpha{_t}) }-1\right) I_{(N_{ti}>0)}+1 \right)},  %& N_{ti}=0 \\ 
%e^{-\phi_{t i}} \mu_t e^{-\phi_{t i} \mu_t}  \frac{\left(\lambda_{ t} v_{t i}\right)^{ N_{t i}} }{N_{t i} !} e^{-\lambda_{t} v_{t i}}& N_{ti}=1,2, \ldots\end{cases}
\end{aligned}
\end{equation}
where the support of the function $f_{\text{ZICPG}}$ is $\left(\{0\}\times\ljj{\{0\}\times}\{0,1\}\times(0,\infty)\right) \cup \left(\mathbb{N}\ljj{\times \mathbb{R}^{+}}\times\{1\}\times(0,\infty)\right)$. It can be shown that \eqref{eq:f_ZICPG3} is the marginal distribution of the above augmented distribution; see \zzz{Section SM.D of the Supplementary Material} for more details. %[\ljj{I think it is benefical to include the calculations in an appendix...}]%see Appendix \ref{ZIP_integral} for more calculation details. 

By conditional arguments, we can also check that $\delta_{ti}$, given 
data $N_{ti}=S_{ti}=0$ and parameters ($\mu_t$ and $\lambda_t$), has a Bernoulli distribution, i.e.,
$ \delta_{ti} \mid  N_{ti}=0, \mu_t, \lambda_{t}\  \sim \  \text{Bern} \left(\frac{\mu_{t}w_{ti} e^{-\lambda_t u_{ti}}}{1+\mu_{t}w_{ti} e^{-\lambda_t u_{ti}}} \right), $
%\yz{$\delta_{ti}$ is independent of $\phi_{ti}$ but related to $v_{ti}$..}
and $\delta_{ti}=1$ if $N_{ti}>0$. 
%\yz{\#given data $N_{ti}>0$}.
Furthermore, \ct{$\phi_{ti} \mid \mu_t \  \sim \  \text{Exp}\left(1+\mu_t w_{ti}\right)$. %OMIT rest $\phi_{ti}$, given $\mu_t$, has an exponential distribution, i.e.,
%$$    \phi_{ti} \mid \mu_t \  \sim \  \text{Exp}\left(1+\mu_t w_{ti}\right).$$
}
%\yz{$\phi_{ti}$ is independent of $\delta_{ti}$ and $\lambda_{t}$, but related to $v_{ti}$..}

For each terminal node $t$, we treat $\alpha_t$ as known, $\mu_t, \lambda_t$ and $\beta_t$ as unknown and apply the above data augmentation. According to the notation used in Section \ref{Sec_BCART} this means $\vk \theta_{t,M}= \alpha_t$ and $\vk\theta_{t,B}=(\mu_t, \lambda_t, \beta_t)$.
Here $\alpha_t$ will be estimated as in \eqref{eq:alpha_gamma2} \jj{using a subset of data with $N>0$.} \zzz{We treat $\mu_t$, $\lambda_t$ and $\beta_t$ as uncertain and use independent conjugate gamma priors, i.e.,} % with hyper-parameters 
$\mu_t\sim \text{Gamma}(\alpha^{(\mu)},\beta^{(\mu)}), 
\lambda_t\sim \text{Gamma}(\alpha^{(\lambda)},\beta^{(\lambda)}), \beta_t\sim \text{Gamma}(\alpha^{(\beta)},\beta^{(\beta)})$, \zzz{where the superscript $(\mu)$ (or $(\lambda)$ and $(\beta)$) indicates this hyper-parameter is assigned for the parameter $\mu_t$ (or $\lambda_t$ and $\beta_t$).}
%respectively, where the superscripts are 
%used as before (see Subsection \ref{sec:CPG_BCART}).
%Besides, $\alpha_t$ can be estimated and updated by using \eqref{eq:alpha_CPG} in each step before updating other parameters.
%With these Gamma priors and the estimated parameter $\hat{\alpha}_t$, t
Then, \zzz{given the estimated parameter $\hat{\alpha}_t$}, the integrated augmented likelihood for terminal node $t$ can be obtained as
\begin{align}
\label{eq:ZICPG3_integrated}
&p_{\text{ZICPG}}\left(\vk{N}_{t},  \vk{S}_{t}, \vk{\delta}_t,\vk{\phi}_t\mid \vk{X}_{t}, %\vk{w}_t, \vk{u}_t, 
\hat{\alpha}_t \right) \nonumber\\
&=\int_0^\infty \int_0^\infty  \int_0^\infty  f_{\text{ZICPG}}\left(\vk{N}_t,  \vk{S}_{t}, \vk{\delta}_t, \vk{\phi}_t \mid \mu_t,\lambda_{t}, \hat{\alpha}_t, \beta_t \right) p(\mu_t) p(\lambda_{t}) p(\beta_{t}) d\mu_t d \lambda_{t} d \beta_t \nonumber\\
&= \iiint  \prod_{i=1}^{n_t} \left( e^{-\phi_{ti}(1+\mu_{t}w_{ti})} \left( \frac{\mu_{t}w_{ti}\left(\lambda_{ t} u_{t i}\right)^{ N_{t i}} }{N_{t i} !} e^{-\lambda_{t} u_{t i}} \right)^{\delta_{t i}}  \right) \jj{\prod_{i:N_{ti}>0}}\frac{\beta_t^{N_{ti} \hat{\alpha}_t} S_{ti}^{N_{ti} \hat{\alpha}_t-1}  e^{-\beta_t S_{ti}} }{\Gamma(N_{ti} \hat{\alpha}_t) }  \nonumber\\
& \ \ \ \times \frac{\beta^{(\mu)\alpha^{(\mu)}}\mu_{t}^{\alpha^{(\mu)}-1} e^{-\beta^{(\mu)} \mu_{t}}}{\Gamma\left(\alpha^{(\mu)}\right)} \frac{\beta^{(\lambda)\alpha^{(\lambda)}} \lambda_{t}{ }^{\alpha^{(\lambda)}-1} e^{-\beta^{(\lambda)} \lambda_{t}}}{\Gamma\left(\alpha^{(\lambda)}\right)} 
\frac{\beta^{(\beta)\alpha^{(\beta)}} \beta_{t}{ }^{\alpha^{(\beta)}-1} e^{-\beta^{(\beta)} \beta_{t}}}{\Gamma\left(\alpha^{(\beta)}\right)} 
d \mu_{t} d \lambda_{t} d \beta_t \nonumber\\ 
&= \frac{\beta^{(\mu)\alpha^{(\mu)}}}{\Gamma\left(\alpha^{(\mu)}\right)} \frac{\beta^{(\lambda)\alpha^{(\lambda)}}}{\Gamma\left(\alpha^{(\lambda)}\right)} 
\frac{\beta^{(\beta)\alpha^{(\beta)}}}{\Gamma\left(\alpha^{(\beta)}\right)}
\prod_{i=1}^{n_t} \left(e^{-\phi_{t i}} w_{t i}^{\delta_{t i}} u_{ti}^{\delta_{t i}N_{t i}} \left({N_{t i} !}\right)^{-\delta_{t i}}\right) \jj{\prod_{i:N_{ti}>0}}\frac{S_{ti}^{N_{ti}\hat{\alpha}_t -1}}{\Gamma(N_{ti} \hat{\alpha}_t)} %\mathbbm{1}\left(Z_{t i}=1 \text { when } N_{t i}>0\right)
 \nonumber\\
&\ \ \ \times \frac{\Gamma\left(\sum_{i=1}^{n_t} \delta_{t i}+\alpha^{(\mu)}\right)}{\left(\sum_{i=1}^{n_t} \phi_{t i}w_{ti}+\beta^{(\mu)}\right)^{\sum_{i=1}^{n_t} \delta_{t i}+\alpha^{(\mu)}}} \frac{\Gamma\left(\sum_{i=1}^{n_t} \delta_{t i} N_{t i}+\alpha^{(\lambda)}\right)}{\left(\sum_{i=1}^{n_t} \delta_{t i} u_{t i}+\beta^{(\lambda)}\right)^{\sum_{i=1}^{n_t} \delta_{t i} N_{t i}+\alpha^{(\lambda)}}}\nonumber\\
& \ \ \ \times \frac{\Gamma(\sum_{i:N_{ti}>0} N_{ti}\hat{\alpha}_t+\alpha^{(\beta)})}{(\sum_{i:N_{ti}>0}S_{ti}+\beta^{(\beta)})^{\sum_{i:N_{ti}>0} N_{ti}\hat{\alpha}_t+\alpha^{(\beta)}}}.
\end{align}
%Moreover, from the above, we see that the posterior distributions of $\mu_t, \lambda_{t}$ are the same as in the ZIP3 model; and the posterior distribution of $\beta_t$ is the same as in the CPG model. 

The integrated augmented likelihood for the tree $\CMcal{T}$ is thus given by
%\yz{$\mu_{t}$ is independent of $v_{ti}$ and $\lambda_{t}$ is independent of $\phi_{t}$...}
\begin{equation*} %\label{eq:ZICPG3_whole}
   p_{\text{ZICPG}}\left(\vk{N}, \vk{S}, \vk{\delta},\vk{\phi}\mid \vk{X}, %\vk{w}, \vk{u}, 
   \vk{\hat{\alpha}}, \CMcal{T} \right)=\prod_{t=1}^{b} p_{\text{ZICPG}}\left(\vk{N}_{t}, \vk{S}_{t}, \vk{\delta}_t,\vk{\phi}_t\mid \vk{X}_{t}, %\vk{w}_t, \vk{u}_t, 
   \hat{\alpha}_t \right).
\end{equation*}

\zzz{We now discuss DIC which can be derived similarly as in Subsection \ref{sec:gamma_ind_bcart} with a three-dimensional unknown parameter $(\mu_t, \lambda_t, \beta_t)$. We first focus on DIC$_t$ of terminal node $t$. It follows that}
%Now, we discuss the DIC for this tree. % which can be derived as a special case of the DIC proposed in \cite{zhang2024bayesian} (see NB models) with a three-dimensional unknown parameter $(\mu_t, \lambda_t, \beta_t)$. We first focus on 
%As before the DIC$_t$ of terminal node $t$ is given as 
%It is derived that 
\begin{eqnarray}\label{eq:ZICPG3_D}
D\left(\bar{\mu}_t, \bar{\lambda}_t, \bar{\beta}_t\right) &&=-2 \log f_{\mathrm{ZICPG}}\left(\boldsymbol{N}_t, \boldsymbol{S}_t \mid \bar{\mu}_t, \bar{\lambda}_t, \hat{\alpha}_t, \bar{\beta}_t\right) \nonumber \\
&& =-2 \sum_{i=1}^{n_t} \log \left(\frac{1}{1+\bar{\mu}_t w_{t i}} I_{\left(N_{t i}=0\right)}+\frac{\bar{\mu}_t w_{t i}}{1+\bar{\mu}_t w_{t i}} \frac{\left(\bar{\lambda}_t u_{t i}\right)^{N_{t i}} e^{-\bar{\lambda}_t u_{t i}}}{N_{t i}!}\right. \nonumber \\
&& \quad \jj{\times\left(\left(\frac{\bar{\beta}_t^{N_{t i} \hat{\alpha}_t} S_{t i}^{N_{t i} \hat{\alpha}_t-1} e^{-\bar{\beta}_t S_{t i}}}{\Gamma\left(N_{t i} \hat{\alpha}_t\right)}-1\right) I_{\left(N_{t i}>0\right)}\zzz{+}1\right)},
\end{eqnarray}
where
\begin{equation}\label{eq:zip3_mu_lambda}
    \bar{\mu}_t= \frac{\sum_{i=1}^{n_t} \delta_{t i}+\alpha^{(\mu)}}{\sum_{i=1}^{n_t} \phi_{t i}w_{ti}+\beta^{(\mu)}}, \qquad \bar{\lambda}_t= \frac{\sum_{i=1}^{n_t} \delta_{t i} N_{t i}+\alpha^{(\lambda)}}{\sum_{i=1}^{n_t} \delta_{t i} u_{t i}+\beta^{(\lambda)}}, \qquad
    \bar{\beta}_t= \frac{\sum_{i:N_{ti}>0} N_{ti}\hat{\alpha}_t+\alpha^{(\beta)}}{\sum_{i:N_{ti}>0}S_{ti}+\beta^{(\beta)}}.
\end{equation}
%and $\bar{\beta}_t$ is given in \eqref{eq:CPG_mean_lambda-beta}. 
Furthermore, direct calculations yield the effective number of parameters for terminal node $t$  given by
\begin{align*}
\label{eq:ZICPG_sdt}
		\jj{p}_{D{t}}&= \jj{1}- 2 \mathbb{E}_{\text{post}}\left( \log f_{\text{ZICPG}}(\vk{N}_{t},\vk{S}_{t}, \vk{\delta}_t,\vk{\phi}_t\mid \mu_t, \lambda_t, \hat{\alpha}_t, \beta_t )\right) \\ 
  & \ \ \ \ +2 \log  f_{\text{ZICPG}}(\vk{N}_{t}, \vk{S}_{t}, \vk{\delta}_t,\vk{\phi}_t\mid \bar{\mu}_t, \bar{\lambda}_t, \jj{\hat{\alpha}_t},\bar{\beta}_t)
    \\
  & =  1 + 2 \left(\log \left(\sum_{i=1}^{n_t} \delta_{t i}+\alpha^{(\mu)}\right)-\psi \left(\sum_{i=1}^{n_t} \delta_{t i}+\alpha^{(\mu)}\right)\right) \sum_{i=1}^{n_t} \delta_{t i} \\ 
  & \ \ \ \ +2 \left(\log \left(\sum_{i=1}^{n_t} \delta_{t i} N_{t i}+\alpha^{(\lambda)}\right)-\psi \left( \sum_{i=1}^{n_t} \delta_{t i} N_{t i}+\alpha^{(\lambda)}\right)\right) \sum_{i=1}^{n_t} \delta_{t i} N_{t i} \\
  & \ \ \ \ + 2 \left(\log \left(\sum_{i:N_{ti}>0} N_{ti}  \hat{\alpha}_t+\alpha^{(\beta)} \right)-\psi \left(\sum_{i:N_{ti}>0} N_{ti} \hat{\alpha}_t+\alpha^{(\beta)}\right)\right) \sum_{i:N_{ti}>0} N_{ti} \hat{\alpha}_t, 
\end{align*}   
and thus % \ct{do not display the following equation to save space?}
$\text{DIC}=\sum_{t=1}^{\zzz{b}} \text{DIC}_t=\sum_{t=1}^{\zzz{b}} \left(D(\bar{\mu}_t,\bar{\lambda}_t,\bar{\beta}_t)+2 p_{Dt}\right).$
% can be derived. %directly from \eqref{eq:ZICPG3_D} and \eqref{eq:ZICPG_sdt}.

%For the above ZICPG model, the DIC of the tree $\CMcal{T}$ is obtained by using \eqref{eq:DIC_chapter4}. 
\COM{With the formulas derived above, we can use the three-step approach proposed in \cite{zhang2024bayesian}, together with Algorithm \ref{Alg:NB} (treat $\vk{\theta_M}=\vk{\alpha}$, $\vk{\theta_B}=(\vk{\mu},\vk{\lambda},\vk{\beta})$, and $\vk{z}=(\vk{\delta},\vk{\phi})$), to search for an optimal tree.
%which can then be used to predict new data.
}

Using these formulas for ZICPG,  we can follow the approach presented in Table 1 of \cite{zhang2024bayesian}, together with Algorithm \ref{Alg:2}  (here $\vk{z}_t=(\vk{\delta}_t,\vk{\phi}_t)$), to search for a tree which can then be used for prediction with \eqref{eq:yhat}. Given a tree, the estimated pure premium per year in  terminal node $t$ is given as
\begin{eqnarray}\label{eq:ZICPG_S}
    \bar S_t=\frac{\bar \mu_t\bar\lambda_t\hat{\alpha}_t}{\bar{\beta}_t(1+\bar \mu_t)},
\end{eqnarray} which can be determined using \eqref{eq:alpha_gamma2} %\eqref{eq:CPG_mean_lambda-beta} 
and \eqref{eq:zip3_mu_lambda}. 

\begin{remark}
%(a) \zyy{For compound distribution with independent $Y_j$ that are independent of $N$, it is implicitly assumed that $\bar S$ and $N$ are non-negatively correlated, as shown in \eqref{eq:fs-cov} and (10). The joint BCART model links different average severity models as discussed in Subsections \ref{sec:gamma_ind_bcart} and \ref{sec:sev_bcart}.}

%(b) 
We observe that the effective number of parameters %for each terminal node has the same expression as in all ZICPG models, 
 does not depend on exposures $w_{ti}$ and $u_{ti}$, illustrating that the way to embed the exposure does not affect the effective number of parameters. This is intuitively reasonable and is in line with the \jj{observations} for NB and ZIP models in \cite{zhang2024bayesian}. 
\end{remark}

\subsubsection{Evaluation metrics for joint models} \label{Sec:EM-JM}

Note that the ultimate goal in insurance rate-making is to set the pure premium based on the estimate of the aggregate \zzz{claim} amount $S$. Thus, for joint models, we focus on evaluation metrics defined via %the accuracy of the prediction of 
the second component $S$ in the bivariate response $(N,S)$. %Below we define some evaluation metrics which are in line with 
We follow the definitions of M1'--M4' in Subsection \ref{sec:eva_two_trees};  here the number of \zzz{cell}s $c$ is the number of terminal nodes $b$. 

%\zzz{Question: I think here we should follow the definition of M1-M4 directly since joint models only have one joint tree, which does not need to consider the combination of two trees. \jj{M1-M4 are for average severity, but below we want to use the formula for total amount $S$, in this case, M1'-M4' look to be more appropriate to refer to. ? M1'-M4' themselves do not involve the combination of trees.}}
 % the evaluation metric ({\bf M1-M5}) introduced for the average severity $\bar S$ in Subsection \ref{sec:eva_two_trees} can also be similarly defined for the aggregate claim amount $S$ and their estimations $\bar S$. 

Suppose we have obtained a tree with $b$ terminal nodes and corresponding %parameter estimates which we will use to obtain the 
predictions $\hat{S}_t$ $(t=1,\ldots,b)$ given in %\eqref{eq:CPG_S} or 
\eqref{eq:ZICPG_S}. Consider a test dataset with $m$ observations. 
%Given $m=\sum m_t$, d
Denote the test data in terminal node $t$  by  $\left(\vk{X}_t, \vk{v}_t, \vk{N}_t,\vk{ S}_t)=((\vk{x}_{t1}, v_{t1}, N_{t1}, S_{t1}), \ldots, (\vk{x}_{t  m_t}, v_{t m_t}, N_{t  m_t}, S_{t m_t})\right)^\top$. % $(t=1,\ldots,b)$. 
The RSS, SE and Lift are defined by M1', M2' and M4' respectively, with $c$ replaced by $b$. The DS is also similarly defined by M3', but with \jjj{$\hat V_t$} being equal to \zzz{$\bar{\lambda}_t \hat{\alpha}_t (1+\hat{\alpha}_t) / \bar{\beta}_t^2$} for the CPG model, and %\zzz{$\bar{\mu}_t \bar{\lambda}_t \hat{\alpha}_t (1+\hat{\alpha}_t) / (1+\bar{\mu}_t) \bar{\beta}_t^2$} 
\jj{$\bar{\mu}_t \bar{\lambda}_t  \hat{\alpha}_t(1+\hat{\alpha}_t+\bar{\mu}_t +\hat{\alpha}_t\bar{\mu}_t +\hat{\alpha}_t\bar{\lambda}_t) / \left((1+\bar{\mu}_t) \bar{\beta}_t^2\right)$} for the ZICPG model.

%%%%%%%%%%%%%%%%%%%%%5

\subsection{Two separate trees versus one joint tree: adjusted rand index}\label{sec:ari}

%\ct{OMIT? In the above subsections, we introduced three types of BCART models for aggregate \zzz{claim} amount, where both the frequency-severity model and the sequential model produce two separate optimal trees while the joint model produces one optimal joint tree. We presented evaluation metrics for these models in Sections \ref{eva:two_trees} and \ref{Sec:EM-JM}, %. This inclusion aims to facilitate the evaluation of model performance, 
%providing a basis for comparisons.} % with joint models using one joint tree. 
In this section, we extend our focus to examine the similarity between the BCART generated optimal trees. This exploration will give us confidence and valuable insights into whether information sharing through one joint tree is essential for model accuracy and effectiveness, compared to 
separate trees.% (from the frequency-severity model or sequential model).

Measuring the similarity of two trees is \jj{generally} challenging, particularly when there are variations in the number of terminal nodes or the structure (balanced/unbalanced) of the two trees;
see \cite{nye2006novel} and the references therein. %There are various methods to measure the similarity between different trees (see, e.g., \cite{nye2006novel}), typically involving considerations of the tree structure. Measuring similarity becomes challenging  
%\jj{As a first step}, 
We propose to explore \zyj{one simple index commonly employed in 
cluster analysis comparison, namely, the %Rand Index (RI) and 
adjusted Rand Index (ARI) which is a widely recognized metric for assessing the similarity of different clusterings; see, e.g., \cite{rand1971objective, hubert1985comparing, gates2017impact}.} We extend its application to evaluate the similarity of two trees. This is a natural application since a tree generates a partition of the covariate space which automatically induces clusters 
\ct{(i.e.\ observations belonging to the same leaf)} of policyholders in the insurance context. %\ct{??Omit mention of RI in this para and the next??} \zyj{Agree.} \jjj{I agree it seems strange to keep both RI and ARI if we only use ARI in the sequel, but we need a proper introduction of ARI then (actually, I did not check ARI, just know that the value can be obtained using fossil package as written here, I believe it is worth giving a clearer concise definition/idea of it, the current description does not seems to be very clear to me how ARI is working.).}

\zyj{The ARI measures the similarity between two data partitions by comparing the number of pairwise agreements and disagreements, adjusting for the possibility of random clustering to ensure that the index values are corrected for chance. This results in a score ranging from $-1$ to $1$, where $1$ indicates perfect agreement, $0$ suggests a similarity no better than random chance, and negative values imply less agreement than expected by chance. The ARI is particularly valued for its ability to account for \ga{%variations in Replace by?
different} cluster sizes and number of clusters, %\ga{adjusting for expected agreement of random clustering,} 
making it a robust metric. %\ct{remove? for comparing different clustering results.}
} \ct{Our results use \yyy{the} \texttt{adj.rand.index} \yyy{function} in the \textsf{R} package \texttt{fossil} \yyy{(see more details in \cite{R:fossil})}.} %\ct{remove as we already have reference [51] in the previous paragraph? See \cite{hubert1985comparing} for a detailed discussion of  RI and ARI.}

\section{Simulation examples}\label{sec:sim}

In this section, we investigate the performance of the BCART models \ga{introduced} in Section \ref{Sec:CF} by using simulated data. %[\jj{A summary on the following two simulations .}..]
\zzz{In Scenario 1, the effectiveness of sequential BCART models in capturing the dependence between the number of \zzz{claims} and average severity is examined, \zzz{along with \lj{their performance when using  claims count $N$ treated as a numeric variable} (or its estimate $\hat N$). % \footnote{$N$ is treated as a numeric variable in the following analysis.}) 
 Full details can be found in Section SM.E of the Supplementary Material; below we \lj{only present the simulation framework} and conclusion for the sake of brevity.} Scenario 2 focuses on the influence of shared information between the number of \zzz{claims} and average severity.}

In the sequel, we use the abbreviation Gamma-CART to denote CART for the Gamma model, %BCART to denote BCART, 
and other abbreviations can be similarly understood (e.g., ZICPG1-BCART denotes the BCART for \yyy{ZICPG1 model}).

\subsection{Scenario 1: Varying dependence between the number of \zzz{claims} and \jj{average severity}}\label{sec:agg_depen_sim}

We simulate  $\left\{(\boldsymbol{x}_i, v_i, N_i, \bar{S}_i)\right\}_{i=1}^n$ with $n=5,000$ independent observations. Here $\boldsymbol{x}_i=\left(x_{i 1}, x_{i 2}\right)$, with independent components $x_{i k} \sim N(0,1)$ for $k=1,2$. We assume exposure $v_i \equiv 1$ for simplicity. %as it is not a key feature in this context. 
Moreover, $N_i \sim \text{Poi}(\lambda(x_{i1},x_{i2}))$, where
\begin{eqnarray} \label{eq:LL}
    \lambda \left(x_{1},x_{2}\right)=\left\{\begin{array}{ll}
	1 & \text { if } x_{1}x_{2} \leq 0, \\
	7 & \text { if } x_{1}x_{2} > 0.
\end{array}\right.
\end{eqnarray}
We obtained $N_i=0$ for 901 occurrences, for which we set $\bar{S}_i=0$. For the remaining 4099 cases, $\bar{S}_i$ is generated from a gamma distribution with a pre-specified and \jj{varying} dependence parameter $\zeta$, i.e., 
$\bar{S}_i \mid N_i \sim  \text{Gamma}\left(1, 0.001+\zeta N_i\right)$,
in which 
%$\beta_{\zeta}=0.001+\zeta N_i$.  %the shape parameter of the Gamma distribution, and for simplicity, it is fixed at 1 since it is also not a key factor here. 
the %basic value for the
\yyy{shape (fixed as 1 for simplicity) and rate parameters are} chosen to maintain the \zzz{empirical} average claim amount $\bar{S}_i$ to be around 500, aligning with \ga{real-world scenarios}. % \jj{what is observed in real data}. %Similarly as before,  we  choose to make variables $x_1, x_2$ as characters for the algorithms (correct ??). 
The data is split into two subsets: a training set with $n-m=4,000$ observations and a test set with $m=1,000$ observations. \yz{In this case, our goal is to %examine
\zzz{investigate} how the dependence modulated by $\zeta$ influences the performance of both frequency-severity models and sequential models, and the performance of incorporating $N$ (or $\hat{N}$) into the sequential models. %The simulation setting avoids designing a tree structure for \jj{average severity} modelling as we would normally do before. Without prior knowledge of the optimal \jj{average severity} tree, we lack information about the splitting rules involved. Therefore, 
If %the models choose 
$N$ (or $\hat{N}$) is selected as a splitting covariate, it would indicate that the claim count plays an important role in \jj{average severity} modeling, and thus sequential models should be preferred.}
%Table \ref{table-S5-data} presents \yyy{a numerical} summary \jj{of the average severity} and conditional correlation coefficients between the number of \zzz{claims} and \jj{average severity} for datasets with different values of $\zeta$. It is obvious that 
\zzz{As $\zeta$ changes,}
%By changing the value of $\zeta$, 
the conditional correlation between the number of \zzz{claims} and \jj{average severity} varies. %For simplicity, we only focus on the case where $\zeta=0.001$, indicating a strong \zzz{negative conditional} dependence between them. 
\zzz{Stronger dependence (e.g., $\zeta$=0.001) is expected to favor sequential models, while weaker dependence (e.g., $\zeta$=0.00001) 
makes both model types perform similarly, as $N$ (or $\hat{N}$) is less likely to be \lj{\ct{used in} the sequential model tree}.}

%%%%%%%%%%%%%%%%%%%%%%%%%%
\COM{
\begin{table}[!t]  
	
	\centering
 \captionsetup{width=.8\linewidth}
	\caption{%Model comparison in test data 
		\zzz{Average iteration times to obtain an ``optimal'' tree (5 terminal nodes) and number of models selected by DIC as the optimal model.} The number in brackets after the abbreviation of the model indicates the number of terminal nodes for this tree. The Gamma1 and Gamma2 models treat the claim count $N_i$ and $\hat{N}_i$ as a covariate respectively, where $\hat{N}_i$ comes from Poisson-BCART. \zzz{(10 runs using 10 different datasets generated with the same data generating scheme)}}
	%\begin{tabular}{lp{1.7cm}p{2.4cm}p{1.9cm}p{2.1cm}p{1.5cm}} 
	\begin{tabular}{lcc}
		
		\toprule   
		
		Model & Average iteration times (s.d.) &  Number of models chosen by DIC \\ 
		\midrule   

		Gamma-BCART (4) & 2534 (182) & 0 \\
		\textbf{Gamma-BCART (5)} & 2569 (175) & 10 \\
		Gamma-BCART (6) & 2562 (190) & 0 \\\hline
		
		Gamma1-BCART (4) & 2489 (176) & 0 \\
		\textbf{Gamma1-BCART (5)} & 2473 (174) & 10  \\
		Gamma1-BCART (6) & 2544 (186) & 0  \\\hline

		Gamma2-BCART (4) & 2502 (165) & 0  \\
		\textbf{Gamma2-BCART (5)} & 2515 (168) & 10 \\
		Gamma2-BCART (6) & 2573 (184) & 0 \\
		
		\bottomrule  
		
	\end{tabular}
	
	\label{table-S5-3}
\end{table}
}

\lj{%Training data results indicate that %show that Gamma2-BCART (with DIC = 2618) performs best, confirming the benefit of including $\hat{N}$ as a covariate. 
The observed relatively large DIC differences in training data} between the \jj{Gamma model without claim count (or its estimate) as a covariate (i.e., Gamma-BCART)} and \zzz{those with it} (i.e., Gamma1-BCART %compared to the difference between Gamma1-BCART 
or Gamma2-BCART) suggests that \zzz{incorporating claim count (or its estimate) as a covariate improves performance in the presence of stronger inherent dependence} in the data, \lj{where Gamma2-BCART (with DIC = 2618) performs best.} \zzz{On test data, Gamma2-BCART also outperforms other models across all evaluation metrics. Additional experiments, repeated with new training and test datasets, confirm the robustness of these results.
%showing that they are not due to random variation.
}

\zzz{Further simulations with different \lj{pairs of values for} $\zeta$ and $\lambda$ in \eqref{eq:LL} reveal that: 1) 
%Even if the dependence between the number of claims and average severity is weak and the claim count is not included as a covariate, sequential models still outperform frequency-severity models. This is reasonable because sequential models inherently account for dependence through the Monte Carlo simulation algorithm used to calculate aggregate claim amounts, and the performance improves as the number of iterations increases;
Weak dependence between the number of claims and average severity results in similar performance between frequency-severity and sequential models, in this case the former is preferred for   computational efficiency; 
2) Strong dependence allows sequential models to perform better by incorporating} the number of claims (or its estimate) as a covariate; 3) Gamma2-BCART consistently outperforms Gamma1-BCART, validating our discussion in Section \ref{sec:sequential}.

\subsection{Scenario 2: 
\jj{Covariate\yyy{s} sharing between the number of \zzz{claims} and average severity}}\label{sec:agg_share_sim}

%\yz{Now we investigate scenarios where identical covariates exhibit similar or distinct impacts on \zzz{claim} frequency and \jj{average severity}. The objective is to assess the effectiveness of employing two trees and one joint tree in such cases and obtain a general conclusion.} 

\jj{In this scenario, we consider two simulations where common covariates are used for parameters representing the number of \zzz{claims} and \jj{average severity}. The objective is to assess the effectiveness of frequency-severity BCART models and joint BCART models, %\ct{omit? in such simulations}
that is, whether it is preferred to share information using one joint tree.} To this end, 
%To simplify and better illustrate the necessity of sharing information, 
we consider CPG distribution in joint BCART models, %in this example. %and the comparison of CPG and ZICPG models is provided in real data analysis, see Chapter 6. 
%Besides, since the CPG model involves Poisson and Gamma distributions, we restrict the use of 
and correspondingly, Poisson distribution and gamma distribution involving $N$ as a model weight in the frequency-severity BCART models to keep consistency for comparison. We first explain these two simulations and then present some %of the 
findings and suggestions.

\begin{figure}[!htb]
	\centering       \includegraphics[width=1\linewidth]{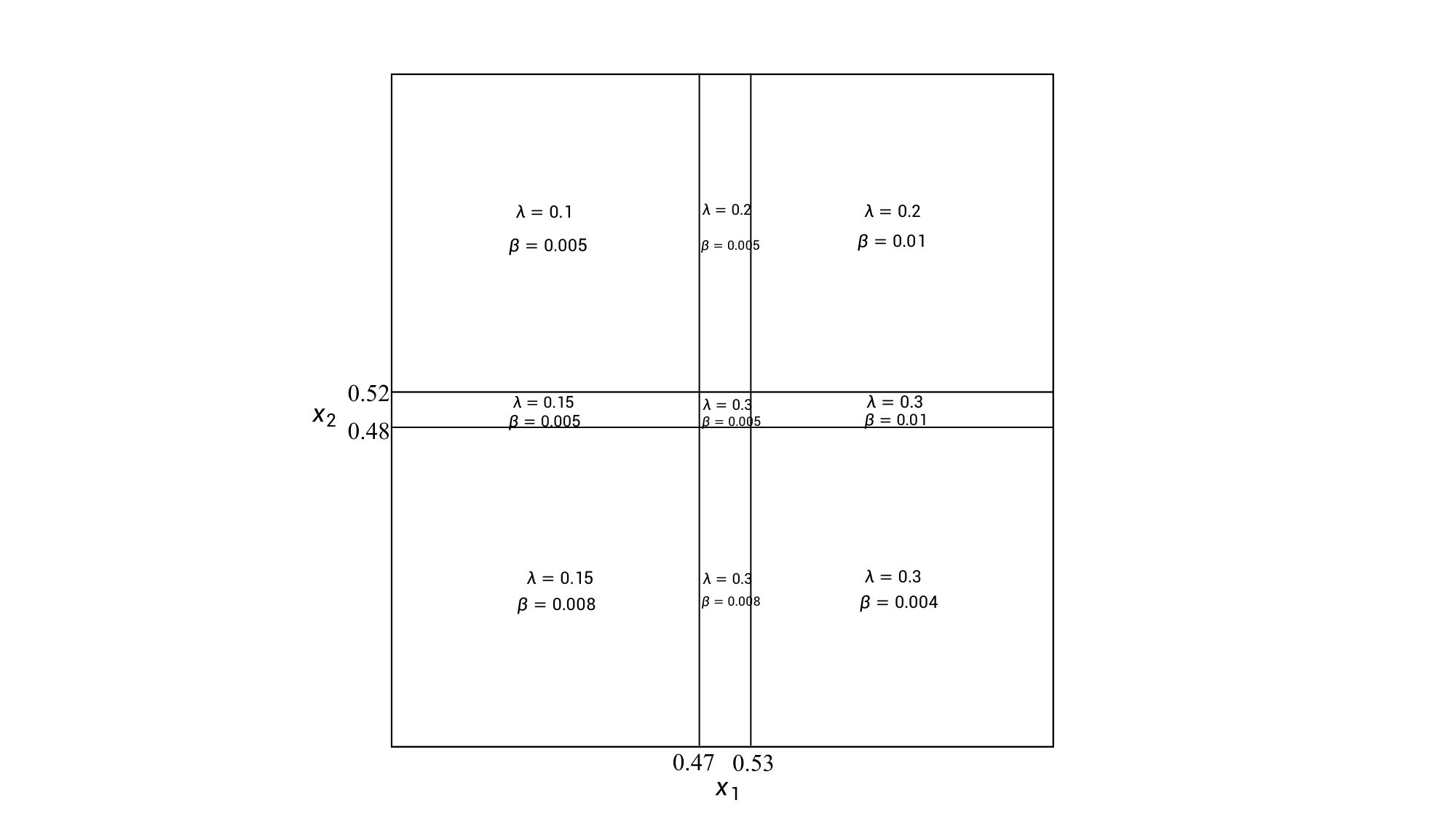}
 \captionsetup{width=.8\linewidth}
	\caption{Covariate space  partition for a \jj{CPG}-distributed simulation. %Two covariates $x_{1}$ and $x_{2}$ follow a Normal and Uniform distribution respectively, i.e., $x_{1} \sim N(0,1)$, $x_{2} \sim U(-1,1)$. 
 The values of parameters $\lambda$ and $\beta$ are provided in each region.}
	\label{Fig:shared_data2}
%\end{figure}
%\begin{figure}[hbtp]
	\centering       
 \includegraphics[width=1\textwidth]{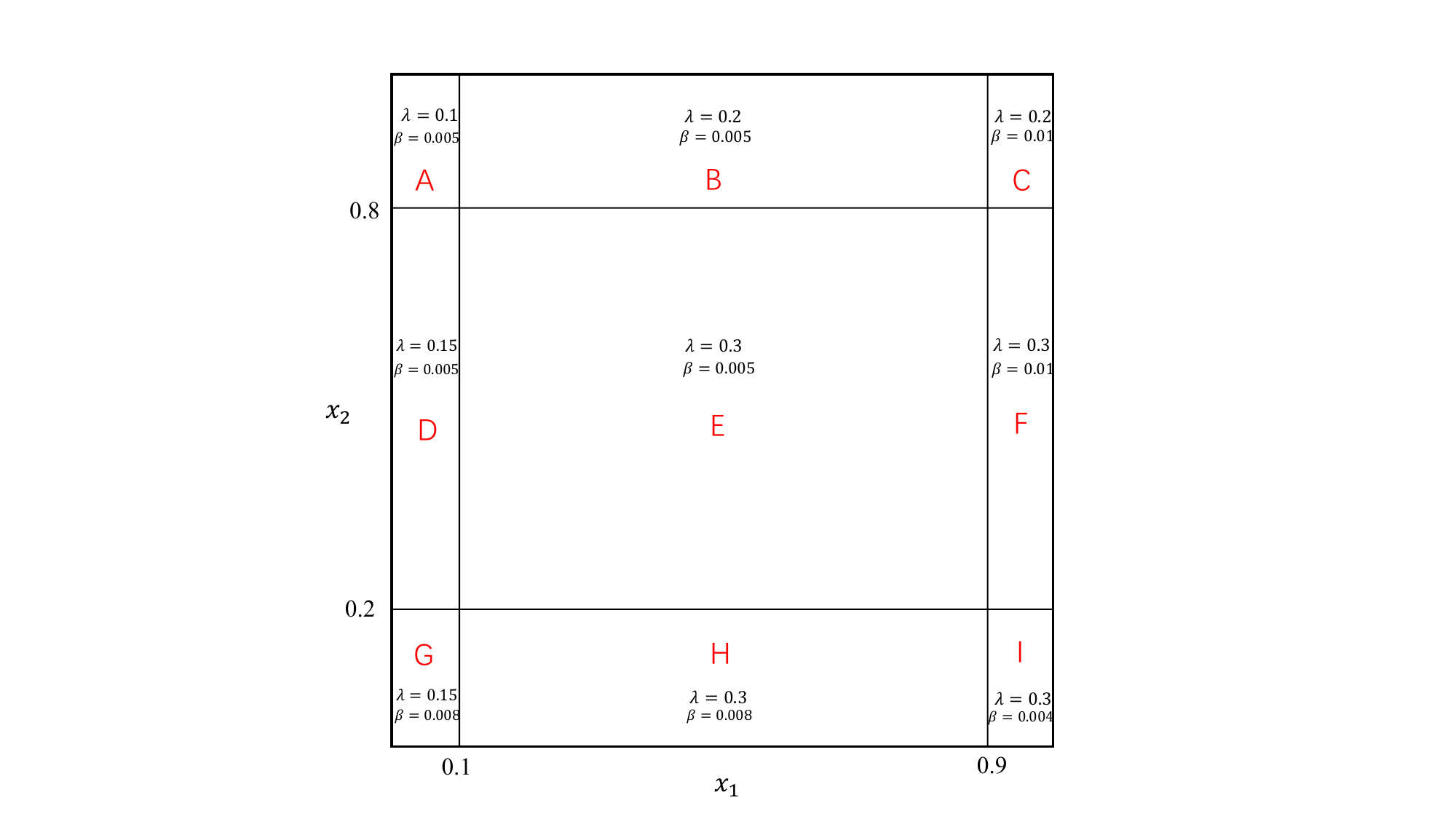}
 \captionsetup{width=.8\linewidth}
	\caption{Covariate space partition for a CPG-distributed simulation. %Two covariates $x_{1}$ and $x_{2}$ follow a Normal and Uniform distribution respectively, i.e., $x_{1} \sim N(0,1)$, $x_{2} \sim U(-1,1)$. 
 The values of parameters $\lambda$ and $\beta$ are provided in each region, \zzz{which has been labelled with names. %\ga{Why L and not I?} \jjj{It may be because $I$ was used for indicator function. But, I do think using $I$ here may be better which wouldn't cause any confusion}
 }}
	\label{Fig:shared_data1}
\end{figure}

%Specifically, the CPG distribution models individual claim amount $Y_{ij}$ first and then obtain the distribution for the aggregate claim amount ${S}_i$. We shall use Gamma-BCART proposed in Subsection \ref{sec:gamma_ind_bcart} for the frequency-severity models, which also models individual claim amount $Y_{ij}$ first rather than modelling $\bar{S}_i$ directly. In the simulation setting, we would model $\bar{S}_i$ involving $N_i$ as model weights, i,e., $\bar{S}_i \sim \text{Gamma}(N_i \alpha, N_i \beta (\vk{x}_i))$.
\underline{\bf Simulation 2.1}:
We simulate a dataset $\left\{(\boldsymbol{x}_i, v_i, N_i, \bar{S}_i)\right\}_{i=1}^n$ with $n=5,000$ independent observations. Here $\boldsymbol{x}_i=\left(x_{i 1}, \ldots, x_{i 5}\right)$, with independent components $x_{i1} \sim N(0,1)$, $x_{i2} \sim U(-1,1)$, $x_{i3}\sim U(-5,5)$, $x_{i4}\sim N(0,5)$, $x_{i5}\sim U\{1,2,3,4\}$ and $v_i \sim U(0,1)$, \jj{where %\ct{ OMIT:  $N(\cdot, \cdot)$ stands for normal distribution and} 
\zzz{$U(\cdot,\cdot)$ (or $U\{\cdot,\cdot\})$ stands for continuous (or discrete-type)} uniform distribution.} Moreover, $N_i \sim \text{Poi}(\lambda(x_{i1},x_{i2})v_i)$, where
$$
\lambda \left(x_{1},x_{2}\right)=\left\{\begin{array}{ll}
		0.1 & \text { if } x_{1}\leq 0.47,\, x_{2}>0.52, \\
		0.2 & \text { if } x_{1}>0.47,\, x_{2}>0.52, \\
		0.3 & \text { if } x_{1}> 0.47,\, x_{2}\leq 0.52, \\
		0.15 & \text { if } x_{1}\leq 0.47,\,x_{2}\leq 0.52.
	\end{array}\right.
$$
If $N_i=0$ then $\bar{S}_i=0$, otherwise %$\bar{S}_i$ follows a gamma distribution, i.e., 
$\bar{S}_i \sim \text{Gamma}\left(N_i \alpha, N_i \beta \left(x_{i1},x_{i2}\right)\right)$, where 
$$
\beta \left(x_{1},x_{2}\right)=\left\{\begin{array}{ll}
		0.005 & \text { if } x_{1}\leq 0.53,\, x_{2}>0.48, \\
		0.01 & \text { if } x_{1}>0.53,\, x_{2}>0.48, \\
		0.004 & \text { if } x_{1}> 0.53,\, x_{2}\leq 0.48, \\
		0.008 & \text { if } x_{1}\leq 0.53,\, x_{2}\leq 0.48.
\end{array}\right.
$$
For simplicity, we assume $\alpha=1$, %the shape parameter of the Gamma distribution, which is specified to be 
%fixed at 1% for simplicity since it is not a key feature here
 %Similarly as before,  we  choose to make variables $x_1, x_2$ as characters for the algorithms (correct ??). 
and the values of $\beta$ are selected  such that the average \zzz{claim} amount $\bar{S}_i$ is around 200, which is close to the situation \jjj{in  real-world scenarios.} See  Figure \ref{Fig:shared_data2} for an illustration of the true covariate space partition and corresponding values of parameters. %, the variables $x_{k}, k=3,4,5$ are all noise variables which are used to illustrate the effectiveness of the proposed models.
%independent of the bivariate response variable ($\vk{N}$,$\vk{\bar{S}}$). 

\underline{\bf Simulation 2.2}: We keep most simulation settings as in Simulation 2.1, except the partition of the covariate space; see Figure \ref{Fig:shared_data1}. Specifically, %e design another case where the values of split points are more distinct, i.e.,}
$$
\lambda \left(x_{1},x_{2}\right)=\left\{\begin{array}{ll}
		0.1 & \text { if } x_{1}\leq 0.1,\, x_{2}>0.8, \\
		0.2 & \text { if } x_{1}>0.1,\, x_{2}>0.8, \\
		0.3 & \text { if } x_{1}> 0.1,\, x_{2}\leq 0.8, \\
		0.15 & \text { if } x_{1}\leq 0.1,\, x_{2}\leq 0.8,
	\end{array}\right.
$$
and for non-zero $N_i$, generate  $\bar{S}_i \sim \text{Gamma}\left(N_i, N_i \beta(x_{i1},x_{i2})\right)$, where
	$$
	\beta \left(x_{1},x_{2}\right)=\left\{\begin{array}{ll}
		0.005 & \text { if } x_{1}\leq 0.9,\, x_{2}>0.2, \\
		0.01 & \text { if } x_{1}>0.9,\, x_{2}>0.2, \\
		0.004 & \text { if } x_{1}> 0.9,\, x_{2}\leq 0.2, \\
		0.008 & \text { if } x_{1}\leq 0.9,\, x_{2}\leq 0.2.
	\end{array}\right.
	$$
 
The specific design here is that both components of the response variable ($N$,$\bar{S}$) are affected by the same covariates $x_1$ and $x_2$. In Simulation 2.1 they share similar split points, while for Simulation 2.2 they have quite different split points. The variables $x_{k}, k=3,4,5$ are all noise variables. % which are used to illustrate the effectiveness of the proposed models.
We aim to compare separate BCART trees versus one joint BCART tree. 
%The intuition is not very clear because, although they share some information (having the same covariates and similar split values), the ideal partitions differ between the case of two trees and one joint tree. 

Each simulation dataset is split into a training set with $n-m=4,000$ observations and a test set with $m=1,000$ observations. The outputs from
%We shall use the comparison indexes in both
training data and test data for the BCART models are presented in Tables \ref{table-S6-3}--\ref{table-S6-4} for Simulation 2.1 and Tables \ref{table-S6-1}--\ref{table-S6-2}  for Simulation 2.2.

\begin{table}[tb] 
	
	\centering
 \captionsetup{width=.8\linewidth}
	\caption{Hyper-parameters, $p_D$  and DIC on training data for Simulation 2.1. The number in brackets after the abbreviation of the model indicates the number of terminal nodes for that tree. Bold font indicates DIC selected model.}
	
	%\begin{tabular}{lp{2cm}p{2cm}p{3cm}p{2cm}} 
	\begin{tabular}{l|cccc} 
		
		\toprule   
		
		Model & $\gamma$ & $\rho$ & $p_D$  & \multicolumn{1}{c}{DIC} \\  
		
		\hline %\midrule   
		
		Poisson-BCART (3) & 0.95 & 15 & 2.98 & 3697 \\
		\textbf{Poisson-BCART (4)} & 0.99 & 13 & 3.98  & \textbf{3572}  \\
		Poisson-BCART (5) & 0.99 & 10 & 4.97 & 3616 \\\hline
		
		Gamma-BCART (3) & 0.95 & 10 & 5.97 &  30586 \\
		\textbf{Gamma-BCART (4)} & 0.99 & 10 & 7.97 & \textbf{30319}  \\
		Gamma-BCART (5) & 0.99 & 8 & 9.96 & 30414 \\\hline
		\bottomrule
		
		CPG-BCART (3) & 0.99 &  5 & 8.92 & 34017 \\
		\textbf{CPG-BCART (4)} & 0.99 & 4 & 11.90 &  \textbf{33582}\\
		CPG-BCART (5) & 0.99 & 3 & 14.89 & 33711 \\\hline
		
		\bottomrule  
		
	\end{tabular}
	
	\label{table-S6-3}
\end{table}

\begin{table}[tb]  
	
	\centering
 \captionsetup{width=.8\linewidth}
	\caption{%Model comparison in test data 
		Model performance on test data with bold entries determined by DIC (see Table \ref{table-S6-3}). $\text{F}_{\text{P}}\text{S}_{\text{G}}$ \ga{denotes} the frequency-severity models by using Poisson and gamma distributions separately. The number in brackets after the abbreviation of the model indicates the number of terminal nodes for those trees. %Particularly, two numbers for frequency-severity models indicate the number of terminal nodes for each tree.
    }
	%\begin{tabular}{lp{1.7cm}p{2.4cm}p{1.9cm}p{2.1cm}p{1.5cm}} 
	\begin{tabular}{l|cccc}
		
		\toprule   
		
		Model & RSS($\vk{S}$) \zyj{$\times 10^{-8}$} & \zzz{SE} & \zzz{DS} & \zzz{Lift} \\  
		\hline %\midrule   

		$\text{F}_{\text{P}}\text{S}_{\text{G}}$-BCART (3/3) &  3.03 & 0.1324 & 0.0833 & 2.13
		 \\
		\textbf{$\text{F}_{\text{P}}\text{S}_{\text{G}}$-BCART (4/4)} & 2.89 & \textbf{0.1245} & \textbf{0.0791} & 2.21 \\
		$\text{F}_{\text{P}}\text{S}_{\text{G}}$-BCART (5/5) &  2.81 & 0.1273 & 0.0812 & 2.23 \\\hline

		CPG-BCART (3) & 3.01 & 0.1319  & 0.0820 & 2.15 \\
		\textbf{CPG-BCART (4)} & 2.84 & \textbf{0.1211} & \textbf{0.0769} & 2.27   \\
		CPG-BCART (5) & 2.78 & 0.1254 & 0.0795 &  2.29 \\
		
		\bottomrule  
		
	\end{tabular}
	
	\label{table-S6-4}
\end{table}

%[\ljj{Is it helpful to check the correlation of the simulated data ($N, \bar S$), etc?}]

\begin{table}[tb] 
	
	\centering
 \captionsetup{width=.8\linewidth}
	\caption{Hyper-parameters, $p_D$  and DIC on training data for Simulation 2.2. The number in brackets after the abbreviation of the model indicates the number of terminal nodes for that tree. Bold font indicates DIC selected model.}
 
	%\begin{tabular}{lp{2cm}p{2cm}p{3cm}p{2cm}} 
	\begin{tabular}{l|cccc} 
		
		\toprule   
		
		Model & $\gamma$ & $\rho$ & $p_D$  & \multicolumn{1}{c}{DIC} \\  
		
		\hline %\midrule   
		
		Poisson-BCART (3) & 0.95 & 15 & 2.97 & 3875\\
		\textbf{Poisson-BCART (4)} & 0.99 & 12 & 3.97  & \textbf{3669}  \\
		Poisson-BCART (5) & 0.99 & 10 & 4.96 & 3724 \\\hline
		
		Gamma-BCART (3) & 0.95 & 10 & 5.97 &  32156 \\
		\textbf{Gamma-BCART (4)} & 0.99 & 10 & 7.96 & \textbf{31798}  \\
		Gamma-BCART (5) & 0.99 & 8 & 9.96 & 31904 \\\hline 
  \bottomrule

		CPG-BCART (8) & 0.99 &  5 & 23.85 & 36174 \\
		\textbf{CPG-BCART (9)} & 0.99 & 3 & 26.81 &  \textbf{35622}\\
		CPG-BCART (10) & 0.99 & 2 & 29.79 & 35781\\\hline
		
		\bottomrule  
		
	\end{tabular}
	
	\label{table-S6-1}
\end{table}

\begin{table}[tb]  
	
	\centering
 \captionsetup{width=.8\linewidth}
	\caption{%Model comparison in test data 
		Model performance on test data with bold entries determined by DIC (see Table \ref{table-S6-1}). $\text{F}_{\text{P}}\text{S}_{\text{G}}$ denotes the frequency-severity models by using Poisson and gamma distributions separately. The number in brackets after the abbreviation of the model indicates the number of terminal nodes for those trees. %Particularly, two numbers for frequency-severity models indicate the number of terminal nodes for each tree.
  }
	%\begin{tabular}{lp{1.7cm}p{2.4cm}p{1.9cm}p{2.1cm}p{1.5cm}} 
	\begin{tabular}{l|cccc}
		
		\toprule   
		
		Model & RSS($\vk{S}$)$\times 10^{-8}$  & \zzz{SE} & \zzz{DS} & \zzz{Lift} \\  
		
		\hline %\midrule   

		$\text{F}_{\text{P}}\text{S}_{\text{G}}$-BCART (3/3) &  3.21 & %0.1521
  0.152 & %0.0912
  0.091 & 2.18
		 \\
		\textbf{$\text{F}_{\text{P}}\text{S}_{\text{G}}$-BCART (4/4)} & 3.04 & \textbf{%0.1394
  0.140} & \textbf{%0.0732
  0.073} &  2.30 \\
		$\text{F}_{\text{P}}\text{S}_{\text{G}}$-BCART (5/5) &  2.95 & %0.1412
  0.141 & %0.0789
  0.079 & 2.32 \\\hline

		CPG-BCART (8) & 3.23 & %0.1598
  0.160 &% 0.0965
  0.097 & 2.15 \\
		\textbf{CPG-BCART (9)} & 3.08 & \textbf{%0.1424
  0.142} & \textbf{%0.0881
  0.088} &  2.24 \\
		CPG-BCART (10) & 3.01 & %0.1461
  0.146 & %0.0902
  0.090 &  2.25 \\
		
		\bottomrule  
		
	\end{tabular}
	
	\label{table-S6-2}
\end{table}

%First, from model choice on training data, based on DIC and WAIC, we can know all models (P-BCART, NB-BCART and ZIP-BCART) can choose four terminal nodes, which is exactly the number we want. Then, we want to compare the performance in test data. 

\jj{We start by looking at the DICs on training data in Tables \ref{table-S6-3} and  \ref{table-S6-1}.} For Simulations 2.1 and 2.2, both Poisson-BCART and Gamma-BCART %, it is easy to check that both of them find 
can find the optimal tree with the correct 4 terminal nodes. %, as suggested by DIC on training data in Table \ref{table-S6-3}. 
However, the \jjj{selected} joint CPG-BCART tree for Simulation 2.1 has %, in the simulation setting, we expect it to still have 9 terminal nodes, %as in the previous example
%but DIC indicates that the tree 
only 4 terminal nodes which is different from its simulation scheme that should result in a covariate space partition with 9 \zyy{cell}s. In contrast, the \jjj{selected} joint CPG-BCART tree for Simulation 2.2 has 9 terminal nodes which is consistent with its simulation scheme. %is the best one. 
This result for the frequency-severity BCART models is expected, based on our earlier discussion of the separated frequency and average severity models. 
Now we look into the details of the \jjj{selected}  joint tree
to explore the reason. 
For Simulation 2.1, we see that both $x_1$ and $x_2$ are used in the tree and the %structure and find the chosen
split points for them are close to 0.5 which is the mean of % two different split values in $\lambda(x_1,x_2)$ and $\beta(x_1,x_2)$, namely, 
0.47 and 0.53 for $x_1$, and  also the mean of 0.52 and 0.48 for $x_2$. Since these split values are very close, it is reasonable for the joint BCART model to select a split value around their mean, resulting in a \jjj{selected} joint tree with 4 terminal nodes. 
For Simulation 2.2 the \jjj{selected} joint tree includes 9 terminal nodes, which is also reasonable because the split values for both variables are far apart. %\jj{[To do - look at the estimation of these trees and compare. I guess, the estimation of the joint tree may be not as good as the frequency-severity trees, because it uses less data to calculate these estimates. I also guess if the number of data increases, the performance of the frequency-severity model and joint model should perform similarly for Simulation 2.2, and thus you might not be able to say which one is preferred.]}
%In doing so, the tree structure can be simplified significantly. 

The results for test data are shown in Tables \ref{table-S6-4} and  \ref{table-S6-2}. For Simulation 2.1 we see that the joint model performs better than the frequency-severity model, while for Simulation 2.2 the opposite is observed. %[\jj{Maybe- finally for Simulation 2.2 we could have some conclusions like if the dataset is small the frequency-severity model works better, but if the dataset is large, they may be similar.} ]
%In addition, in terms of interpretability, one tree would be better. 

In the above, we \zzz{focused on using evaluation metrics to assess model performance.} %used evaluation metrics to compare the performance of different models. 
%\ct{OMIT? Below we employ ARI as an auxiliary tool to explore why two trees or one joint tree may be more effective.} 
We \ct{now} calculate the ARI for these three trees, \jj{using the test data}. First, for Simulation 2.1 we have
\begin{eqnarray*}
   && \text{ARI(Poisson-BCART (4), Gamma-BCART (4))}=%0.8667466
0.87,\\
&&
\text{ARI(Poisson-BCART (4), CPG-BCART (4))}=%0.940392
0.94,\\
&&
\text{ARI(Gamma-BCART (4), CPG-BCART (4))}=%0.9235653
0.92.
\end{eqnarray*}
This confirms the preference of joint models in Simulation 2.1, as the ARI values 
%for all the trees are relative \zzz{high}, % close to 1, 
%indicating 
indicate \zzz{strong} similarity. \zzz{This suggests that information sharing can avoid redundant use of similar information, which is evident in the similarities between the two trees.} %However, determining a specific ARI threshold that indicates when sharing information becomes more effective requires further research. In the real data analyses in Subsection \ref{sec:rd}, we use evaluation metrics to compare different models first. ARI is then employed as an auxiliary tool to explore why two trees or one joint tree may be more effective.
\jj{%Subsequently, we calculate the ARI between the three optimal trees for each type of model, namely Poisson-BCART (4), Gamma-BCART (4), and CPG-BCART (9).
Next, for Simulation 2.2 we have %(are the following correct?)
}
\begin{eqnarray*}
 &&   \text{ARI(Poisson-BCART (4), Gamma-BCART (4))}=0.28, %0.2797073,
 \\
&&
\text{ARI(Poisson-BCART (4), CPG-BCART (9))}= 0.77, %0.7650106.
\\
&&
\text{ARI(Gamma-BCART (4), CPG-BCART (9))}=0.73. %0.7294593.
\end{eqnarray*}
This indicates that the frequency-severity model should be preferred in Simulation \zzz{2.2} as the ARI values are smaller, especially the first. \jj{It is not obvious} how to 
determine a specific ARI threshold that indicates when sharing information becomes worthwhile. This requires further research.

%To explore this further, we hypothesize that the greater the dissimilarity between two trees (using the same splitting variables but different split values/categories), the less necessary it is for them to share information in one joint tree. %to avoid redundant use of the same or similar information. 

\zzz{Building on the above findings in Simulation 2.2, our investigation shows that both the frequency-severity model and joint model identify the optimal trees as expected, indicating that information sharing may not be necessary. Further exploration reveals that the parameter estimates of the joint tree are not as accurate as those for the frequency-severity trees. This discrepancy \jj{arises because the joint tree uses less data for the estimates in some of the 9 terminal notes, compared to the separate two trees, each of which only has 4 terminal nodes.  We suspect that, with fewer observations the differences will increase, \yyy{and vice versa}}.
%and with a larger number of data observations, the differences will be getting smaller.} %performance of the frequency-severity model and the joint model would perform similarly. 
To investigate this intuition, we use the same data generation scheme with different sample sizes (ranging from 1,000 to 50,000) to conduct 10 repeated experiments for each size. %Labelling the regions in Figure  \ref{Fig:shared_data1} as $A, B, C, \ldots, \jjj{I}$, Tables \ref{table-S6-5} and \ref{table-S6-6} respectively present the average absolute parameter estimation errors for $\lambda$ and $\beta$, along with their standard deviations. %, based on the training data for different cases. 
We find %observe from these tables 
that as the amount of data %available to both the frequency-severity models and the joint models in each region 
increases, \jj{the differences between the two model estimates become smaller (\zzz{see Supplementary Material SM.F for details}). This investigation suggests if less data is available the frequency-severity models may be preferred as they produce more accurate parameter estimates than the joint models, while if more data is available the joint models may be preferred \jjj{to save computation time}.}
%as they produce similar parameter estimates as the frequency-severity models using just a one joint tree.} % and their performance becomes more similar.
}

We also run several other simulation examples which are not shown here. From our results, we conclude that: when two trees have similar splitting rules (high ARI), one joint tree is more effective through information sharing. Conversely, if all covariates affecting \zzz{claim} frequency and \jj{average severity} are different (ARI is close to $-1$), two trees outperform one joint tree. This conclusion aligns with our intuition and can be generalized to a wider field; see also \cite{linero2020semiparametric}. %[\jj{some parts here can be moved to an earlier place, Introduction?}]

\COM{
\begin{enumerate}
    \item  If all covariates have the same splitting rules for separate \zzz{claim} frequency and \jj{average severity} trees, it is easy to check that the same covariate would have the same strong or weak correlation with \zzz{claim} frequency and \jj{average severity} respectively. In this case, one joint tree is better since they can use the shared covariates more efficiently. ARI can verify this discovery: ARI between two separate trees is close to one, indicating they are significantly similar, and one joint tree should be used. 
    
    \item When half of the covariates affect both \zzz{claim} frequency and \jj{average severity}, and the remaining can only influence one of them individually, ARI decreases, making it difficult to determine whether the frequency-severity models using two separate trees or joint models using one tree are better. There is not a big difference between them, and we would recommend one tree due to time-saving and better interpretability. However, this is not necessarily true, and we believe this area is worth further exploration.
    
    \item If all covariates affecting \zzz{claim} frequency and \jj{average severity} are different, ARI is close to 0, which means that there is a random agreement between these two separate trees, i.e., they are significantly different. In this case, frequency-severity models using two separate trees outperform joint models using one tree, which is consistent with our intuition.
\end{enumerate}
}

\section{Real data analysis}\label{sec:rd}
We illustrate our methodology with the insurance dataset \textit{dataCar}, available from the library \texttt{insuranceData} in \textsf{R}; see \cite{R:insurance} for details.  This dataset is based on one-year vehicle insurance policies taken out in 2004 or 2005. There are 67,856 policies of which 93.19\% made no \zzz{claims}. A description of the variables is given in Table \ref{table_cardata}. We split this dataset into training (80\%) and test (20\%) datasets such that the proportion of non-zero \zzz{claims} remains the same in both training and test datasets. 

%\textcolor{red}{can we remove table 11 and just refer to it from our previous paper} -- \ljj{I think it is good to have it here, where is the claim amount?}

\begin{table}[ht]

 \centering
 
\caption{Description of variables (\textit{dataCar})}

\begin{tabular}{p{2.4cm}|p{9cm}|p{1.3cm}} 

\toprule   

  Variable   & Description & Type \\  

\hline %\midrule   
 numclaims (\zzz{$N$})  & number of claims & numeric\\
 exposure (\zzz{$v$})  &in yearly units, between 0  and 1 & numeric\\
 claimscst0 (\zzz{$S$}) & total claim amount for each policyholder & numeric \\
   veh\_value  &vehicle value, in \$10,000s & numeric\\  

  veh\_age   & vehicle age category, 1 (youngest), 2, 3, 4 & numeric \\
   
agecat & driver age category, 1 (youngest), 2, 3, 4, 5, 6 & numeric\\

veh\_body  &vehicle body, one of: 
\ji{HBACK, UTE, STNWG, HDTOP, PANVN, 
SEDAN, TRUCK, COUPE, MIBUS,  MCARA, 
BUS, CONVT, RDSTR} &character\\

gender  & Female or Male &character\\

area  & coded as A B C D E F  &character\\

  \bottomrule  

\end{tabular}
\label{table_cardata}

\end{table}

\begin{table}[htbp]  
	\centering
   \captionsetup{width=.8\linewidth}
	\caption{\zzz{Numerical summary} of the \jj{average severity} in \emph{dataCar}.}
	
	\begin{tabular}{l|cccccc} 
		
		\toprule   

  \zzz{Statistics}  & Min & Mean  & Max & Standard Deviation & Skewness & Kurtosis\\
		
		\hline %\midrule

		\jj{Average severity} & \zzz{200} &% 130.58 
  \zzz{1916} & %55922.13
  55922 & %1024.46
  \zzz{3461} & %18.39
  \zzz{5} & %533.96 
  \zzz{48} \\

\bottomrule
	\end{tabular}
 \label{table-cardata-seve}
	
\end{table}

\subsection{\jj{Average severity} modeling}\label{sec:data1_sev}

\COM{
\begin{figure}[!htb]
	\centering       \includegraphics[width=0.85\linewidth, height=0.45\textheight]{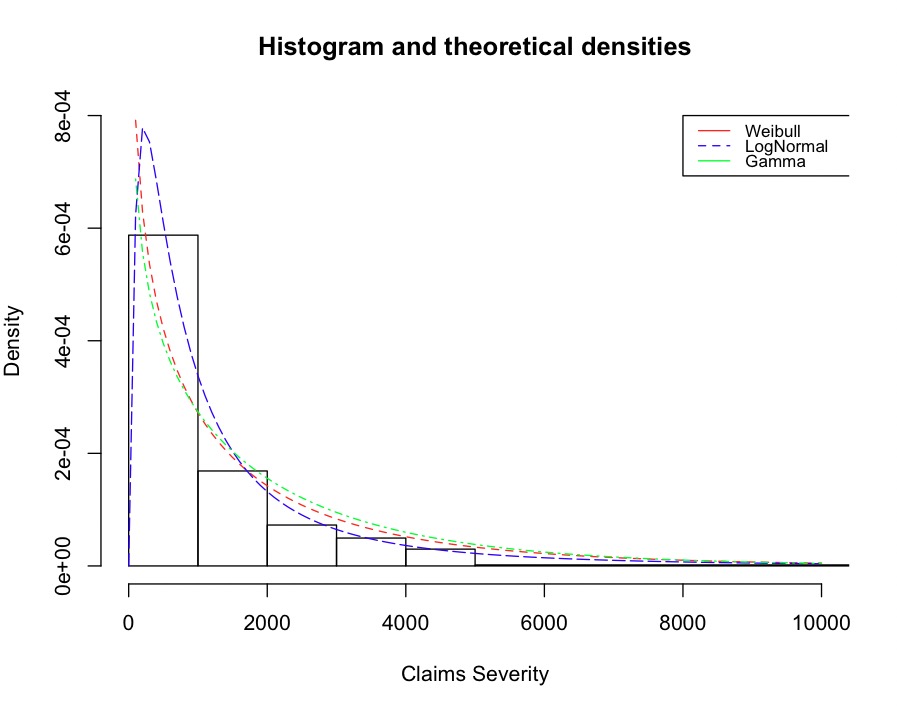}
 \captionsetup{width=.8\linewidth}
	\caption{Histogram and theoretical densities of gamma, lognormal, and Weibull distributions for \jj{average severity} data. Parameters used to generate the plot are estimated by using the \textit{``fitdist''} function in the \textsf{R} package \texttt{fitdistrplus} (see more details in \cite{R:fitdistrplus}).} 
	\label{Fig:data1_severity_histgram}
\end{figure}

\begin{figure}[!htb]
	\centering       \includegraphics[width=0.85\linewidth, height=0.45\textheight]{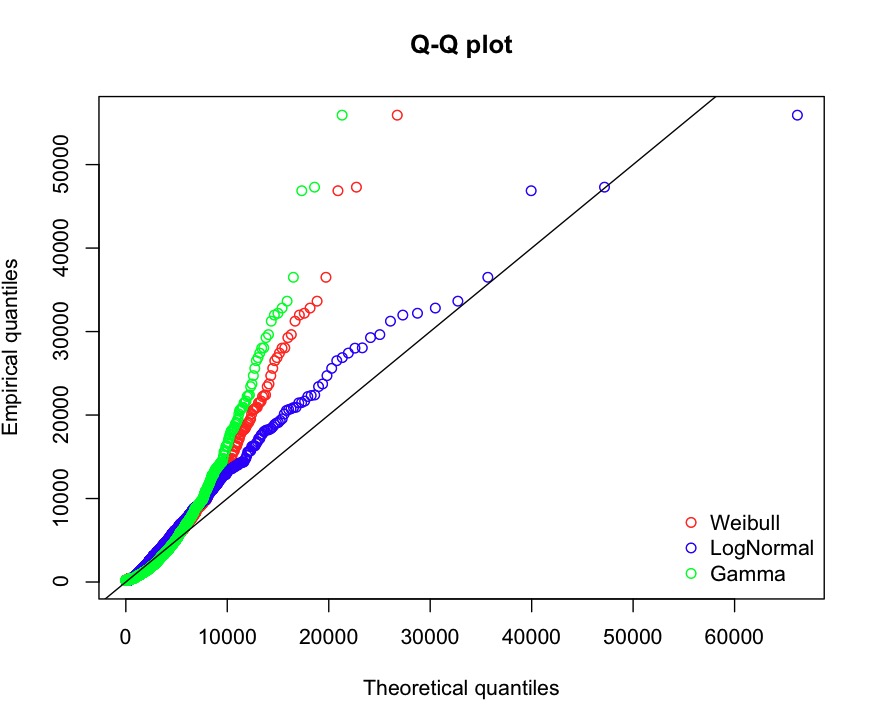}
 \captionsetup{width=.8\linewidth}
	\caption{Q-Q plot of gamma, lognormal, and Weibull distributions for \jj{average severity} data. Parameters of the distributions are estimated by using the \textit{``fitdist''} function in the \textsf{R} package \texttt{fitdistrplus} (see more details in \cite{R:fitdistrplus}).}
	\label{Fig:data1_severity_qqplot}
\end{figure}
}

For average severity modeling, we  
%Based on Subsection \ref{sec:sev_bcart}, we aim to directly model the average claim amount $\bar{S}$ for \jj{average severity}. The \jj{average severity} model considered here only applies to the 
consider a subset of the data with $N>0$. 
%This is a traditional and commonly used way to deal with \jj{average severity} data; see, e.g., \cite{henckaerts2021boosting} and \cite{frees2014predictive}. 
Among all 67,856 policies, 4,624 policies satisfy this requirement (3,699 in the training data, and 925 in the test data). %\ct{OMIT: The data do not include the average severity directly (see Table \ref{table_cardata}); therefore, we}
\ct{We} calculate the average severity by dividing the total claim amount by the number of claims for each policyholder. A numerical summary of the average severity data is displayed in Table \ref{table-cardata-seve}, \zzz{indicating that the average severity data exhibit right-skewness and \zyj{heavy} tails.} %[\jj{Do you need to keep 0 as Min? also in the plots?}] right-skewed and heavy-tailed [\jj{how to check these?}].\ct{ Are we using the average claim of each policy holder? (say so)}
We start with some exploratory analysis, fitting gamma, lognormal and Weibull distributions to the whole data. %Two graphs are used to illustrate which distribution better fits the \jj{average severity} data in 
From \zzz{the histogram and Q-Q plots (see Supplementary Material SM.G),}
%Figures \ref{Fig:data1_severity_histgram} and \ref{Fig:data1_severity_qqplot}
we \zzz{find} %see 
that 
%At first glance, distinguishing their performance is challenging in Figure \ref{Fig:data1_severity_histgram} since 
all distributions can capture the right-skewed feature, however, %in Figure \ref{Fig:data1_severity_qqplot}, 
none of them correctly captures the heavy right tail of the distribution. It appears that the lognormal distribution fits slightly better when the data are treated as IID. However, as we will see below the lognormal distribution is not the best choice in the BCART models for this data.
%might be preferred, as the points from the LogNormal distribution form a line that is closest to the 45-degree reference line. \yz{This finding is somewhat surprising and goes against our expectations. 
%We expected that the Weibull distribution would be the best among these three}, as discussed in Subsection \ref{sec:sev_bcart}. The reason may be that the plots are drawn by using parameters estimated once with the entire \jj{average severity} data, whereas in our proposed BCART models, we update the parameters in the Bayesian framework at every step using subsets of data in each node. The approach we use not only provides more accurate parameter estimates but also allows the model to better capture data characteristics and fit the data more effectively in each group (terminal node). 

For comparison, we first run some benchmark CART models. We have tried to fit the ANOVA-CART using both the original and log-transformed training data. Neither of them gives us any reasonable result since no split is identified, resulting in only a root node tree. 
%Thus, we will not include them below.
%First, when running ANOVA-CART on training data, the result \yz{indicates there is no potential split that would lead to improvement.  ANOVA-CART is not capable of capturing this right-skewed and heavy-tailed insurance data (see Table \ref{table-cardata-seve}). Another approach involves running ANOVA-CART on $\log(\bar{S})$ and transforming the results back to $\bar{S}$.  This method is intuitive and straightforward to implement. However, when examining the results, there is still no split identified, resulting in a root node tree. 
%Consequently, we do not include both of them in the following context.} 
We use the \textsf{R} package \texttt{distRforest} (cf. \cite{R:distRforest}) to fit Gamma-CART and LN-CART and both of the trees, after cost-complexity pruning, have 5 terminal nodes. %This package extends the applicable distributions \zz{for decision trees to include Gamma and LogNormal distributions, allowing the analysis of light/heavy-tailed data.} It builds a random forest in the ensemble consisting of individual CARTs based on the package \texttt{rpart}; see more details in \cite{R:distRforest}. Since our focus is on one tree, we restrict the number of trees to one in the random forest, allowing us to obtain Gamma-CART. 
%We use cross-validation to select the tree size, which has 5 terminal nodes. Similarly, we run LN-CART in the same way, again resulting in a tree with 5 terminal nodes. 
As far as we are aware there is no \textsf{R} package with the Weibull distribution implemented for regression trees.%, %no can be used to implement it in decision trees currently, 
%so we do not include it in the following discussion. % Further research may be needed to include more distributions with different characteristics in decision trees. 

We then apply the proposed Gamma-BCART, LN-BCART, and Weib-BCART to the same data. %Similar to \zzz{claim} frequency modelling in \cite{zhang2024bayesian}, we tune the hyper-parameters $\gamma, \rho$. 
The DICs in Table \ref{table-Car-sev_test} indicate that all these BCART models choose a tree with 4 terminal nodes.

\begin{table}[!t]  
	
	\centering
 \captionsetup{width=.8\linewidth}
	\caption{Hyper-parameters, $p_D$ and DIC on training data (\emph{dataCar}), and model performance on test data for \jj{average severity} models with bold entries determined by DIC. The number in brackets after the abbreviation of the model indicates the number of terminal nodes for this tree.}
 
	\begin{tabular}{@{}l|cccc|cccc} 

		\toprule   

 & \multicolumn{4}{c|}{Training data} & \multicolumn{4}{c}{Test data}\\
\hline
		Model &$\gamma$ & $\rho$ & $p_{D}$ & DIC & RSS($\vk{S}$) $\times 10^{-10}$ & SE & DS & Lift
   \\  
		
		\hline %\midrule   
		
		Gamma-GLM & - & - & - & - & 1.4335 & - & %13769.12
		-  & - \\\hline
  \bottomrule
		
		Gamma-CART (5)& - & - & - & -& 1.4173 & %5740.582
		464  & 0.00171 & 1.625 \\
		
		LN-CART (5)& - & - & - & - & 1.4168 & %5740.582
		458  & 0.00168 & 1.629 \\\hline

   \bottomrule
		
		Gamma-BCART (3) & 0.99 & 4 & 5.97 & 78061 & 1.4201
		& 486  & 0.00181 & 1.567  \\
		\textbf{Gamma-BCART (4)} & 0.99 & 3.5 & 7.97 & \textbf{77779} & 1.4176
		& \textbf{%457.1635
			457}& \textbf{%0.001537635
			0.00154}    & %1.615207 
		1.615  \\
		Gamma-BCART (5) & 0.99 & 2 & 9.95 & 77982 &1.4158 & 472  & 0.00167  & 1.643 \\
  \hline

		LN-BCART (3) & 0.99 & 5 & 5.97 & 78022  & \zzz{1.4193} & \zzz{483} &  \zzz{0.00178}  & \zzz{1.570}  \\
		\textbf{LN-BCART (4)} & 0.99 & 4 & 7.97 & \textbf{77741} & \zzz{1.4171} & \zzz{\textbf{449}}  &  \zzz{\textbf{0.00149}} & \zzz{1.628}  \\
		LN-BCART (5) & 0.99 & 3 & 9.96 & 77893 &  \zzz{1.4153} & \zzz{456}  &  \zzz{0.00161}  & \zzz{1.649}  \\\hline

		Weib-BCART (3) & 0.99 & 7 & 5.98 & 77932 & 1.4177 & 473  &  0.00164 & 1.604  \\
		\textbf{Weib-BCART (4)} & 0.99 & 5 & 7.98 & \textbf{77646} & 1.4154 & \textbf{433}  &  \textbf{0.00131}  & 1.661  \\
		Weib-BCART (5) &  0.99 & 4 & 9.98 & 77821 & 1.4136 & 446 &  0.00144 & 1.693 \\\hline

  \bottomrule

	\end{tabular}
	
	\label{table-Car-sev_test}
\end{table}

\COM{ %%%%%%%%%%%%%%%%%%%%%%%%%

\begin{figure}[!htb]
	\centering       \includegraphics[width=0.7\linewidth, height=0.4\textheight]{figures/data1_gamma_trace_plots.jpg}
 \captionsetup{width=.8\linewidth}
	\caption{Trace plots from MCMC with 3 restarts  for Gamma-BCART ($\gamma$=0.99, $\rho$=3.5).}
	\label{Fig:data1_gamma_trace}
\end{figure}
}%%%%%%%%%%%%%%%%%%%%%%%%%%%

\begin{figure}
	\centering
	\includegraphics[width=0.6\linewidth, height=0.3\textheight]{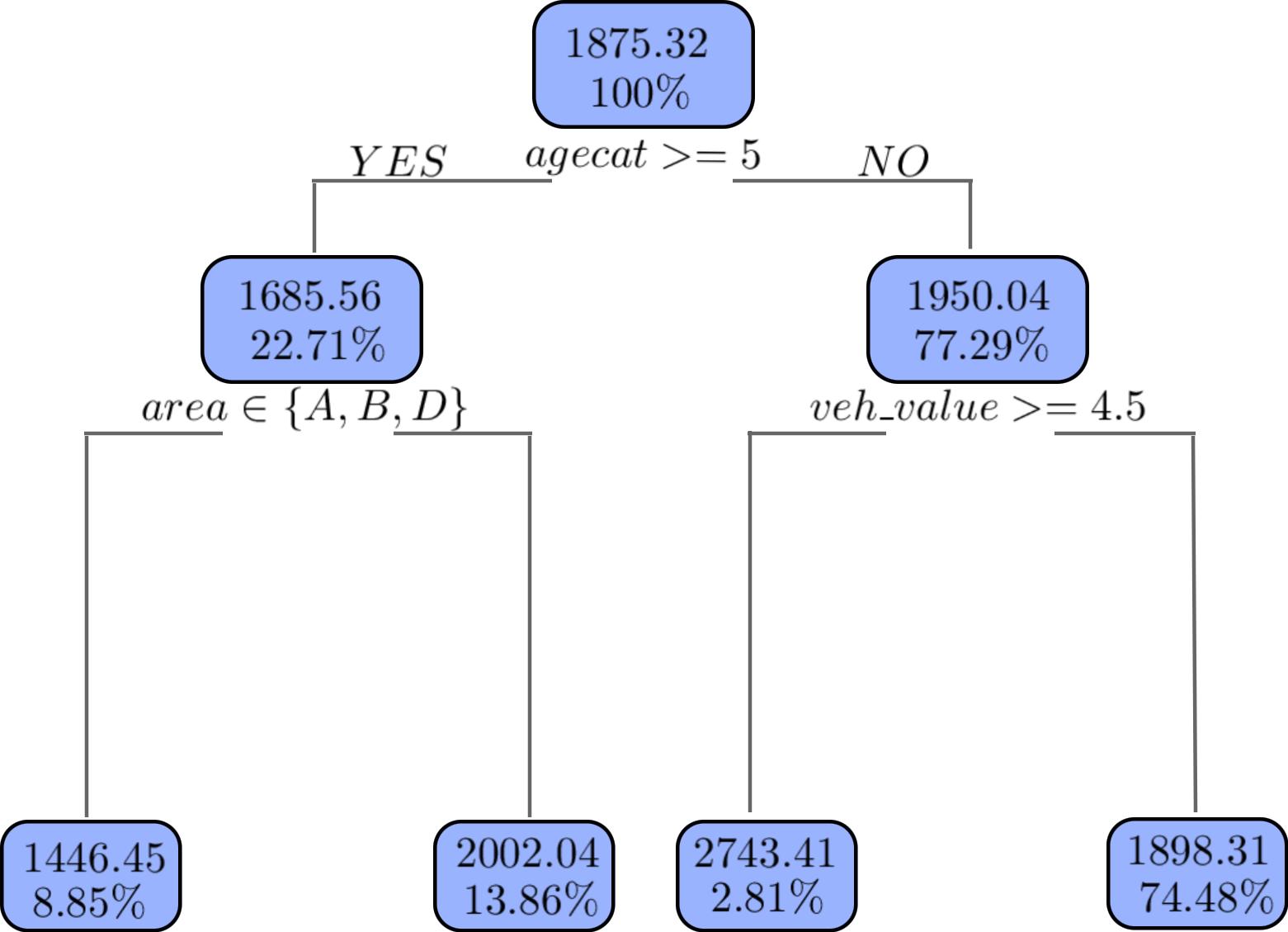}
 \captionsetup{width=.8\linewidth}
	\caption{Optimal tree from Weib-BCART. Numbers at each node give the estimated \zzz{average} severity and the percentage of observations.}
	\label{Fig:data1_weib_bcart}
\end{figure}

We also examine the splitting rules used in each tree. Gamma-CART uses both ``\textit{agecat}'' and ``\textit{veh\_value}'' twice, with the first one being ``\textit{agecat}''. In contrast, LN-CART uses three different variables, ``\textit{veh\_value}'' first, followed by ``\textit{veh\_body}'' and ``\textit{area}''. All trees from BCART models, i.e., Gamma-BCART, LN-BCART, and Weib-BCART, have the same tree structure and splitting variables (``\textit{agecat}'', ``\textit{veh\_value}'', and ``\textit{area}''), while the split values/categories are slightly different. Weib-BCART, in particular, \jj{can identify} a more risky \zzz{cell} (i.e., the one with estimated \zzz{average} severity equal to 2743.41); see Figure \ref{Fig:data1_weib_bcart}. This may be because, as discussed in Subsection \ref{sec:sev_bcart}, Weib-BCART can flexibly control the shape parameter to adapt to data with different tail characteristics, allowing it to handle cases where some \zzz{cell}s (terminal nodes) have lighter tails, and others have heavier tails. \zzz{We generate Q-Q plots of the optimal Weib-BCART tree for average severity data in each terminal node (see Supplementary Material SM.G). The plots reveal that}
%In Figure \ref{Fig:data1_weibull_4}, 
%we observe that 
although all shape parameters are smaller than one, indicating heavy tails for \jj{average severity} data within each terminal node, the \jlp{selected} Weib-BCART tree shows improved data fitting compared to \lj{the initial model where covariates are not taken into account.} %Figure \ref{Fig:data1_severity_qqplot}. 
Similar improvements are obtained from Gamma-BCART and LN-BCART. % (with their Q-Q plots not shown). %Particularly, for Gamma-BCART, the range of shape parameters is 0.62--0.89, whereas Weib-BCART can distinguish between 0.48--0.96, allowing for better tail control and data fitting. 
We also use a standard Gamma-GLM for this dataset. We find that only the variable ``\textit{gender}'' is significant, \yz{and thus no interaction is considered in the Gamma-GLM. Interestingly, ``\textit{gender}'' does not appear in any of the CART and BCART models.} 
In summary, though the variables used for different models may differ, there seems to be a consensus that ``\textit{agecat}'' is still significantly important for \jj{average severity} modeling, as Gamma-CART and all BCART models use it in the first split, and “\textit{veh\_value}” is another relatively important variable. This observation aligns, to some extent, with our initial analysis of the relationship between covariates and \jj{average severity}; see Table \ref{table_data1_cor} \jlp{below}. Particularly, in comparison to CARTs, BCART models reveal another important variable, ``\textit{area}”. %\zzz{which was identified after the initial numerical transformation (see %Table \ref{table:data1_tran_cor_area} in Appendix C).} 
%Notably, Weib-BCART precisely \zz{chooses \textit{areas} A, B, and D (identified as having the three smallest \jj{average severity} after the initial numerical transformation) for splitting.} %\textcolor{red}{if we keep this, we need to mention the initial numerical transformation in detail, which will increase the content again...}

\COM{
\begin{figure}[!htb]
	\centering       \includegraphics[width=0.85\linewidth, height=0.5\textheight]{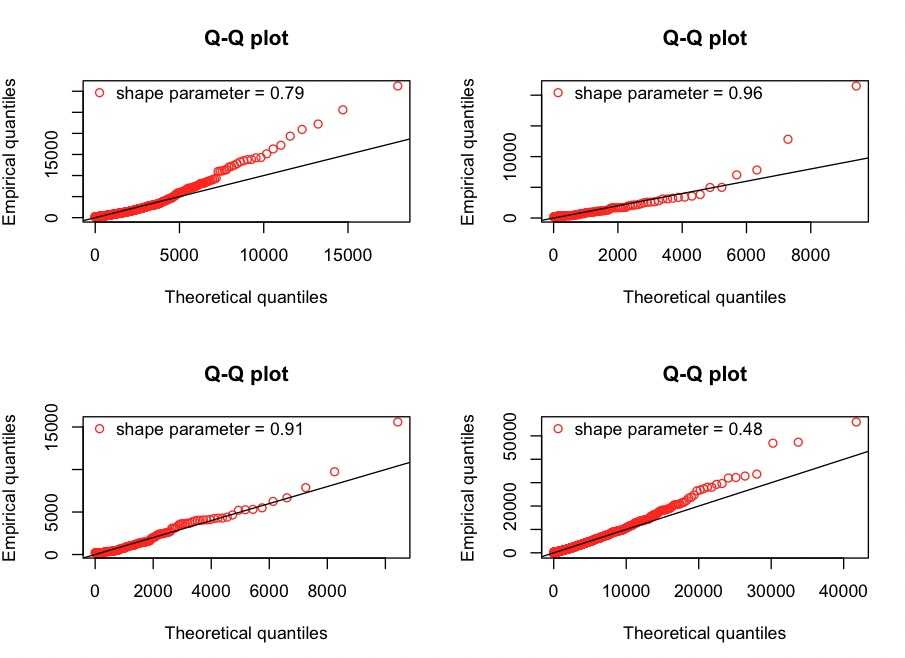}
 \captionsetup{width=.8\linewidth}
	\caption{Q-Q plots of the Weibull distribution for \jj{average severity} data in each terminal node of the optimal Weib-BCART tree. \yz{The shape parameter used to generate the plot is estimated by using \zzz{MME} (see Table \ref{table_Sev_pred} in Section \ref{sec:sev_bcart}), and the scale parameter is estimated using the posterior distribution (see Appendix A)}.} %\eqref{eq:Weib_beta}
 
	\label{Fig:data1_weibull_4}
\end{figure}
}

%Now, we apply the identified optimal trees %obtained from training %\yz{data, lji-maybe not needed}
%to the test data. 
The performance on the test data of the selected tree models above is given in Table \ref{table-Car-sev_test}. It is evident that the Gamma-GLM is not as good as the tree models, \zzz{as reflected in RSS($\vk{S}$).} %From the table, w
\jjj{The study} yields the following model ranking: % \zzz{based on the measures of SE and DS}: 
\text{Weib-BCART}, \text{LN-BCART}, \text{Gamma-BCART}, \text{LN-CART}, \text{Gamma-CART}. %\text{Gamma-GLM.} 
This ranking is consistent with %the conclusions from the simulation example in Section \ref{sec:sev_sim} and 
our expectations. \jj{First, it is common that average severity data is heavy-tailed. Second, the Weibull distribution is advantageous, because it can effectively handle varying tail characteristics in different tree nodes.}
%Although we lack information about the exact distribution of real insurance \jj{average severity} data, we are aware of its right-skewed and heavy-tailed nature. Specifically, in tree models, data within groups (terminal nodes) may exhibit different tail characteristics, highlighting the advantage of the Weibull distribution, which can effectively handle varying tail characteristics. %Moreover, when comparing LogNormal and Gamma distributions, the former demonstrates better adaptability to data with heavier tails than the latter \textcolor{blue}{be careful when you use this comparison because you applied the distribution fit for all the data before, but now the data is split into several sub-groups...}, justifying their relative performance ranking. 
% Concerning time and memory, similar discussions can be considered as in \cite{zhang2024bayesian}. 

%For stability analysis, the same conclusions can be drawn for \jj{average severity} modelling: BCART models are more stable than CART models. To avoid repetition, we omit the details here.

\subsection{Aggregate \zzz{claim} modeling}

\subsubsection{Model fitting and comparison} \label{sub:real-data-1}
We now fit the three BCART models with aggregate \zzz{claim} data, namely, frequency-severity models, sequential models, and joint models. For frequency-severity models, numerous combinations of \zzz{claim} frequency and \jj{average severity} models are possible; see  \cite{zhang2024bayesian} for the frequency models and Section \ref{sec:data1_sev} for \zzz{average} severity models. %. In our implementation, there are 5 models for \zzz{claim} frequency (1 Poisson, 2 NB, and 2 ZIP; see \cite{zhang2024bayesian}) and 4 models for \jj{average severity} (2 Gamma, 1 LogNormal, and 1 Weibull), resulting in 20 frequency-severity models. 
%\ga{Remove? Instead of running all these models,} 
\zzz{Here,} we choose ZIP2-BCART and Weib-BCART as the optimal tree \jjj{models} for frequency-severity models (see \cite{zhang2024bayesian} and Section \ref{sec:data1_sev}). Note that although ZIP2-BCART and Weib-BCART are identified as the best for claim frequency and \jj{average severity} separately, it is unclear whether they remain optimal when combined. This will be examined and discussed below. % in the real insurance dataset. 
For joint models, we discuss the CPG-BCART model and three types of ZICPG-BART models. %distributions in joint models are compared to assess their capability to capture data characteristics with a high proportion of zeros. 
%Additionally, given that the proposed joint models employ a CPG distribution, 
Because of these choices, we also include the frequency-severity BCART model with Poisson and gamma distributions for comparison.
For sequential BCART models, we consider %ensure consistency of \zzz{claim} frequency modelling as the other models by using 
Poisson-BCART (or called P-BCART) and ZIP2-BCART for claim frequency. Subsequently, we treat the claim count $N$ (or $\hat{N}$) as a covariate in the corresponding Gamma-BCART and Weib-BCART for \jj{average severity}. The resulting models are called Gamma1-BCART (or Gamma2-BCART, with $\hat{N}$ from P-BCART) and Weib1-BCART (or Weib2-BCART, with $\hat{N}$ from ZIP2-BCART). 

Table \ref{table_Car_agg_train} presents the DICs for the average severity part in the sequential models and for the joint models. We see that in the average severity modeling with $N$ (or $\hat{N}$) as a covariate, all of them choose an optimal tree with 4 terminal nodes. Upon inspecting the tree structure, $N$ (or $\hat{N}$) is indeed used in the first step in all those optimal trees. % for each model (Gamma1-BCART, Gamma2-BCART, Weib1-BCART, and Weib2-BCART). 
\yz{All of them replace the previously used variable ``\textit{agecat}'' by the covariate $N$ (or $\hat{N}$).} %, with the only difference being in the split values used. This observation aligns with the results obtained in Subsection \ref{sec:agg_depen_sim}. 
We suspect this may be due to a strong relationship between the covariates $N$ and ``\textit{agecat}'', as verified in the claim frequency analysis (see \cite{zhang2024bayesian} \lj{and also Table \ref{table_data1_cor} below}). %\zyy{[also verified by Table 12? the correlation between claim frequency and agecat]} %, where the optimal \zzz{claim} frequency tree selects ``\textit{agecat}'' in the first split, indicating a strong relationship between $N_i$ and ``\textit{agecat}''. Consequently, it is reasonable to replace ``\textit{agecat}'' with $N_i$ (or $\hat{N}_i$) to avoid multicollinearity. 
Furthermore, by comparing the DICs of all Gamma-BCART and Weib-BCART models in Tables \ref{table-Car-sev_test} and \ref{table_Car_agg_train} (with/without $N$ or $\hat{N}$ as a covariate), we \zzz{find} that the model performance improves when considering $N$ (or $\hat{N}$) as a covariate, especially when using $\hat{N}$. %This approach has a practical advantage as there is no direct information about $N_i$ itself for new customers. 

%The difference in DIC between the CPG model and ZICPG models is \zz{significantly} larger than the difference between ZICPG models themselves, illustrating the necessity of considering the zero mass part. \zz{Additionally, there is no big difference between ZICPG2-BCART and ZICPG3-BCART models. %aligning with the conclusion obtained in ZIP models (see Remark 3.5 (b) in Subsection \ref{sec:zip4}). 
%This observation implies that embedding the exposure into both the Poisson part and the zero mass part does not yield substantial improvement; embedding the exposure into the zero mass part is sufficient. However, it cannot be denied that ZICPG3-BCART still exhibits the best performance.}

\begin{table}[!t]  
	
	\centering
 \captionsetup{width=.8\linewidth}
	\caption{Hyper-parameters, \zzz{$p_D$} and DIC on training data (\textit{dataCar}) for aggregate \zzz{claim} models. The number in brackets after the abbreviation of the model indicates the number of terminal nodes for this tree. The Gamma1/Weib1 and Gamma2/Weib2 models treat the claim count $N$ and $\hat{N}$ as a covariate respectively, where $\hat{N}$ for Gamma2 comes from \jj{P-BCART  and that for Weib2 comes from ZIP2-BCART}. Bold font indicates DIC selected model.}
	
	%\begin{tabular}{p{4cm}p{2cm}p{2cm}p{2.5cm}p{2.5cm}}
	\begin{tabular}{l|cccc}
		
		\toprule   
		
		Model & $\gamma$ & $\rho$ & \zzz{$p_D$} & \multicolumn{1}{c}{DIC} \\  
		
		\hline %\midrule   
		
		Gamma1-BCART (3) & 0.99 & 4 & 5.97 &  78032 \\
		\textbf{Gamma1-BCART (4)} & 0.99 & 3.5 & 7.97 & \textbf{77750}  \\
		Gamma1-BCART (5) & 0.99 & 2 & 9.96 & 77854  \\\hline
		
		Gamma2-BCART (3) & 0.99 & 4 & 5.98 & 78024  \\
		\textbf{Gamma2-BCART (4)} & 0.99 & 3.5 & 7.97 & \textbf{77743}  \\
		Gamma2-BCART (5) & 0.99 & 2 & 9.97 & 77849  \\\hline
		
		Weib1-BCART (3) & 0.99 & 7 & 5.98 & 77911 \\
		\textbf{Weib1-BCART (4)} & 0.99 & 5 & 7.98 & \textbf{77619}  \\
		Weib1-BCART (5) & 0.99 & 4 & 9.98 & 77804
		\\\hline
		
		Weib2-BCART (3) & 0.99 & 7 & 5.98 & 77893 \\
		\textbf{Weib2-BCART (4)} & 0.99 & 5 & 7.98 & \textbf{77608} \\
		Weib2-BCART (5) & 0.99 & 4 & 9.98 & 77787 \\\hline
		
		\bottomrule

		CPG-BCART (4) & 0.99 & 10 & 11.96 & 105710 \\
		
		\textbf{CPG-BCART (5)} & 0.99 & 8 & 14.93 & \textbf{105626}  \\
		
		CPG-BCART (6) & 0.99 & 7 & 17.92 & 105643 \\\hline

		ZICPG1-BCART (4) & 0.99 & 11 & 15.97 & 102314 \\

		\textbf{ZICPG1-BCART (5)} & 0.99 & 10 & 19.95 & \textbf{102198}   \\
		
		ZICPG1-BCART (6) & 0.99 & 7.5 &  23.92 & 102225 \\\hline

		ZICPG2-BCART (4) & 0.99 & 12 & 15.95 & 102265 \\

		\textbf{ZICPG2-BCART (5)} & 0.99 & 11 & 19.94 & \textbf{102134}   \\
		
		ZICPG2-BCART (6) & 0.99 & 8 &  23.92 & 102167 \\\hline

		ZICPG3-BCART (4) & 0.99 & 14 & 15.94 & 102247 \\
		
		\textbf{ZICPG3-BCART (5)} & 0.99 & 12 & 19.93 & \textbf{102120}   \\
		
		ZICPG3-BCART (6) & 0.99 & 9 &  23.90 & 102158 \\\hline

		\bottomrule  
		
	\end{tabular}
	
	\label{table_Car_agg_train}
\end{table}

\begin{figure}
	\centering
	\includegraphics[width=0.6\linewidth, height=0.3\textheight]{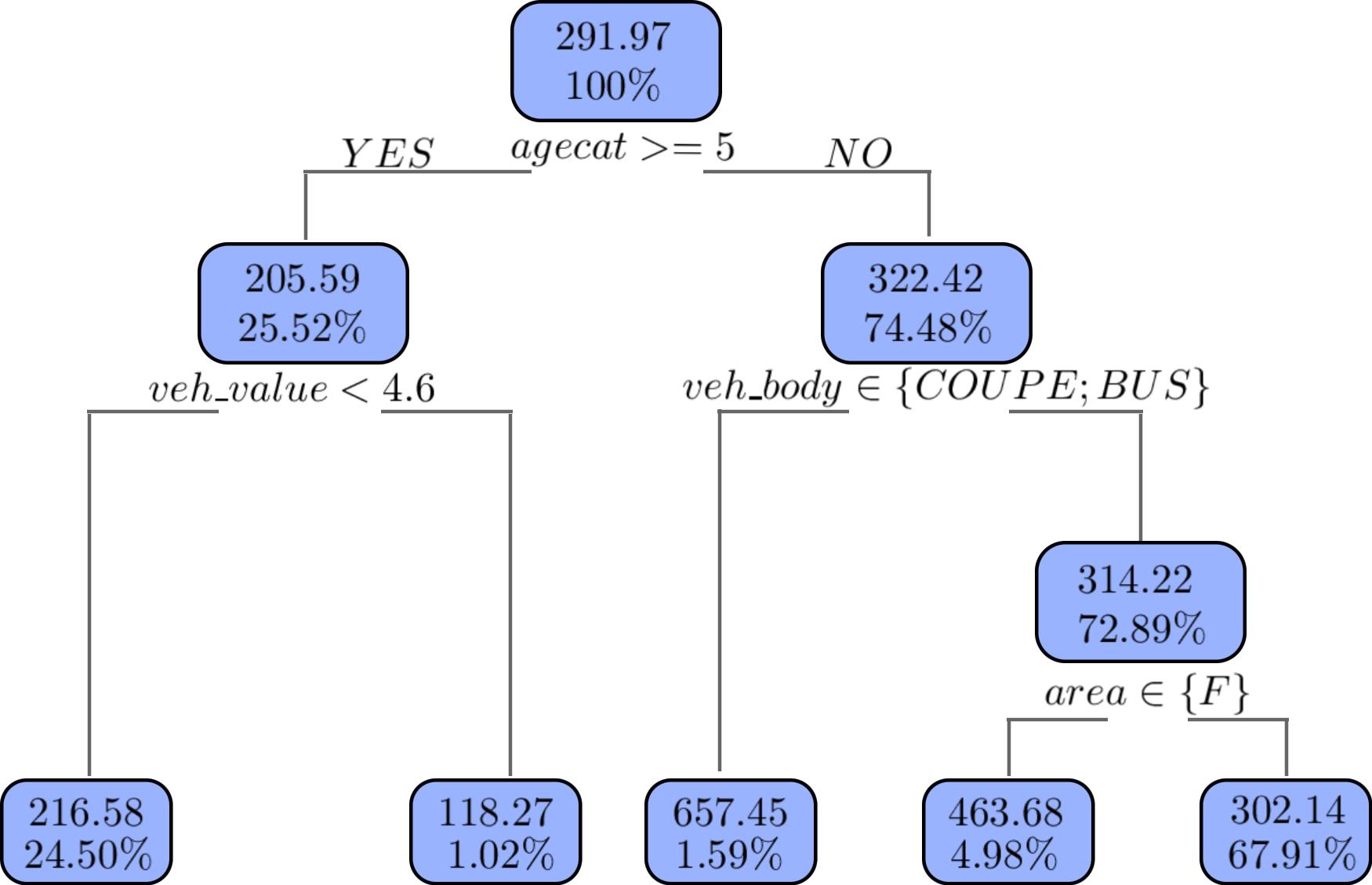}
 \captionsetup{width=.8\linewidth}
	\caption{Optimal tree from ZICPG3-BCART. Numbers at each node give the estimated premium and the percentage of observations.}
	\label{Fig:data1_zicpg3_bcart}
\end{figure}

For joint models, i.e., CPG-BCART and three ZICPG-BCART models, all of them choose optimal trees with 5 terminal nodes. Among them, ZICPG3-BCART, with the smallest DIC ($=102120$), is deemed to be the best. We again examine the splitting rules used in the selected trees \zzz{among joint models}. All \zzz{trees} %models 
use the same splitting variables (``\textit{agecat}'', ``\textit{veh\_value}'', ``\textit{veh\_body}'', and ``\textit{area}''), but the order of use and the tree structures vary. Notably, ``\textit{agecat}'' is consistently the first variable. Among them, ZICPG3-BCART demonstrates the ability to identify a riskier \zzz{cell} (i.e., the one with an estimated pure premium equal to 657.45; \zz{see Figure \ref{Fig:data1_zicpg3_bcart}}), possibly due to the same reason as discussed in \cite{zhang2024bayesian} for the outstanding performance of ZIP2-BCART for claim frequency. %Both ZIP2 and ZICPG3 models exhibit the capacity to handle datasets with a substantial number of zeros by incorporating exposure in the zero mass part. 
\zz{Besides, we observe that the tree structure of ZICPG3-BCART is quite similar to ZIP2-BCART. However, ZICPG3-BCART identifies another important variable ``\textit{area}'', which was recognized as important for \jj{average severity} before (see Section \ref{sec:data1_sev}).} We also fit a CPG-GLM to the data. We find that only the variable ``\textit{agecat}'' is significant, aligning with its consistent selection as the first splitting variable in almost all the BCART models. %Since only one variable is deemed significant in the CPG-GLM, no interactions were included. 
It is also worth mentioning that CART is not included in this analysis \yz{due to the absence of \textsf{R} packages that can directly use CPG \zzz{(or ZICPG)} to process the data.}

The performance of the selected trees for the test data is given in Table \ref{table-Car_agg_test}. As before, GLM exhibits poorer performance compared to %tree-based
\yyy{tree} models, as evidenced by RSS($\vk{S}$). \zzz{\lj{Below} we discuss the three types of aggregate claim models from various perspectives.}
%However, for other models, drawing a clear and unified conclusion is difficult, and so we discuss this from various perspectives. 
The meaning of the abbreviations can be found in the captions of Tables \ref{table_Car_agg_train}\jlp{--\ref{table-Car_agg_test}}.
%and %\ref{table-Car_agg_test}.

\begin{table}[!t]  
	
	\centering
 \captionsetup{width=.8\linewidth}
	\caption{Model performance on test data (\textit{dataCar}) for aggregate \zzz{claim} models with bold entries determined by DIC (see Table \ref{table_Car_agg_train}). %The Gamma1/Weib1 and Gamma2/Weib2 models treat the claim count $N_i$ and $\hat{N}_i$ as a covariate respectively, where $\hat{N}_i$ comes from Poisson-BCART and ZIP2-BCART respectively. 
$\text{F}_{\text{P}}\text{S}_{\text{G}}$ denotes frequency-severity models using Poisson and gamma distributions separately; other abbreviations can be explained similarly referring to Table \ref{table_Car_agg_train}. The number in brackets after the abbreviation of the model indicates the number of terminal nodes for those trees. %Particularly, two numbers for frequency-severity models indicate the number of terminal nodes for each tree.
}
	
	\begin{tabular}{l|c|c|c|c} 
		
		\toprule

		Model & \zzz{RSS ($\vk{S}$)\zzz{$\times 10^{-10}$}} & \zzz{SE} &  \zzz{DS} \zzz{$\times 10^4$}  & \zzz{Lift} \\  
		
		\hline %\midrule   
		
		CPG-GLM & 1.5187 & - & - &  - \\\hline

  \bottomrule
		
		$\text{F}_{\text{P}}\text{S}_{\text{G}}$-BCART (5/4) & 1.4874 & %19170.86
  242.12 & 8.32 & 2.532 \\
		$\text{F}_{\text{P}}\text{S}_{\text{G1}}$-BCART (5/4) & %1.4813 
        1.4796 & %19144.24
  %238.14
  237.51 & %8.19
  8.17 & %2.538
  2.540 \\
		$\text{F}_{\text{P}}\text{S}_{\text{G2}}$-BCART (5/4) & %1.4798
        1.4792 & %19140.17
  %237.29
  237.44 & 8.16& %2.540 
  2.541 \\\hline
		
		$\text{F}_{\text{ZIP2}}\text{S}_{\text{Weib}}$-BCART (5/4) & 1.4844 & %19152.01 
  240.89 & 8.27 & 2.534 \\
		$\text{F}_{\text{ZIP2}}\text{S}_{\text{Weib1}}$-BCART (5/4) & %1.4790
        1.4783 & %19138.49
  %237.01
  236.03 & %8.14
  8.01 & %2.541
  2.545 \\
		$\text{F}_{\text{ZIP2}}\text{S}_{\text{Weib2}}$-BCART (5/4) & %1.4779
        1.4770 & %19135.19
  %236.45 
  235.83 & %8.09
  7.92 & %2.544
  2.549 \\\hline
		
		\bottomrule
  
		CPG-BCART (4) & 1.4791 & %19139.01
  237.42 & 8.15 & 2.542 \\
		\textbf{CPG-BCART (5)} & 1.4781 & %19135.45
  \textbf{235.98} & \textbf{7.93} & 2.547   \\
		CPG-BCART (6) & 1.4778 & %19134.88
  236.29 & 8.02 & 2.549 \\\hline

		ZICPG1-BCART (4) & 1.4670 & %19110.57
  232.87 & 7.85 & 2.560 \\
		\textbf{ZICPG1-BCART (5)} & 1.4497 & %19101.23
  \textbf{229.73} & \textbf{7.56} & 2.584 \\
		ZICPG1-BCART (6) & 1.4478 & %19095.67
 231.35 & 7.79 & 2.587 \\\hline
		
		ZICPG2-BCART (4) & 1.4612 & %19108.33
  232.15 & 7.81 & 2.563 \\
		\textbf{ZICPG2-BCART (5)} & 1.4434 & %19093.56
  \textbf{229.41} & \textbf{7.52} & 2.595 \\
		ZICPG2-BCART (6) & 1.4417 & %19091.45
  231.10 & 7.77 & 2.597  \\\hline

		ZICPG3-BCART (4) & 1.4598 & %19105.02
  231.24 & 7.79 & 2.570 \\
		\textbf{ZICPG3-BCART (5)} & 1.4415 & %19086.42
  \textbf{228.88} & \textbf{7.45} & 2.601 \\
		ZICPG3-BCART (6) & 1.4409 & %19084.87
  229.53 & 7.69 & 2.604 \\\hline
		
		\bottomrule
	\end{tabular}
	
	\label{table-Car_agg_test}
\end{table}

\begin{enumerate}
	\item A comparison of our frequency-severity models suggests using the combination of two best models for \zzz{claim} frequency and \jj{average severity} respectively \zzz{based on all evaluation metrics}, i.e., $\text{F}_{\text{ZIP2}}\text{S}_{\text{Weib}}$-BCART $>$ $\text{F}_{\text{P}}\text{S}_{\text{G}}$-BCART. %This leads to the conclusion that the strategy for choosing the optimal combination within frequency-severity models is to individually select the best model for \zzz{claim} frequency and \jj{average severity}. 
 In the sequential models, the same conclusion as in Section \ref{sec:agg_depen_sim} is reached: using the \zzz{estimate} of the claim count $\hat{N}$ is superior to using $N$ itself when treating them as a covariate in the \jj{average severity} tree. Regarding joint models, ZICPG models outperform the CPG model, with ZICPG3-BCART being the best. %These conclusions align with those obtained in the training data.
	
	\item When comparing frequency-severity models and sequential models, it is evident that adding $N$ (or $\hat{N}$) as a covariate improves performance, \zzz{as shown by all evaluation metrics}, i.e., $\text{F}_{\text{P}}\text{S}_{\text{G2}}$-BCART $>$ $\text{F}_{\text{P}}\text{S}_{\text{G1}}$-BCART $>$ $\text{F}_{\text{P}}\text{S}_{\text{G}}$-BCART. The same ranking is observed for another combination, i.e., $\text{F}_{\text{ZIP2}}\text{S}_{\text{Weib2}}$-BCART $>$ $\text{F}_{\text{ZIP2}}\text{S}_{\text{Weib1}}$-BCART $>$ $\text{F}_{\text{ZIP2}}\text{S}_{\text{Weib}}$-BCART. This is reasonable, as real data often \llj{exhibit some} correlation between the number of \zzz{claims} and \jj{average severity}, % (see \cite{garrido2016generalized}), 
 favouring sequential models that consider this \llj{feature directly} %\zyy{conditional} correlation 
 over frequency-severity models assuming independence.

\item In comparing frequency-severity models and joint models, %, $\text{F}_{\text{P}}\text{S}_{\text{G}}$-BCART and CPG-BCART (or ZICPG-BCART) are intuitive to examine as they use the same (or similar) distributions. 
\zzz{all evaluation metrics indicate that} the \zzz{optimal} CPG-BCART (or ZICPG-BCART) \zzz{chosen by DIC} consistently outperforms \zzz{frequency-severity models, }%$\text{F}_{\text{P}}\text{S}_{\text{G}}$-BCART,
suggesting that sharing information is beneficial for this dataset, i.e., one joint tree exhibits better performance. Exploration of the reasons is provided below. %However, for $\text{F}_{\text{ZIP2}}\text{S}_{\text{Weib}}$-BCART and CPG-BCART (or ZICPG-BCART), there is no clear intuition due to the use of different distributions. % Performance rankings, based on evaluation metrics, conclude that joint models are superior.

	\item 

As for sequential models and joint models, they address dependence in different ways. The former uses two trees, treating the number of \zzz{claims} \yyy{(or its estimate)} as a covariate in \jj{average severity} modeling to address the dependence. In contrast, the latter uses one joint tree, potentially hiding some dependence in the common variables used to split the nodes and incorporating the number of \zzz{claims} as a model weight in the aggregate claim amount distribution. %Their performance ranking is obtained through evaluation metrics. 
\yz{Joint models employing ZICPG distributions} perform better than all the sequential models \zzz{as demonstrated by all evaluation metrics}, possibly due to a small \zzz{negative} conditional correlation between the number of \zzz{claims} and \jj{average severity} ($-0.0336$) and the dataset involving a high proportion of zeros (93.19\%). %\zzz{Based on the SE and DS, joint models employing CPG models consistently outperform sequential models. This superior performance is likely because joint models not only share information but also capture hidden dependence.} %There is no unified conclusion for \yz{joint models employing CPG distributions} and sequential models (for example, $\text{F}_{\text{ZIP2}}\text{S}_{\text{Weib2}}$ is better than CPG-BCART but $\text{F}_{\text{ZIP2}}\text{S}_{\text{Weib1}}$ is worse \jj{SE, DS - CPG-BCART is always better??}). %, indicating the need for further exploration, especially for data with high dependence between the number of \zzz{claim} and \jj{average severity}.
\end{enumerate}

\begin{remark}\label{Rem:PNCor}
\llji{Our primary motivation for the joint models was to incorporate the association between the number of claims and average severity through splitting variables. To evaluate its effectiveness, we examine the conditional dependence within each terminal node of the selected joint tree from ZICPG3-BCART.
%Unlike previous studies which typically assess the correlation between the number of claims and average severity % by conditioning on positive claim counts 
%without accounting for covariates (see \cite{shi2015dependent,garrido2016generalized}), we calculate correlations within each terminal node, inherently incorporating the influence of covariates. 
%optimal trees: the claims frequency tree (ZIP2-BCART, see \cite{zhang2024bayesian}), the average severity tree (Weib2-BCART), and the joint tree (ZICPG3-BCART). 
%This leads to some interesting findings. %In the claims frequency tree (ZIP2-BCART), we observe a positive conditional correlation when the estimated claim frequency is at its highest. In the average severity tree (Weib2-BCART), all conditional correlations are negative, with the correlation strength increasing as the estimated average severity rises. 
%In the selected joint tree, w
We identify two cells of positive conditional correlation (out of a partition consisting of 5 cells), occurring when the estimated premium is at its minimum and maximum levels, respectively. 
This finding reveals the nuanced dependence between the number of claims and average severity, emphasizing the role of covariates, %The positive correlation at extreme frequency and premiums suggests segmentation effects, while the consistently negative correlation in the average severity model may reflect risk mitigation or underwriting influences. 
and provides valuable implications for insurance pricing and risk management. } 
\end{remark}

\subsubsection{Information sharing - why is it beneficial?}

\jj{From the above data analysis, we have seen the advantages of being able to share information between claim frequency and severity (through common covariates in the tree) in the joint modeling of this dataset. In this section, we will further}
investigate the superior performance of the joint models by looking at the correlation
%tree in this dataset, we discuss two aspects. The first perspective 
%focuses on the relationship 
between covariates and claim frequency and \jj{average severity}. 

The correlation coefficients are displayed in Table \ref{table_data1_cor}. 
%On the other hand, we can calculate the correlation coefficient numerically. However, 
Note that some of the covariates are categorical, for which % they cannot be used directly in our calculations. We use 
we use some transformations to replace them with numerical values for the correlation calculation; see \zzz{Supplementary Material SM.H} for further details. % categorical variables %(\textcolor{red}{if we include this, we need to provide more details, which will increase the content again...}) 
%to do this; see Table \ref{table_data1_cor}. 
For claim frequency,  variables with the strongest correlation coefficients include ``\textit{agecat}''  and ``\textit{veh\_value}'' which are used in the tree selected by the frequency BCART models in \cite{zhang2024bayesian}. For \jj{average severity}, we observe the same two variables exhibit the strongest correlation.
Furthermore,``\textit{veh\_age}'' has the third strongest correlation, but the selected average severity tree (cf. Figure \ref{Fig:data1_weib_bcart}) does not include it. We suspect this is due to a strong relationship between covariates ``\textit{veh\_age}'' and ``\textit{veh\_value}'' 
\zzz{(see Supplementary Material SM.H)}
%\yyy{(validated in Figure \ref{fig:data_age_value} in Appendix C)}, %(see Appendix C), 
and thus one of them is dropped %, which has a stronger correlation with the response variable 
to avoid multicollinearity. Additionally, both claim frequency and \jj{average severity} show the strongest correlation with \textit{``agecat''} which is the first splitting variable in the selected optimal frequency, \zzz{average} severity and joint trees, illustrating the effectiveness of BCART models for variable selection. 
\jj{The above discussion suggests that some common covariates exhibit strong relationships with both claim frequency and \jj{average severity}, it is thus beneficial to share this information using joint modeling. This validates the conclusion of Subsection \ref{sub:real-data-1}.} 
%In \cite{zhang2024bayesian} and Subsection \ref{sec:data1_sev}, we discovered that the same two variables are used in optimal \zzz{claim} frequency and \jj{average severity} trees, namely, ``\textit{agecat}'' and ``\textit{veh\_value}''. These two variables account for 2/3 of all splitting variables, indicating a high rate. Moreover, both trees choose ``\textit{agecat}'' in the first step, illustrating its importance for both \zzz{claim} frequency and \jj{average severity}.

\begin{table}[!t]  
	\centering 
 \captionsetup{width=.8\linewidth}
	\caption{Correlation coefficients between \zz{covariates (numerical ones and transformed categorical ones)} and claim frequency and average severity. Bold font indicates the \zzz{strongest} %largest 
    correlation %coefficient 
    (though generally small) in each row.}
	\begin{tabular}{l|rrcrcc}  
		\toprule   
		& \multicolumn{1}{c}{veh\_value} & veh\_age & agecat & veh\_body & gender & area \\  
		\hline   
		Claim frequency & %-0.004693165
		$-0.0047$ & %0.001279792
		$0.0013$ & %-0.01306942 
            \bm{$-0.0131$} & 
		%0.002218552
		$0.0022$ & %0.0008322709
		$0.0008$ & %-0.002136128
		$-0.0021$ \\
		%\yao{Standard deviation of Iteration times} & 168 & 187 & 177 & 161 \\
		\jj{Average severity} & %0.01346906
		$0.0135$ & %-0.005870604
		$-0.0059$ & \bm{$-0.0274$} & %-0.003498326
		$-0.0035$ & %0.0003214578
		$0.0003$ & %-0.003351879
		$-0.0034$ \\ 
		\bottomrule 
	\end{tabular}
 \label{table_data1_cor}
\end{table}

%The second perspective involves using 
Next, we consider ARI (see Section \ref{sec:ari}) % and \ref{sec:agg_share_sim})
to assess the similarity between different trees and determine whether information sharing is necessary; see Table \ref{table_data1_ari}. As discussed in Section \ref{sec:agg_share_sim}, although a specific threshold of ARI for making a direct judgment about the necessity for information sharing is unknown, it is evident that ARI values between all claim frequency and average severity trees are greater than 0.5. This suggests that significant intrinsic similarities of these models cannot be ignored. This also validates the preference of adopting a joint model from Subsection \ref{sub:real-data-1}.
%Therefore, based on these two aspects discussed, \yz{we conclude that one joint tree performs better than two trees (in frequency-severity models) in this dataset.}

\begin{table}[!t] 
	\centering
 \captionsetup{width=.8\linewidth}
	\caption{Values of ARI between different trees. The number in  brackets after the abbreviation of the model indicates the number of terminal nodes.}
	\begin{tabular}{@{}l|p{1.55cm}p{1.55cm}p{1.55cm}p{1.55cm}p{1.55cm}p{1.55cm}}  
		\toprule   
		& Poisson-BCART\,(5) & ZIP2-BCART\,(5) & Gamma-BCART\,(4) & Weib-BCART\,(4) & 
                 CPG-BCART\,(5) & ZICPG3-BCART\,(5) \\  
		\hline   
		Poisson-BCART\,(5) & 1 & 0.7396 & 0.5398 & 0.5163 & 0.8585 & 0.7823  \\
		ZIP2-BCART\,(5) & - & 1 & 0.5599 & 0.5179 &  0.8587 &  0.8152 \\
		Gamma-BCART\,(4) & - & - & 1 & 0.7430 &  0.6921 & 0.6423 \\
		Weib-BCART\,(4) & - & - & - & 1 & 0.6491 & 0.6022 \\
		CPG-BCART\,(5) & - & - & - & - & 1 & 0.6351 \\
		ZICPG3-BCART\,(5) & - & - & - & - & - & 1\\
		
		\bottomrule  
		
	\end{tabular}

	\label{table_data1_ari}
\end{table}

After conducting a thorough analysis of this dataset, we suggest that insurers need to pay more attention to policyholders who are younger and have vehicles with higher values since they are more likely to have higher risks. %In addition, policyholders with specific vehicle body types (such as ``\textit{COUPE}'' and ``\textit{BUS}''), also require additional attention. 

\begin{remark}

We \ga{have also applied} the proposed BCART models to other datasets (\textit{dataOhlsson} included in the library \texttt{insuranceData} in \textsf{R}, \textit{freMTPL2freq} and \textit{freMTPL2sev} included in the library \texttt{CASdatasets}; see more details in \cite{charpentier2014computational}). Due to the similarity in analysis methods and the consistency of conclusions, we omit the details.

\end{remark}

\section{Summary}

%\jj{Note - We may not want to reveal all the ideas to others. Maybe call this section ``Summary" and only keep the first paragraph. This can be discussed later together with the Abstract and Introduction.}

This work develops BCART models \jjj{for} insurance pricing, building upon the foundation of previous \zzz{claim} frequency analysis (see \cite{zhang2024bayesian}).
%These tree-based models can automatically perform variable selection and detect non-linear effects and possible interactions among explanatory variables. The obtained optimal trees are relatively accurate, stable and straightforward to interpret by a visualization of the tree structure. These are desirable aspects of insurance pricing. %It is also shown that these models can obtain tree structures that can characterize the  
In particular, for \jj{average severity}, we incorporated BCART models \jjj{with} gamma, lognormal, and Weibull distributions, which have different abilities to handle data with varying tail characteristics. \yz{We found that the Weib-BCART performs better than \zzz{Gamma-BCART or LN-BCART} since it can deal with cases where some \zzz{cell}s have lighter tails, while others have heavier tails. Besides, in the comparison between Gamma-BCART and LN-BCART, the former is preferable for data with a lighter tail and the latter is more suitable for data with a heavier tail.} This finding provides us with a practical strategy for choosing models. Concerning aggregate \zzz{claim} modeling, we proposed three types of models. First, we found that the sequential models treating the number of \zzz{claims} \yyy{(or its estimate)} as a covariate  in the \jj{average severity} modeling perform better than the standard frequency-severity models when the underlying true dependence between the number of \zzz{claims} and \jj{average severity} is stronger. \yz{Second, we explored the choice between using two trees or one joint tree. In particular,} when there are indeed common covariates affecting %or similar splitting rules between 
\zzz{claim} frequency and \jj{average severity and there is a relatively large amount of data available, it may be beneficial to use one joint tree}, which supports the conclusion in \cite{linero2020semiparametric} by illuminating the benefits of information sharing. Third, \zz{we provided details of evaluation metrics in the case of two trees and proposed the use of ARI to quantify the similarity between two trees}, which can assist in explaining the necessity of information sharing. Finally, in the analysis of various joint models, especially the three ZICPG models employing different ways to embed exposure, we found that ZICPG3-BCART, which embeds exposure in both the zero mass component and Poisson component, delivers the most favourable results in real insurance data. We address their similarities to the analysis of ZIP models discussed in \cite{zhang2024bayesian}. Furthermore, we introduced %another new type of DIC and 
a more general MCMC algorithm for BCART models. These enhancements extend the applicability of BCART models to a broader range of applications. \zyy{Although this paper does not fully resolve the challenge %of explicitly modeling the 
of dependence modeling between the number of claims and average severity, it highlights the importance and a need for further research, with copula-based approaches emerging as a promising direction. %; see, e.g., \cite{shi2020regression}. %In particular, copula-based approaches could be explored as a promising direction for explicitly capturing this dependence structure; see, e.g., \cite{shi2020regression}.
}

\bigskip

\noindent{\bf Acknowledgement}:
\zyy{We are thankful to the anonymous referees for their constructive suggestions which have led to a significant improvement of the manuscript.}

\medskip

\noindent{\bf Supplementary material}:
\zyy{Supplementary material related to this article can be found online.}
\medskip

\noindent{\bf Declaration of competing interest:} \zyy{There is no competing interests.}

\medskip
\noindent{\bf Declaration of generative AI and AI-assisted technologies in the writing process:} \zyy{There is no use of generative AI and AI-assisted technologies in the writing process.}

\printbibliography

\end{document}